\documentclass[longauth]{aa} 
\usepackage[colorlinks=true,linkcolor=blue,citecolor=blue]{hyperref}

\usepackage{graphicx}
\usepackage{multirow}
\usepackage{tabularx}
\usepackage{longtable,tabu}
\usepackage{lscape}
\usepackage{subfigure}
\usepackage{longtable}
\usepackage{pdflscape}
\usepackage{placeins}
\usepackage{capt-of}

\def\sun{\odot}
\bibpunct{(}{)}{;}{a}{}{,} 
\usepackage{multirow}

\usepackage{amssymb}
\usepackage{pifont}
\usepackage{url}
\usepackage{threeparttable}
\usepackage{lipsum,multicol}
\usepackage{natbib}
\usepackage[varg]{txfonts}
\def \kms{km\,s$^{-1}$}
\def \ms{m\,s$^{-1}$}
\def \cms{cm\,s$^{-2}$}

\def \mjup{M$_\mathrm{jup}$}
\def \rjup{R$_\mathrm{jup}$}
\def \Hei{\ion{He}{I}\,\,}
\def\sun{\odot}
\def\logrhk{\ensuremath{\log \mathrm{R}^{\prime}_{\mathrm{HK}}}}
\newcommand{\cmark}{\ding{51}}%
\usepackage{verbatim}

\begin{document}
	
	\title{The GAPS Programme at TNG}
	\subtitle{LIV. A \Hei survey of close-in giant planets hosted by M-K dwarf stars with GIANO-B\thanks{Based on observations made with the Italian Telescopio Nazionale Galileo (TNG) operated on the island of La Palma by the Fundacion Galileo Galilei of the INAF at the Spanish Observatorio Roque de los Muchachos of the IAC in the frame of the programme Global Architecture of the Planetary Systems (GAPS).} }
	
	\titlerunning{GIANO-B \Hei survey: the M-K dwarf sample}
	
	\authorrunning{Guilluy et al.}

	\author{G. Guilluy \inst{1}
		\and 
		M. C. D'Arpa \inst{2, 3}
		\and 
		A.~S. Bonomo \inst{1}
		\and
		R.~Spinelli \inst{2}
		\and
		F.~Biassoni \inst{4,5}
		\and
		L.~Fossati \inst{6}
		\and 
		A.~Maggio \inst{2}
		\and 
		P.~Giacobbe\inst{1} 
		\and
		A.~F.~Lanza\inst{7}
		\and
		A.~Sozzetti\inst{1} 
		\and
		F.~Borsa\inst{5} 
		\and
		M.~Rainer\inst{5} 
		\and
		G.~Micela\inst{2}
		\and
		L.~Affer \inst{2} 
		\and
		G.~Andreuzzi \inst{8,9}
		\and
		A.~Bignamini \inst{10}
		\and
		W.~Boschin \inst{8,11,12}
		\and
		I.~Carleo \inst{11,12}
		\and
		M.~Cecconi  \inst{8}
		\and
		S.~Desidera \inst{13}
		\and
		V.~Fardella \inst{2, 3}
		\and
		A.~Ghedina \inst{8}
		\and
		G.~Mantovan \inst{13,14}
		\and
		L.~Mancini \inst{1,15,16} 
		\and
		V.~Nascimbeni \inst{13}
		\and
		C.~Knapic \inst{10}
		\and
		M.~Pedani  \inst{8}
		\and
		A.~Petralia \inst{2}
		\and
		L.~Pino \inst{17}
		\and
		G.~Scandariato \inst{7}
		\and
		D.~Sicilia \inst{7}
		\and
		M.~Stangret \inst{13}
		\and
		T.~Zingales \inst{13,14}
	}

	\institute{
		INAF -- Osservatorio Astrofisico di Torino, Via Osservatorio 20, 10025, Pino Torinese, Italy 
		\and
		INAF -- Osservatorio Astronomico di Palermo, Piazza del Parlamento, 1, I-90134 Palermo, Italy 
		\and
		University of Palermo, Department of Physics and Chemistry “Emilio Segrè, Via Archirafi 36, Palermo, Italy  
		\and
		DISAT, Università degli Studi dell’Insubria, via Valleggio 11, I-22100 Como, Italy 
		\and
		INAF – Osservatorio Astronomico di Brera, Via E. Bianchi 46, 23807 Merate, Italy 
		\and
		Space Research Institute, Austrian Academy of Sciences, Schmiedlstrasse 6, 8042 Graz, Austria 
		\and
		INAF -- Osservatorio Astrofisico di Catania, Via S. Sofia 78, I-95123, Catania, Italy 
		\and
		Fundación Galileo Galilei-INAF, Rambla José Ana Fernandez Pérez 7, 38712 Breña Baja, TF, Spain 
		\and
		INAF - Osservatorio Astronomico di Roma, Via Frascati 33, 00078 Monte Porzio Catone, Italy 
		\and
		INAF – Osservatorio Astronomico di Trieste, via Tiepolo 11, 34143 Trieste 
		\and
		Instituto de Astrofísica de Canarias (IAC), Calle Vía Láctea s/n, 38200, La Laguna, Tenerife, Spain  
		\and
		Departamento de Astrofísica, Universidad de La Laguna (ULL), 38206 La Laguna, Tenerife, Spain 
		\and
		INAF – Osservatorio Astronomico di Padova, Vicolo dell'Osservatorio 5, 35122, Padova, Italy 
		\and
		Dipartimento di Fisica e Astronomia ``Galileo Galilei'', Università di Padova, Vicolo dell'Osservatorio 3, IT-35122, Padova, Italy 
		\and
		Department of Physics, University of Rome ``Tor Vergata'', Via della Ricerca Scientifica 1, 00133, Roma, Italy 
		\and
		Max Planck Institute for Astronomy, Königstuhl 17, 69117, Heidelberg, Germany 
		\and
		INAF -- Osservatorio Astrofisico di Arcetri, Largo E. Fermi 5, 50125, Firenze, Italy 
	}

	\date{Received date ; Accepted date }

	\abstract
	{Atmospheric escape plays a fundamental role in shaping the properties of exoplanets. The metastable near-infrared (nIR) helium triplet at 1083.3 nm (\ion{He}{I}) is a powerful proxy of extended and evaporating atmospheres.}
	{We used the GIARPS (GIANO-B + HARPS-N) observing mode of the Telescopio Nazionale Galileo to search for \Hei absorption in the upper atmospheres of five close-in giant planets hosted by the K and M dwarf stars of our sample, namely WASP-69\,b, WASP-107\,b, HAT-P-11\,b, GJ\,436\,b, and GJ\,3470\,b. }
	{We focused our analysis on the nIR \Hei triplet,  performing high-resolution transmission spectroscopy by comparing the in-transit and out-of-transit observations. In instances where nightly variability in the \Hei absorption signal was identified, we investigated the potential influence of stellar magnetic activity on the planetary absorption signal by searching for variations in the
		H$\alpha$ transmission spectrum.}
	{We spectrally resolve the \Hei triplet and confirm the published detections for WASP-69\,b (3.91$\pm$0.22\%, 17.6$\sigma$), WASP-107\,b (8.17$^{+ 0.80 }_{ -0.76 }$\%, 10.5$\sigma$), HAT-P-11\,b (1.36$\pm$0.17\%, 8.0$\sigma$), and GJ\,3470\,b (1.75$^{+ 0.39 }_{ -0.36 }$\%, 4.7$\sigma$). We do not find evidence of extra absorption for GJ\,436\,b.   We observe night-to-night variations in the \Hei absorption signal for WASP-69\,b, associated with variability in H$\alpha$, which likely indicates the influence of pseudo-signals related to stellar activity. Additionally, we find that the \Hei signal of GJ\,3470\,b originates from a single transit observation, thereby corroborating the discrepancies found in the existing literature. An inspection of the H$\alpha$ line reveals an absorption signal during the same transit event.}
	{By combining our findings with previous analyses of GIANO-B \Hei measurements of planets orbiting K dwarfs, we explore potential trends with planetary and stellar parameters that are thought to affect the absorption of metastable \ion{He}{I}. Our analysis is unable to identify clear patterns, thus emphasising the necessity for additional measurements and the exploration of potential additional parameters that may be important in controlling \Hei absorption in planetary upper atmospheres.}
	\keywords{planets and satellites: atmospheres – techniques: spectroscopic - methods: observational - infrared: planetary systems}

	\maketitle
	\section{Introduction}\label{sec:intro}
	
	The atmospheres of exoplanets orbiting close to their host stars can be significantly influenced by stellar irradiation,        which can cause the upper atmospheric layers to expand and, in some cases, even evaporate \citep[e.g.][]{Bourrier2018_lib}. Atmospheric loss due to photo-evaporation could have played a roll in shaping the close-in exoplanets' demographics and could be responsible for both the Neptunian Desert \citep[i.e. the dearth of planets with sizes between terrestrial and Jovian close to host stars;][]{Lecavelier2007, Davis2009, Szabo2011, Beauge2013} and the radius gap, which separates the dense super-Earths from the larger
	and puffier mini-Neptunes \citep{Fulton2017, Fulton2018}.
	Pioneering investigations of atmospheric escape were carried out two decades ago \citep[e.g.][]{Vidal-Madjar2003, Vidal-Madjar2004}. These studies focused on the Ly-$\alpha$ transition of hydrogen, which is the dominant element in hot gas giant atmospheres. The atomic hydrogen resulting 
	from the thermal dissociation of H$_2$ absorbs the stellar X-ray and extreme ultraviolet (EUV; together XUV) radiation in the thermosphere, significantly increasing the local temperature; this then can lead to atmospheric expansion and escape. However, Ly-$\alpha$ observations are contaminated by both interstellar medium (ISM) absorption and geocoronal emission, which pose challenges to the interpretations of the observations. 
	
	\citet{Seager2000} and \citet{Oklop2018} identified the \Hei 2$^{3}$S triplet at 1083.3~nm (referred to as \ion{He}{I}) as a robust diagnostic for studying atmospheric expansion and possibly mass loss. This tracer resides in a region of the near-infrared (nIR) spectrum affected by neither ISM nor geocoronal emissions. Since the groundbreaking discovery of an extended atmosphere surrounding the super-Neptune WASP-107\,b \citep{Spake2018}, the field has made tremendous strides, with nearly 40 planets having been searched for the presence of \ion{He}{I} to date  \citep{DosSantos2023}. 
	However, despite this large sample, the underlying factors responsible for triggering the detection or non-detection of \Hei remain unclear. 
	Being aware of the potential limitations arising from differences in instrumentation and data analysis techniques that may affect the identification of trends related to \Hei detection, we decided to undertake a uniform survey, searching for \Hei in the atmospheres of all the exoplanets available within the atmospheric sample of the Global Architecture of Planetary Systems (GAPS) project \citep[we described our GAPS-ATMO sample in][]{Guilluy2022}, which includes hot and warm Jupiter- and Neptune-like planets.
	
	We divided the GAPS targets into two distinct subsamples. The first sample comprises planets orbiting M-K dwarf planet hosts, which we analyse in this paper. The second dataset, consisting of planets around G-F-A main-sequence stars, will be the subject of a future study (Guilluy et al., in prep). We decided to focus this first paper on M-K dwarf stars as they are the best suited to host planets with an escaping \ion{He}{I} atmosphere. Indeed, their high XUV flux ionises the \ion{He}{I} ground state, which, recombining, populates the metastable 2${^3}$S ground level (from which the 1083.3~nm absorption originates), and their moderately low mid-UV flux prevents the 2${^3}$S ionisation \citep[e.g.][]{Oklop2019, Biassoni2023}.
	
	Past observations have revealed that stellar activity can contaminate \Hei measurements, leading to inaccurate estimations of  \Hei absorption and the mass-loss rate \citep[e.g.][]{Salz2018, Guilluy2020}. 
	Specifically, the plage-like active regions are darker than the rest of the stellar disc in the helium channel; a planet transiting over quiescent regions of the star would enhance the contribution of active regions to the observed flux, resulting in a stronger absorption at 1083.3 nm. Conversely, if the planet obscures an active region, the net effect would be a reduction in \Hei absorption. One approach to addressing these pseudo-signals is to simultaneously monitor stellar activity diagnostics in the optical (such as H$\alpha$, \ion{Ca}{ii} H\&K, and \ion{Na}{i}) that exhibit an opposite behaviour compared to \Hei \citep{Guilluy2020}.
	
\begin{figure}
		\centering
		\includegraphics[width=\linewidth]{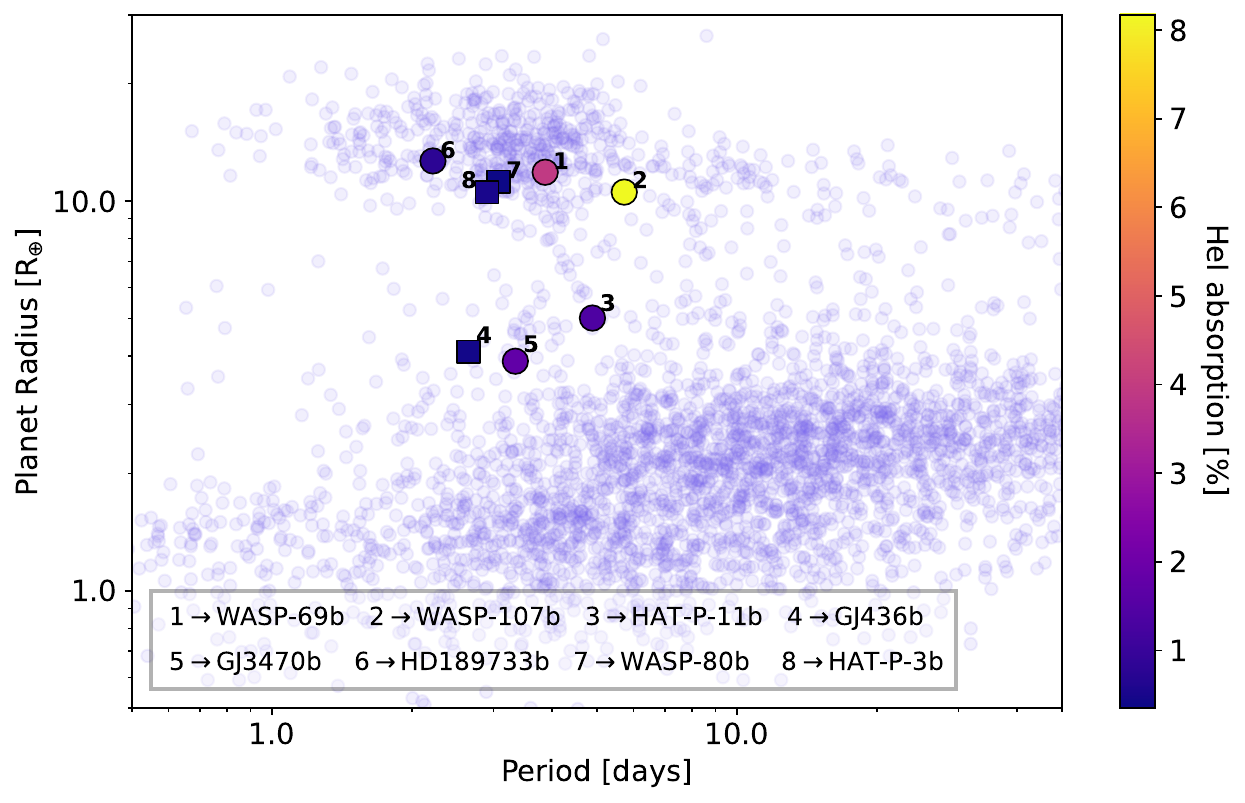}
		\caption{Known exoplanets as a function of their radius and period from the NASA Exoplanet Archive \citep{Akeson2013}. Targets analysed in this work and previous GIANO-B papers \citep{Guilluy2020, Fossati2022, Guilluy2023} are highlighted. The marker colour reflects the \Hei absorption signal we found in this work. Circles and squares represent detections and non-detections, respectively. The non-detections are reported at the 1$\sigma$ level}.
		\label{fig:my_label}
	\end{figure}

	In this paper we present an investigation of the upper atmosphere of WASP-69\,b, WASP-107\,b, HAT-P-11\,b, GJ\,436\,b, and GJ\,3470\,b (see Fig.~\ref{fig:my_label}) using high spectral resolution observations in both the nIR with GIANO-B\citep{Oliva2006} and in the optical using the High Accuracy Radial velocity Planet Searcher for the Northern hemisphere (HARPS-N; \citealt{Cosentino2012}) spectrographs.
	
	The paper is organised as follows. We describe the sample in Sect.~\ref{sample} and the observations in Sect.~\ref{observations}. We detail the data reduction procedures
	in Sect.~\ref{data_analysis}. We highlight our findings and place our results in the context of previous works in Sect.~\ref{results}. We finally draw our conclusions in Sect.~\ref{conclusions}.
	\section{Case history}\label{sample}
	Our survey encompasses five targets, namely WASP-69\,b, WASP-107\,b, HAT-P-11\,b, GJ\,436\,b, and GJ\,3470\,b. In this section, we provide a brief overview of the key characteristics of each system and summarise the existing \Hei studies presented in the literature. The stellar and planetary parameters we adopted in the analysis are reported in Table~\ref{tab_par}.
	
	WASP-69\,b is a warm Saturn orbiting a K5 star \citep{Anderson2014}. Due to its large expected atmospheric signal 2R$_\mathrm{P}$H$_\mathrm{eq}$/R$_\star^2 \sim$ 283~ppm (where $H_\mathrm{eq}$ is the atmospheric scale height, and R$_\mathrm{P}$ and R$_\star$ are the planetary and stellar radius, respectively, \citealt{Brown2001}),  WASP-69\,b represents a prominent target for performing atmospheric studies. \citet{Tsiaras2018} and \citet{Estrela2021} found clues to the presence of water in the planet's atmosphere. 
	We recently reported the presence of five molecules, and a possible hint of disequilibrium in its lower atmosphere \citep{Guilluy2022}.  Sodium was also detected in its upper atmosphere (see e.g. \citealt{Casasayas-Barris2017,Khalafinejad2021,Langeveld2022}). Evidence of an extended helium atmosphere surrounding the planet has been reported by \citet{Nortmann2018} with CARMENES, by \citet{Vissapragada2020,Vissapragada2022} with Palomar/WIRC, by \citet{Allart2023} with SPIRou, and by \citet{Tyler2023} with  Keck/NIRSPEC.
	
	WASP-107\,b is a super Neptune orbiting a K6 star located at the upper radius border of the Neptunian desert \citep{Anderson2017}.
	Because of its large atmospheric scale height, it represents an excellent target for atmospheric studies. \citet{Kreidberg2018} detected water using the Wide Field Camera 3 (WFC3) of the \textit{Hubble} Space Telescope (HST), finding evidence for a methane-depleted atmosphere and high-altitude condensates.   \citet{Spake2018} also detected water using a wider wavelength coverage than \citet{Kreidberg2018}. 
	Recently, photochemically produced sulphur dioxide (SO$_2$) was detected in its atmosphere \citep[][]{Dyrek2023}. Detections of the extended helium atmosphere of WASP-107\,b have been reported multiple times \citep[e.g.][]{Spake2018, Allart2018, Kirk2020,Spake2021}.
	
	HAT-P-11\,b is a transiting warm Neptune-class exoplanet orbiting a K4V star, and located at the edge of the evaporation desert \citet{bakos2010}.
	\citet{Fraine2014}, \citet{Tsiaras2018}, and \citet{Chachan2019} detected the presence of water vapour in the atmosphere of HAT-P-11 b with low-resolution observations by using data from  HST and \textit{Spitzer}. \citet{Chachan2019} also suggested the presence of methane. More recently, at high-spectral resolution \citet{Basilicata2023} reported the detection of H$_2$O, NH$_3$,  and a tentative one of CO$_2$ and CH$_4$. \citet{Jaffel2022} by studying the upper atmosphere found a phase-extended transit absorption of neutral hydrogen and singly ionised carbon, while several works (i.e. \citealt{Allart2018}, \citealt{Mansfield2018}, and \citealt{Allart2023}) reported the presence of metastable helium surrounding the planet.
	
	GJ\,436\,b is a warm Neptune in the lower-mass edge of the Neptune desert orbiting a quiet M2.5V star \citet{Gillon2007}. Its atmosphere has been extensively studied through various observations. According to low-resolution studies, there appears to be a scarcity of methane and a surplus of CO and CO$_2$ in the atmosphere, along with a slight deficiency of H$_2$O when compared to the predicted amounts based on equilibrium chemistry assuming solar metallicity \citep[e.g.][]{Stevenson2010, Knutson2014}.
	Observations of the upper atmosphere of GJ\,436\,b n Ly-$\alpha$ line of neutral hydrogen revealed that the planet is surrounded by a giant coma of \ion{H}{I} extending tens of planetary radii \citep[e.g.][]{Kulow2014, Ehrenreich2015, Lavie2017}, while \citet{Nortmann2018} did not detect any evidence of a helium-extended atmosphere.
	
	GJ\,3470\,b is a warm Neptune orbiting an M-dwarf and located very close to the Neptunian desert \citet{Bonfils2012}.  Previous investigations of its atmosphere, based on observations from the  HST, have indicated the presence of a cloudy, hydrogen-rich atmosphere \citep{Ehrenreich2014}. Additionally, studies analysing visible range observations by \citet{Nascimbeni2013} and \citet{Chen2017} have detected a Rayleigh slope in the atmosphere. \citet{Benneke2019} obtained a robust detection of water absorption (>5$\sigma$) by combining  HST and \textit{Spitzer} observations.  They revealed a low-metallicity, hydrogen-dominated atmosphere similar to a gas giant, but strongly depleted in methane gas. In terms of upper atmospheric layers,  \citet{Bourrier2018} have reported the existence of a giant hydrogen exosphere while \citet{Palle2020} and \citet{Ninan2020} have identified evidence of \Hei absorption during transit. On the other hand, \citet{Allart2023} have reported a non-detection of \Hei in this planet’s atmosphere.
	

	\section{Observations}\label{observations}

	\begin{table*}
		\caption{Observations log.}
		\centering
		\begin{tabular}{c |c | c | c  | c |c|c }
			\hline \hline
			&       \textbf{Date{$^+$}} & \textbf{N$_{\mathrm{obs}}$} & \textbf{Exp Time [s]} & \textbf{S/N$_{\mathrm{avg}}$} & \textbf{$\sigma_\mathrm{T_\mathrm{c}}$ [min]} & \textbf{Telluric/stellar flag}\\                       
			\hline
			\multirow{4}{*}{\textbf{WASP-69\,b}}  & 24 July 2019 &  60 & 200.0 &  53.42  & 0.62 & \cmark \\
			& 09 August 2020 &  54 & 200.0 &  54.39  &  0.72 & \\
			& 28 October 2021&  46 & 200.0 &  41.90  &  0.85  & \cmark \\
			& 14 September 2022 &  42 & 200.0 &  51.88  &  0.95  & \\
			\hline 
			\multirow{2}{*}{\textbf{WASP-107\,b}} & 07 February 2019 &  58 &  200.0 &  28.21 & 0.24 & \\
			& 04 May 2019  &  60 &  200.0 &  24.55 & 0.26& \cmark \\
			\hline
			\multirow{3}{*}{\textbf{HAT-P-11\,b}}   & 07 July 2019 &  60  & 200.0  & 59.31 & 0.15 & \\
			&18 June 2020 &  60  & 200.0  & 65.22 & 0.16 & \\
			& 19 September 2020 &  58  & 200.0  & 60.79 & 0.17 & \\
			\hline
			\multirow{5}{*}{\textbf{GJ-436b}} & 16 April 2018 &  44  & 200.0 & 67.34 & 0.17 & \\
			&  19 February 2020 &  38  & 200.0 & 87.39  & 0.20 & \\
			&   27 February 2020 &  38  & 200.0 & 93.26  & 0.20 & \\
			&03 March2023 &  30  & 200.0 &  78.45 & 0.24 & \\
			&11 March2023 &  46  & 200.0 &  90.75  & 0.24 & \\
			&24 December 2022&  42  & 200.0 &  81.6 1& 0.25 &  \\
			\hline
			\multirow{5}{*}{\textbf{GJ-3470b}} & 13 January 2018 &  38 &  200.0 &  20.16 & 0.27 & \cmark \\
			&04 February 2019 &  52 &  200.0 &  21.41 & 0.32 &\cmark\\
			&28 December 2019 &  64 &  200.0 &  31.23 & 0.36 & \\
			&27 January 2020 &  48 &  200.0 &  36.26 & 0.36 & \cmark \\
			&23 December 2022 &  54 &  200.0 &  32.59 & 0.49 & \\
			\hline
		\end{tabular}
		\tablefoot{From left to right: the observing night, the number of observed spectra, the exposure time, the average S/N across the selected spectral range (1082.49-1085.5\,nm), the uncertainty on the mid-transit time for the observed transits calculated as $\sigma_\mathrm{T_\mathrm{c}}=\sqrt{\sigma_\mathrm{T_\mathrm{0}}^2+n^2 * \sigma_\mathrm{P}^2}$ (where ${T_\mathrm{0}}$ and $P$ are the mid-transit time and orbital period in the adopted ephemerids reported in Table~\ref{tab_par} and $n$ is the number of orbits between times ${T_\mathrm{c}}$  and ${T_\mathrm{0}}$), and flag for significant telluric overlap/possible stellar activity issues.  [$^+$]Beginning of the night. }
		\label{log}
	\end{table*}
	
	The systems in our sample were observed using the GIARPS observing mode \citep{GIARPS_claudi} of the Telescopio Nazionale Galileo (TNG). In this configuration, the TNG is capable of simultaneously acquiring high-resolution (HR) spectra in the optical range (0.39-0.69~$\mu$m) and nIR range (0.95-2.45~$\mu$m) using the HARPS-N ($R \approx 115,000$) and GIANO-B ($R \approx 50,000$) spectrographs.
	
	For the GIANO-B observations, we employed an ABAB nodding pattern, which allows for optimal subtraction of thermal background noise and telluric emission lines. Each target was scheduled for observations $\sim$1~h before, during, and $\sim$1~h after the planetary transit. Figure~\ref{snr_ph_am} displays the signal-to-noise ratio (S/N) averaged over the region of interest (1082.49-1085.5 nm) and the airmass as a function of the planet's orbital phase for each night and target considered. GIANO-B covers four spectral bands in the nIR (Y, J, H, and K) divided into 50 orders. For our analysis, we focused on order~\#39 in the Y-band, where the \Hei triplet is located.
	One transit observation of WASP-69\,b on UT 28 October 2021, was affected by thin clouds (cirri), prompting us to discard the AB pairs of observations with very low S/N compared to the others (lowest S/N in the pair less than 15, as shown in Fig. 2 of \citealt{Guilluy2022}). Furthermore, we discarded the last transit of WASP-107\,b, which took place on UT 09 April 2023, as the entire night was affected by seeing conditions of approximately 2 arcseconds. This resulted in a lower S/N compared to the data from the other observing nights (see Fig.~\ref{snr_ph_am}), thereby preventing us from analysing this dataset. 
	A detailed log of the GIANO-B observations is provided in Table~\ref{log}.
	
	The HARPS-N observations were carried out using the objAB observational setup, with fibre A on the target and fibre B on the sky.  The light collected through the fibres is directed to a 4k $\times$ 4k charge-coupled device (CCD). The CCD is responsible for capturing the spectra from 69 different orders for each fibre, utilising the echelle spectrograph design. To process the HARPS-N data, the standard Data Reduction Software (DRS) was employed, specifically version 3.7 \citep{Cosentino2012}.


	\section{Data analysis}\label{data_analysis}
	In this section, we discuss the main steps of the analysis we performed on both GIANO-B (Sect.~\ref{data_analysis_nir}) and HARPS-N data (Sect.~\ref{data_analysis_opt}).
	
	\subsection{Analysis in the nIR }\label{data_analysis_nir}
	Currently, the most widely used technique for determining whether an exoplanet is surrounded by an extended or evaporating atmosphere is transmission spectroscopy. During a transit, the outgassed atoms produce additional absorption features superimposed on the stellar spectrum. Here, we outline the procedures we implemented to extract the planetary transmission spectra from the raw data obtained by GIANO-B.
	
	\begin{figure}
		\centering
		\includegraphics[width=0.47\textwidth]{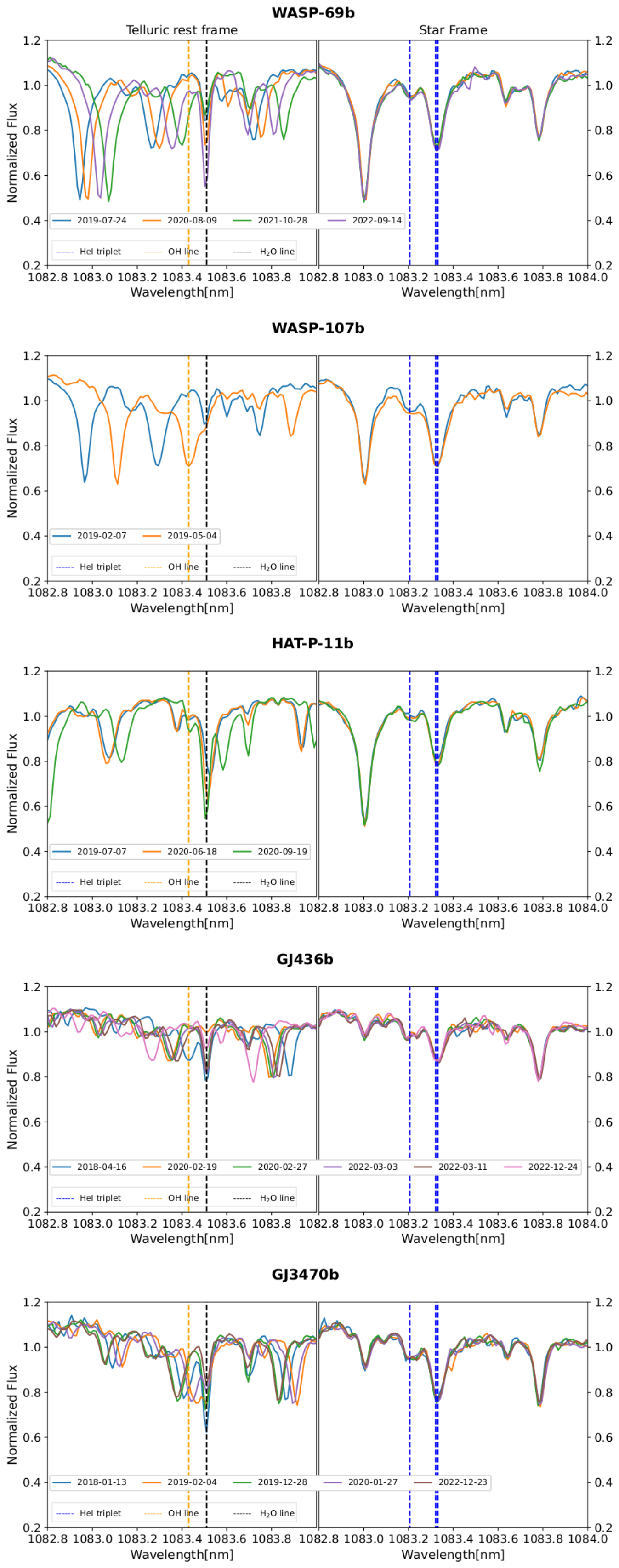}\\
		\caption{Time-averaged spectra before and after the telluric lines removal. Left Panels: Averaged spectrum for each investigated night in the telluric rest frame, with the position of the H$_2$O telluric transmission line at $\sim$1083.51\,nm and the OH telluric emission line at $\sim$1083.43~nm overplotted in black and orange, respectively.
			Right Panels: Averaged spectrum for each investigated night after the telluric removal and the shift in the stellar rest frame. Vertical blue lines correspond to the \Hei triplet. The averaged spectra are plotted after being divided by their median value for visualisation purposes. }
		\label{rest_frame}
	\end{figure}
	\begin{figure*}
		\includegraphics[width=\linewidth]{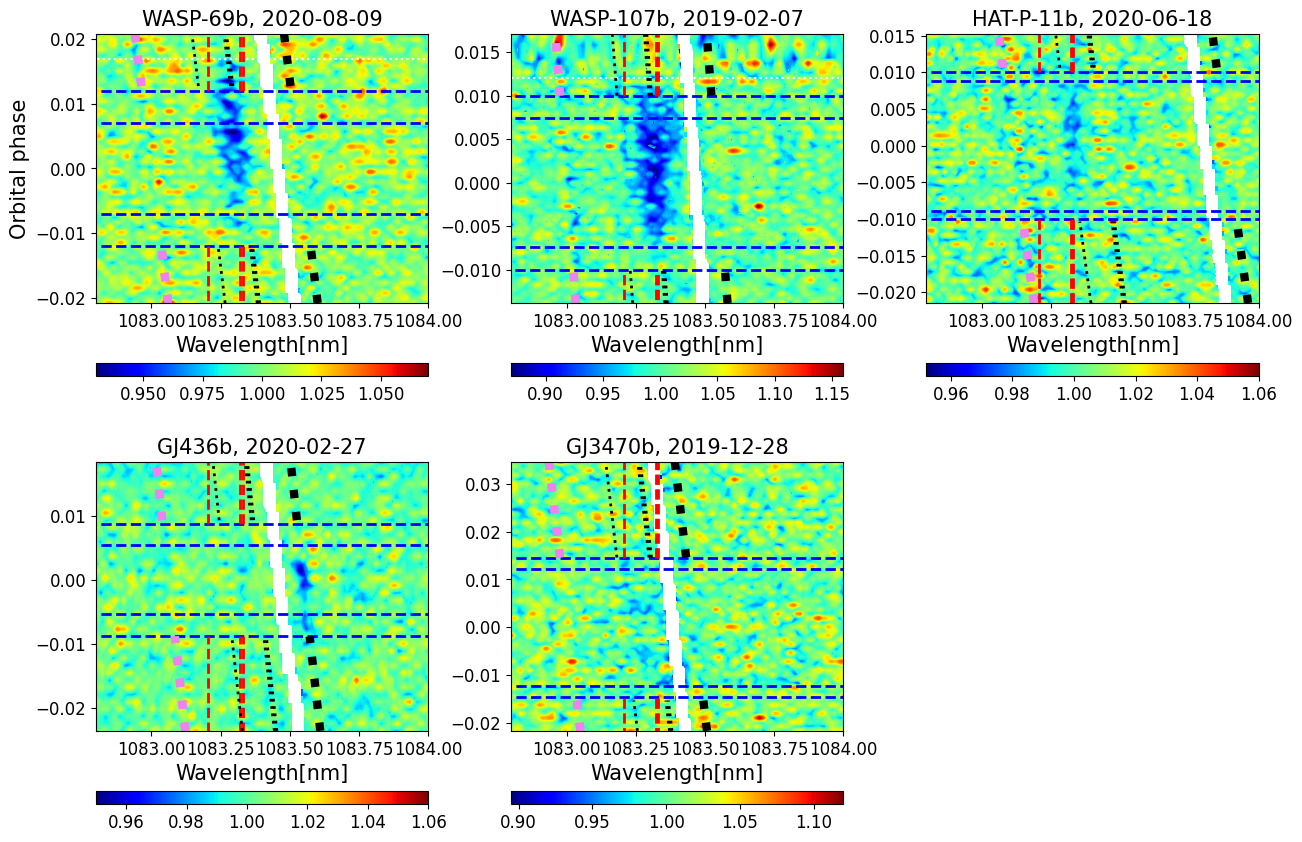}
		\caption{2D transmission spectra maps. For each target, an example of transmission spectra is shown in tomography in the planetary rest frame in the region of the \Hei triplet, as a function of wavelength and planetary orbital phase. The contact points t1, t2, t3, and t4 are marked with
			horizontal blue lines. The regions affected by OH contamination are masked.  For some planets, some residuals are left at the position of the Si$\sim$1083~nm line (highlighted in pink). This is due to the depth of the line (see e.g. \citealt{Krishnamurthy2023, Zhang2023}). Dotted black and red lines mark the position of the \Hei lines in the stellar and planet rest frame, respectively.  Lines with black squares mark the position of the strong H$_2$O telluric line at around 1083.51~nm. The corresponding 2D maps for all the investigated nights are shown in Fig~\ref{MAPS_app}.}
		\label{MAPS}
	\end{figure*}
	
	\begin{figure*}[h!]
		\centering
		\includegraphics[height=20cm]{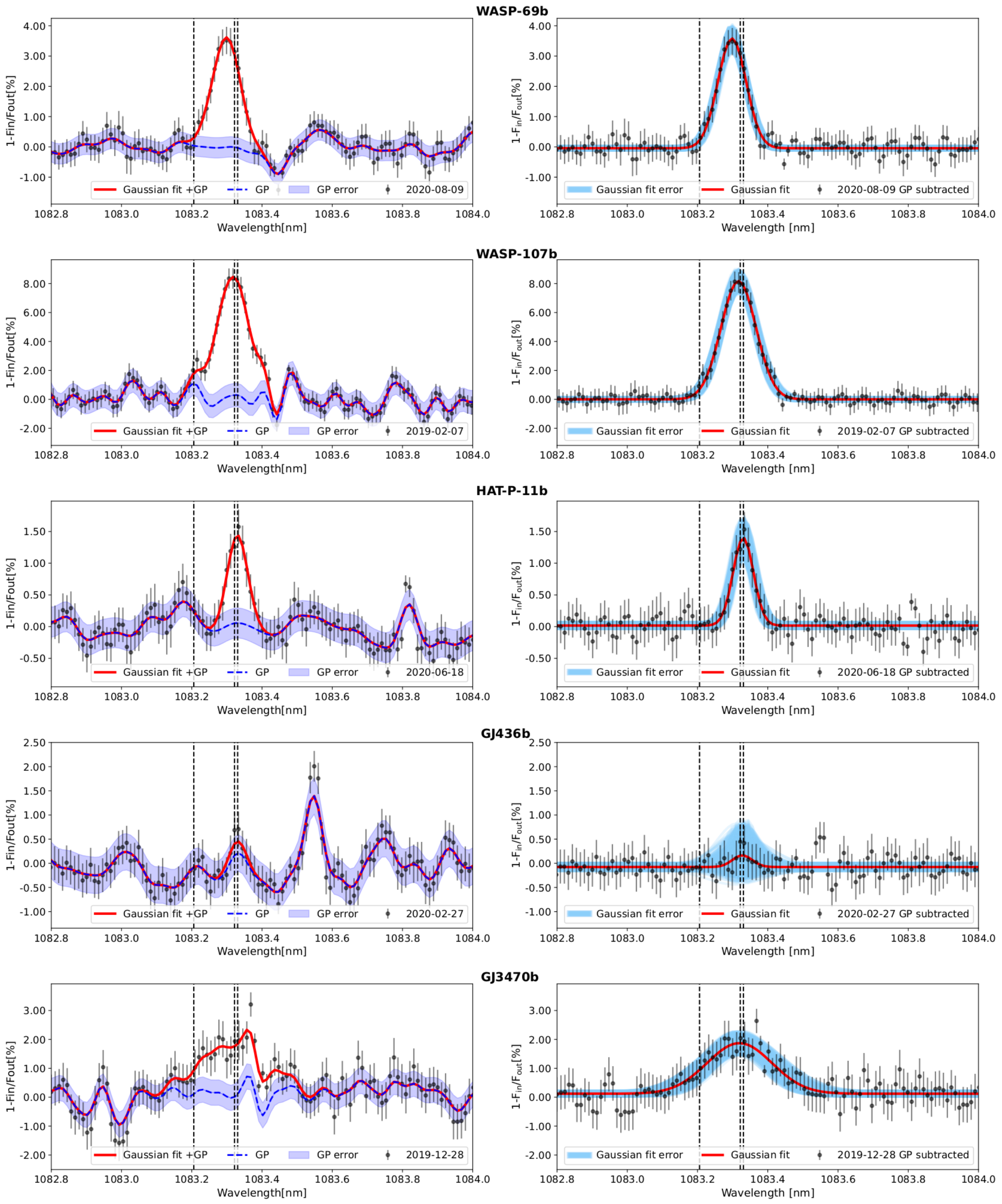}
		\caption{GP correction. For each target, an example of GP correction is shown. Left panel: Transmission spectrum centred on the \ion{He}{I} triplet (in the planet rest frame) with overplotted the GP regression model, along with the 1$\sigma$ uncertainty intervals, (in blue) and the Gaussian+GP model (in red). Right Panel: Final transmission spectrum after removing the GP model. Vertical black dotted lines indicate the position of the \ion{He}{I} triplet.  The spike in the right wing of the \Hei triplet in the GJ\,3470\,b transmission spectrum is a residual due to the OH emission line. The error intervals for the Gaussian fit were computed by displaying 1000 Gaussian fits within the 1$\sigma$ uncertainties of the derived parameters, spanning the 16\%-84\% quantiles.} 
		\label{GP}
	\end{figure*}
	
	\begin{figure}[h!]
		\centering
		\includegraphics[height=19cm]{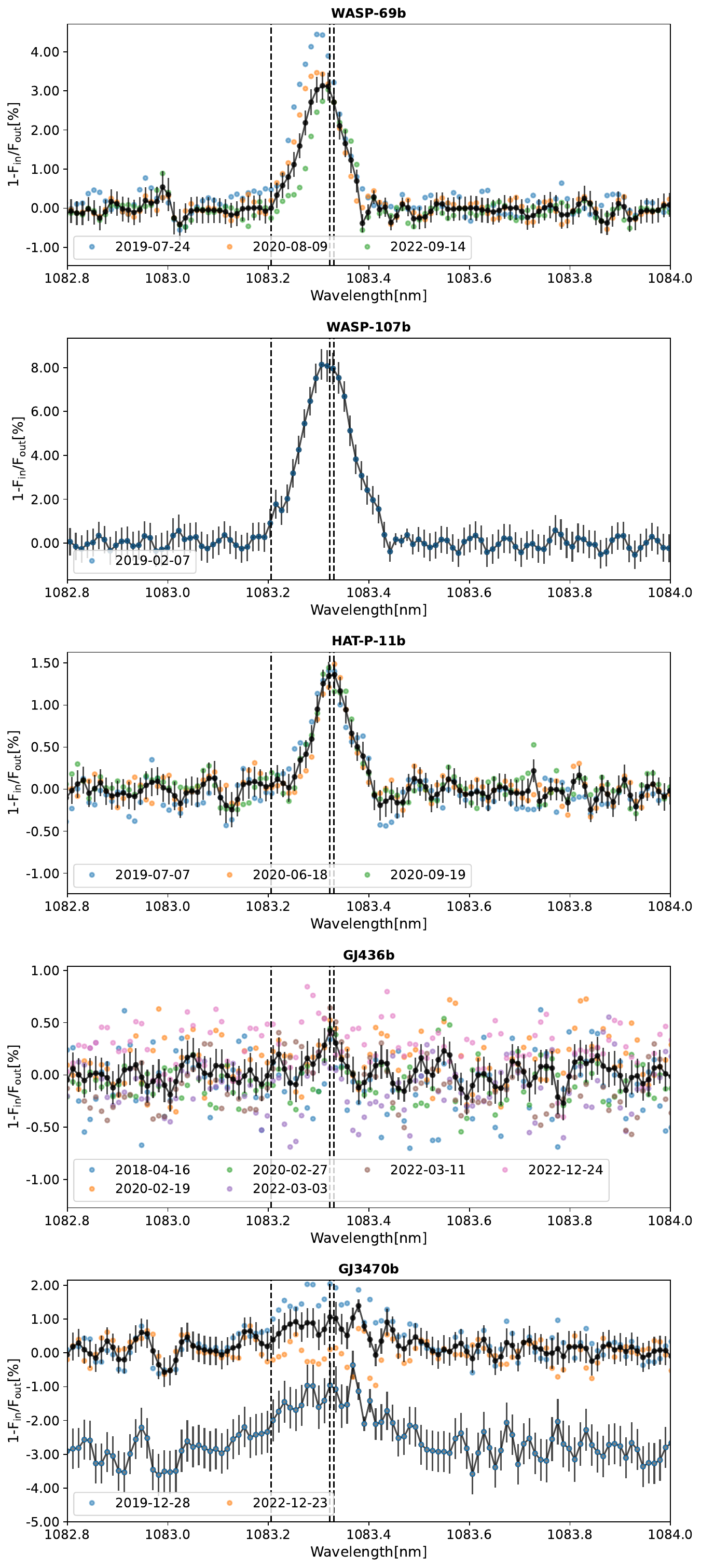}
		\caption{Transmission spectra centred on the \ion{He}{I} triplet (in the planet rest frame) after removal of the GP model. For each target, different colours refer to the different visits, and the black points show the weighted average over the observations. Vertical dotted black lines indicate the position of the \ion{He}{I} triplet. For GJ\,3470\,b two different transmission spectra are reported, one by considering the two nights not affected by the OH contamination and one by considering only the night responsible for the \Hei signal. For WASP-69\,b, the weighted averaged spectrum has been obtained by excluding the first night (i.e. UT 2019-07-24) as likely affected by stellar contamination.} 
		\label{tra_sp}
	\end{figure}
	
	\subsubsection{Spectra extraction}
	We applied several data processing steps to the GIANO-B raw data using the GOFIO pipeline \citep{Rainer2018}. These steps included dark subtraction, flat field correction, removal of bad pixels, spectra extraction (without considering the blaze function correction), and wavelength calibration using a U-Ne lamp spectrum as a template in the vacuum wavelength frame. We used the ms1d spectra, the multi-spectral 1D output, where the echelle orders are kept separated.
	To improve the initial wavelength solution, we employed the same approach described in our previous works \citep[e.g.][]{Giacobbe2021, Guilluy2022}. We aligned all the spectra to the Earth's atmospheric rest frame, assuming it as the frame of the observer (neglecting any $\sim$10 m s$^{-1}$ differences due to winds), by measuring any shifts relative to an average spectrum taken as a template for the night. Subsequently, we refined the wavelength solution by utilising an Earth's atmospheric transmission spectrum generated using the Sky Model Calculator\footnote{\url{https://www.eso.org/observing/etc/bin/gen/form?INS.MODE=swspectr+INS.NAME=SKYCALC}}. 
	
	For the remaining analysis, we focused exclusively on the order~\#39, which encompasses the \Hei triplet. The magnitude of these wavelength calibration refinements, for the considered nights and in the region around the \Hei triplet, is $\sim$0.7 km/s, approximately one-fourth of a pixel.\\
	
	\subsubsection{Telluric correction}
	As our spectra were obtained from ground-based observations, it was necessary to account for the contribution of Earth's atmosphere. The telluric spectrum contains both emission and transmission lines. To correct for the transmission telluric lines, particularly the H$_2$O line at around 1083.51~nm (vacuum wavelength), we followed the approach proposed by \citet{Guilluy2023} and utilised the {\fontfamily{pcr}\selectfont Molecfit} ESO software \citep{Smette2015,Kausch2015}. 
	An example of telluric removal is shown in the right panels of Fig.~\ref{rest_frame}.
	
	Within the spectral region of interest, there are two OH emission line doublets located close to the \ion{He}{I} triplet. The first doublet is at approximately 1083.2103\,nm, and 1083.2411\,nm, while the second one is at 1083.4241\,nm and 1083.4338\,nm in vacuum wavelengths\citep{Oliva2013}. The GIANO-B resolving power is not able to distinguish between the two separate components of this second doublet, resulting in them being observed as a single, intense line in the spectra. The GIANO-B nodding acquisition mode automatically corrects for these emission lines, as described in \citealt{Guilluy2020}. However, due to variations in atmospheric seeing during the observing nights, the A-B subtraction may leave residual signals at the deepest OH doublet (i.e. around 1083.43 nm, as seen in the left panels of Fig.~\ref{rest_frame}). To account for this, we visually inspected the GIANO-B raw spectra to identify the location of this OH line in our spectrograph and created a mask to exclude any possible residuals from our final transmission spectra \citep{Guilluy2023}.
	
	\subsubsection{Transmission spectra calculation}
	To separate the stellar contribution from the potential planetary signal, we took the following steps:\\
	(i) First, we shifted the spectra into the star rest frame using Eq.~(1) from \citet{Guilluy2023} and the parameters listed in Table~\ref{tab_par}. As depicted in the right panels of Fig.~\ref{rest_frame}, the \Hei triplet lines align in the stellar rest frame.\\
	(ii) Next, we normalised each spectrum to the continuum by dividing it by its median value (neglecting the spectral region around the \Hei triplet), excluding spectra with significantly lower S/Ns compared to the other exposures.\\
	(iii) We created a master-out stellar spectrum, $S_\mathrm{out}(\lambda)$, by averaging the out-of-transit spectra (i.e. with an orbital phase smaller than t$_1$ or greater than t$_4$\footnote{For those planets where there was a hint of a tail we discarded the corresponding spectra to compute the master stellar spectrum.}). Individual transmission spectra, $T\mathrm{(\lambda,i)}$, were derived by dividing each spectrum by $S_\mathrm{out}(\lambda)$. \\
	(iv) Finally, we linearly interpolated the transmission spectra in the planet's rest frame using Eq.~(2) from \citet{Guilluy2023}. Fig.\ref{MAPS} displays an example of the transmission spectroscopy 2D maps in the planet's rest frame for each planet, while the 2D maps for all the investigated nights are presented in Fig.\ref{MAPS_app}.
	For each investigated planet and each observed night, we then derived the fully in-transit transmission spectrum in the planet's rest frame T$_\mathrm{mean}$, which is computed by averaging the transmission spectra with an orbital phase between t$_2$ and t$_3$. 
	We did not consider the influence of centre-to-limb variations (CLVs) or the Rossiter-McLaughlin (RM) effect as previous studies have indicated a minimal impact on the \Hei \citep[e.g.][]{Allart2018, Allart2019, Nortmann2018}.
	
	To estimate the contrast $c$ of the excess absorption at the position of the \Hei triplet, we fitted a Gaussian profile to each individual in-transit mean transmission spectrum with the differential evolution (DE) Markov chain Monte Carlo \citep[MCMC;][]{TerBraak2006} method, varying the peak position, the full-width half maximum (FWHM), the peak value ($c$), and an offset for the continuum.
	Correlated noise in the transmission spectra (mainly caused by systematic effects such as fringing, variations in the instrumental profile, changes in the position of the star in the slit, etc) was modelled through the Gaussian process (GP) regression within the same DE-MCMC tools, using a covariance matrix described by a squared exponential kernel (see Eq.~2 in \citealt{Bonomo2023})\footnote{We also ran the DE-MCMC analysis without the GP, and our findings remain compatible within 1-$\sigma$. However, the error bars on the model-free parameters (including the \Hei absorption) from the GP analysis are usually slightly larger, and thus more conservative, than those without the GPs, because they account for the correlated noise.}. We finally accounted for possibly uncorrelated noise with a jitter term $\sigma_\mathrm{jit}$. We imposed several priors on the model parameters as well as on the GP hyper-parameters (see Table~\ref{priors_tab}).
	For each target, Fig.~\ref{GP} shows an example of correction with the GP regression model, the posterior distributions for the same nights are reported in Appendix in Fig.~\ref{Cornerplots}.
	The best-fit parameters from the DE-MCMC Gaussian analysis, for each night are listed in Table~\ref{tab_result_single}. We determined the values and the 1$\sigma$ uncertainties of the derived parameters from the medians and the 16\%-84\% quantiles of their posterior distributions. In cases of a \Hei non-detection, we reported 1$\sigma$ upper limits on the excess absorption at the positions of the stellar \Hei lines \citep{Guilluy2023}.
	
	For each investigated target, the transmission spectra from each night were then weight-averaged after subtracting the GP regression model to create the final transmission spectra  (see Fig. \ref{tra_sp}). 
	Table~\ref{tab_result_combined} reports for each planet the weight-averaged \Hei contrast $c$. \\ 
	\subsubsection{Light-curve computations}
	To monitor the variation of the \Hei signal during transit, we additionally performed spectrophotometry of the helium triplet within a passband of 0.075~nm centred at the peak of excess absorption in the planet rest frame \citep{Allart2019}. The computed transit light curves are presented in Fig.~\ref{LC}.

	\subsection{Analysis in the optical} \label{data_analysis_opt}
	To indicate the host star activity we derived the chromospheric emission from \ion{Ca}{ii} H and K lines of HARPS-N spectra (\logrhk). We used the definition of \citet{Noyes1984} and the implementation of \citet{Lovis2011} through the offline version of the HARPS-N DRS available through the Yabi workflow \citep{yabi} hosted at IA2 Data Center \footnote{\url{https://www.ia2.inaf.it/}}. The only exceptions are GJ436 and GJ3470, for which the Noyes relations are not applicable as they have a colour index B-V>1.1; we thus utilised the equations provided in \citet{suarez2015}. The derived \logrhk~are reported in Table~\ref{logrhk_table}.\\
	
	We then derived transmission spectra in H$\alpha$ similarly as described for the  \Hei triplet. We focused on raw data already processed by the DRS and we used {\fontfamily{pcr}\selectfont Molecfit} to remove telluric contamination. We obtained transmission spectra (2D maps are in the upper panels of Fig.~\ref{tra_sp_halpha} and Fig.~\ref{tra_sp_halpha_GJ}) by dividing the full in-transit spectra by the master out and we modelled CLVs and the RM effect similarly to \citet{yan2017effect}. However, instead of using models computed at different limb darkening angles $\mu$, we used an analytical approach using ExoTethys \citep{morello2020exotethys} to retrieve the limb darkening coefficients and then computing the stellar intensity profile $I(\mu)$ by adopting a quadratic limb darkening law \citep{Darpa}. 
	After shifting the transmission spectra into the planetary rest frame, we computed the weighted average of the full in-transit spectra (bottom-right panel of Fig.~\ref{tra_sp_halpha} and Fig.~\ref{tra_sp_halpha_GJ}), and we fitted a Gaussian to evaluate the absorption/emission depth, FWHM, and the velocity shift. 
	Along with the Gaussian fit, we also performed a linear fit and we computed the Bayesian information criterion (BIC) to compare the two models \citep{kass_bic}. We considered the Gaussian model as preferable over the linear one only when accompanied by a lower BIC and a difference in BIC ($\Delta$BIC) > 10. We did not use the GP correction in the optical as the HARPS-N spectra are much less affected by systematics, and so the use of GPs was not really necessary.\

	\section{Results}\label{results}

	\begin{table*}[h]
		\centering
		\caption{Results combined.}
		\begin{tabular}{c c c c c c c }
			\hline \hline
			Target      &  $c$ &                                  Significance  & Effective radius &g$_\mathrm{P}$ & H   & $\delta_\mathrm{R_P}$/H$_{\mathrm{eq}}$  \\
			&  [\%]&                       [$\sigma$]    & [R$_\mathrm{p}$] &   [m\,s$^{-2}$]        & [km] &                                   \\
			\hline
			WASP-69\,b                                & 3.91$\pm$0.22                         & 17.6 & 1.79$\pm$0.08 & 5.8$\pm$0.6   &    1060  &          56$\pm$8\\
			WASP-69\,b (without 24 July 2019) & 3.46$\pm$0.32                     &10.7 & 1.71$\pm$0.08 & 5.8$\pm$0.6    &    1060  &        51$\pm$8\\ 
			WASP-107\,b (1 night)                     & 8.17 $^{+ 0.80 }_{ -0.76 }$   & 10.5 &  2.19$\pm$0.11& 3.37$\pm$0.31 &  1450 &           55$\pm$7\\
			HAT-P-11\,b                                & 1.36$\pm$0.17                          & 8.0       & 2.17$\pm$0.11 &    10.2$\pm$0.6 &   530  &   71$\pm$9\\
			GJ\,436\,b                               & <0.42(0.52)$^+$                &                      &  <1.27(1.33)       &    14.8$\pm$1.7 &  294   &      < 24(29)\\ 
			GJ\,3470\,b (1 night)                    & 1.75 $^{+ 0.39 }_{ -0.36 }$   &4.7      & 2.05$\pm$0.25 &    8.2$\pm$1.6  &  470   &      55$\pm$19\\ 
			\hline
		\end{tabular}
		\tablefoot{From left to right: the investigated target, the contrast from the DE-MCMC analysis, the significance of the detection, the effective \Hei radius, 
			the planet's gravity, the atmospheric scale height (computed by assuming a mean molecular weight of 1.3), and the ratio between the equivalent height of the \Hei atmosphere and the atmospheric scale height. [$^+$] 1$\sigma$(2$\sigma$) upper limits. For WASP-69\,b, we also present results by excluding the first observing night (i.e. 24 July 2019), as it was likely affected by stellar contamination.}
		\label{tab_result_combined}
	\end{table*}

	In this paper we performed a \Hei HR transmission spectroscopy survey of five gas giants namely WASP-69\,b, WASP-107\,b, HAT-P-11\,b, GJ\,436\,b, and GJ\,3470\,b. In this section, we summarise our findings. The last column of Table~\ref{log} highlights the nights with significant telluric overlap and possible stellar activity issues.\\
	\subsection{Individual analyses}
	\subsubsection{WASP-69\,b}
	We analysed four observing nights for WASP-69\,b, namely 24 July 2019, 09 August 2020, 28 October 2021, and 14 September 2022. However, 28 October 2021 was affected by OH contamination, we thus decided to exclude data from this night when computing the final contrast value\footnote{When we also took into account that for 28 October 2021 an absorption signal of \Hei is still visible, albeit affected by OH contamination, in the 2D map in Fig.~\ref{MAPS_app}, we measured a contrast of 3.88$\pm$0.21\% (18.8$\sigma$).}.
	We measured a contrast of 3.91$\pm$0.22\% (17.6$\sigma$), compatible with the value reported by \citet{Nortmann2018}, \citet{Vissapragada2020}, \citet{Vissapragada2022}, and \citet{Tyler2023}, while slightly higher than the one reported by \citet{Allart2023} with SPIRou (where they measured a maximum excess absorption of $\sim$3.1\%). 
	From the 2D map in Fig.~\ref{MAPS} we can confirm the existence of a cometary tail following the planet \citep{Nortmann2018,Tyler2023}. Our \Hei signal appears to persist for approximately $\sim$50~minutes after egress. \citet{Nortmann2018} reported a  22~minutes of post-transit absorption, while \citet{Tyler2023} observed a longer duration of 1.28~hours after egress. The variations in tail lengths might be attributed to intrinsic atmospheric variability or to insufficient baseline coverage and lower S/N in our study and \citet{Nortmann2018} compared to \citet{Tyler2023}.
	From Fig.~\ref{tra_sp}, UT 24 July 2019 appears to show a more pronounced absorption compared to the other transit events, with a compatibility of $\sim$1.9$\sigma$ with the contrast value of 3.46 $\pm$ 0.32\% (10.7$\sigma$) obtained by averaging the other three nights.
	We thus analysed the H$\alpha$ diagnostic in the visible to investigate whether this disparity in the \Hei absorption levels was attributable to stellar contamination.
	
\begin{figure*}
		
		\includegraphics[scale=0.6]{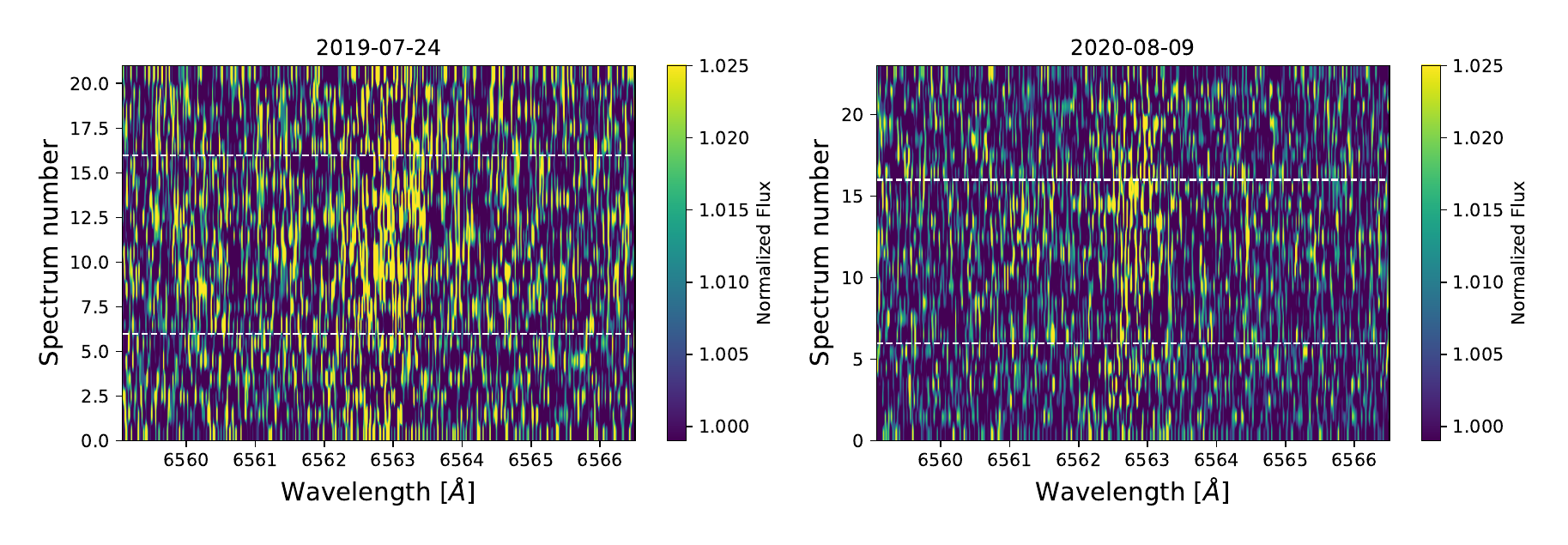}
		\includegraphics[scale=0.65]{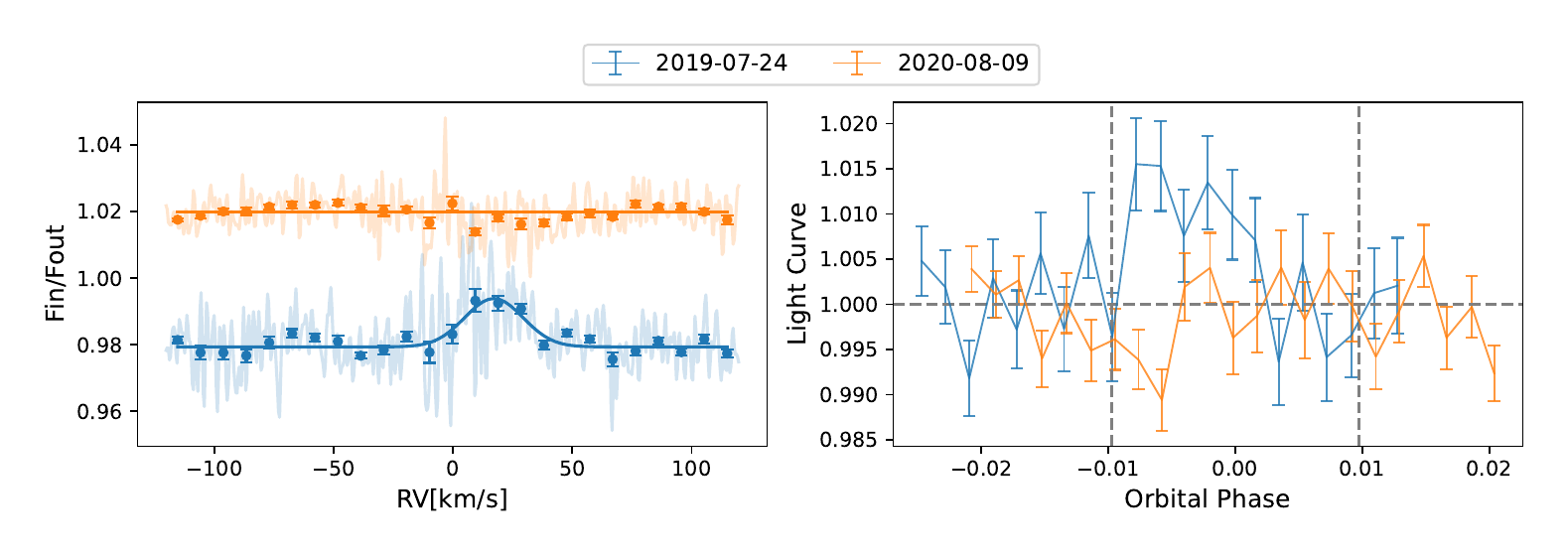}
		\caption{H$\alpha$ transmission spectra and light curves for WASP-69\,b. Top:  2D transmission spectroscopy maps in the stellar rest frame for UT 24 July 2019 (top left) and UT 09 August 2020 (top right). Dotted red  lines mark the position of the H$\alpha$ line in the planet rest frame. Bottom:  Spectroscopic light curves (bottom left) and weighted average of the full in-transit spectra in the planet rest frame (bottom right; the equivalent in the stellar rest frame is in the left panel of Fig.~\ref{tra_sp_halpha_W69star}). Light colours indicate the not binned transmission spectra, while dots represent the transmission spectra binned 20$\times$(in RV). We superimposed the Gaussian profile derived from the best-fit parameters on the unbinned spectra.}
		\label{tra_sp_halpha}
	\end{figure*}

	HARPS-N observations were available only for UT 24 July 2019 and UT 09 August 2020. Our findings are shown in Fig.~\ref{tra_sp_halpha}, with the final transmission spectra in the bottom-right panel. For UT 09 August 2020, we obtained a BIC of 268.6 for the Gaussian fit and 269.8 for the linear fit, resulting in a $\Delta$BIC of 1.8.
	According to \cite{kass_bic}, the linear fit is the model that better describes our data, hence ruling out the presence of features in our transmission spectrum. This is also reinforced by the absence of features in the light curve shown in the bottom-left panel of Fig.~\ref{tra_sp_halpha}. On the other hand, we detected an emission signal on UT 24 July 2019 (1.45$\pm$0.19\%, 7.6$\sigma$) for which the Gaussian fit was heavy favoured over the linear one (Gaussian BIC: 324.2, linear BIC: 352.2). Our detection is corroborated by the spectroscopic light curve, computed with a width of 20 km/s, which shows a clear emission feature during the first half of the transit (bottom-left panel of Fig.~\ref{tra_sp_halpha}). 
	This finding was further corroborated by the \texttt{SLOPpy} routine \citep{Sicilia2022}, which also revealed an excess of absorption during UT 24 July 2019 in the Sodium doublet \citep{Siciliainprep}, in comparison to other nights.
	
	This case appears similar to what was reported for HD189733b in \citet{Guilluy2020} on the third night of observation, where an additional absorption in \Hei was observed in correspondence with an emission signal in H$\alpha$, suggesting that the planet was transiting over quiescent regions of the stellar surface. Given the activity of WASP-69 (as indicated by the \logrhk~measurements in Table~\ref{logrhk_table}) we would expect to have a non-uniform stellar disc, and thus a possible occultation during the transit of quiescent stellar features.  However, this signal appears redshifted in the stellar rest frame (17.4$\pm$3.0 km/s; see Fig.~\ref{tra_sp_halpha_W69star}), while a pseudo signal caused by a plage-like region should be at rest in the star reference frame. An alternative explanation could be the presence of a flare. 
	In this case, we would be observing the ejected material falling back to the star and hence moving away from us. This would account for the behaviour in \Hei and H$\alpha$, as well as the redshift in H$\alpha$ in the stellar reference frame.
	The H$\alpha$ line could be produced within a moving structure, for example, material falling towards the chromosphere or material moving away from us with the flare occurring near the edge of the star (thus explaining the redshift), while the \Hei absorption could be produced in a region outside the flare but irradiated by the XUV rays of the flare, which contribute to populating the atomic level from which the absorption that produces the line originates.
	
	It is worth noting that \citet{Nortmann2018} also reported significantly different depths in the two CARMENES nights analysed in their study. This raises the question of whether a similar mechanism might be at play in their data.
	
	In Fig.~\ref{MAPS}, the \Hei signal appears to be slightly tilted, indicating a different atmospheric K$_\mathrm{P}$ compared to the one obtained from the radial velocity curves (see Table~\ref{tab_par}). We investigated a lag vector corresponding to possible atmospheric K$_\mathrm{P}$ in the range 0$\leq$K$_\mathrm{P}$$\leq$250 km s$^{-1}$, in steps of 1 km s$^{-1}$. We considered the two nights where the \Hei signal was less affected by the OH telluric emission line (i.e.  UT 24 July 2019 and 09 August 2020). For each K$_\mathrm{P}$ value, we derived a 2D map (similar to those shown in Fig.~\ref{MAPS}) and a T$_\mathrm{mean}$ that we fitted with a Gaussian to determine the peak of \Hei absorption. We see that the \Hei absorption peak is maximised for an atmospheric K$_\mathrm{P}$ of approximately 80 km s$^{-1}$ (lower than the one reported in the literature; see Table~\ref{tab_par}). We have to stress that given the width of the single line we are investigating, this result has to be taken just as a maximisation but without any statistical significance. 3D hydrodynamic simulations suggest that the \Hei signal in robust outflows may not precisely follow the expected K$_\mathrm{P}$. Once the gas starts getting outside the planetary Roche lobe the velocity dynamics can differ (see e.g. \citealt{Nail2023} and \citealt{Gully2023}).

	\subsubsection{WASP-107\,b}
	We analysed two transits of WASP-107\,b, 07 February 2019 and 04 May 2019. However, the second observing night was strongly affected by the OH emission line. If we consider only the first transit (i.e. UT 07 February 2019) we obtained an absorption of 8.17 $^{+ 0.80 }_{ -0.76 }$\% (10.5$\sigma$) compatible. If we consider only the first transit (i.e. UT 07 February 2019) we obtained an absorption of 8.17 $^{+ 0.80 }_{ -0.76 }$\% (10.5$\sigma$) compatible 
	with 7.92$\pm$1.00\% and 7.26$\pm$0.24\% reported by \citet{Allart2019} and \citet{Kirk2020}, respectively. Our 2D maps in Fig.~\ref{MAPS} and the spectroscopic light curve in Fig.~\ref{LC} indicate the possible presence of a tail following the planet, in agreement with Fig. 3 of \citet{Kirk2020} and \citet{Spake2021}. However, our tail appears to be shorter than found in the latter study; this difference may be attributed to intrinsic atmospheric variability, as highlighted for WASP-69\,b. 
	
	\subsubsection{HAT-P-11\,b} 
	We gathered three transit observations for HAT-P-11\,b.
	We reported an extra absorption of 1.36$\pm$0.17\%  (8.0$\sigma$) in agreement with $\sim$1.2\% and $\sim$1.3\% reported by \citet{Allart2018} and \citet{Allart2023}, respectively. According to Table~\ref{tab_result_single}, all the helium absorption measurements are compatible within 0.5$\sigma$.\\
	
	\subsubsection{GJ\,436\,b}
	We collected six transits of GJ\,436\,b. We did not report any \Hei extra absorption for GJ\,436\,b, as our findings are consistent with zero absorption. Neither the single nights taken individually (see Fig.~\ref{MAPS}) nor the light curve (see Fig.~\ref{LC}) show evidence of helium absorption with a 1$\sigma$ upper limits of  0.42\% (0.52\% at 2$\sigma$). The small \Hei feature in our final transmission spectrum (Fig.~\ref{tra_sp}) could be due to correlated noise falling at the position of the stellar helium triplet not perfectly corrected with the GP regression model. Our result is in agreement with \citet{Nortmann2018}, who did not find evidence of \Hei in the upper atmosphere of GJ\,436\,b (they reported a 90\% confidence upper limits of 0.41\% calculated from the standard deviation of the spectrum).   \\
	
	\subsubsection{GJ\,3470\,b}     
	Here, we analysed five nights of observations, namely 13 January 2018, 04 February 2019, 28 December 2019, 27 January 2020, and 24 December 2022, which we refer to as transit~1, transit~2, transit~3, transit~4, and transit~5 for simplicity.
	As the 2D maps in Fig.~\ref{MAPS} show, three (i.e. transit~1, transit~2, and transit~4) were strongly contaminated by the OH telluric emission line, we thus decided to discard these nights. The 2D maps (Fig.~\ref{MAPS}) suggest that the majority of the signal originates from transit~3. If we account for only this transit, we obtained a tentative detection of the extra absorption of 1.75$^{+ 0.39 }_{ -0.36 }$\% (4.7$\sigma$),  in agreement with \citet{Ninan2020}, $\sim$1.5\% from their Figure~5, and \citet{Palle2020}, 1.5$\pm$0.3\%. \citet{Allart2023} could not replicate the detection with SPIRou and only placed a 3$\sigma$ upper limit on the presence of helium at 0.63\%. Despite our detection being consistent at 3$\sigma$ with no absorption, tomography reveals a hint of absorption at the helium triplet position. 
	
	It is worth noting that \citet{Palle2020} observed variability across the analysed nights (see their Table 3) and attributed it to differences in the S/N. However, in our investigation, we find comparable S/N between transit~3 and transit~5. The tension in the literature is not so surprising, considering our result. 
	
\begin{figure*}
		\centering
		\includegraphics[scale=0.41]{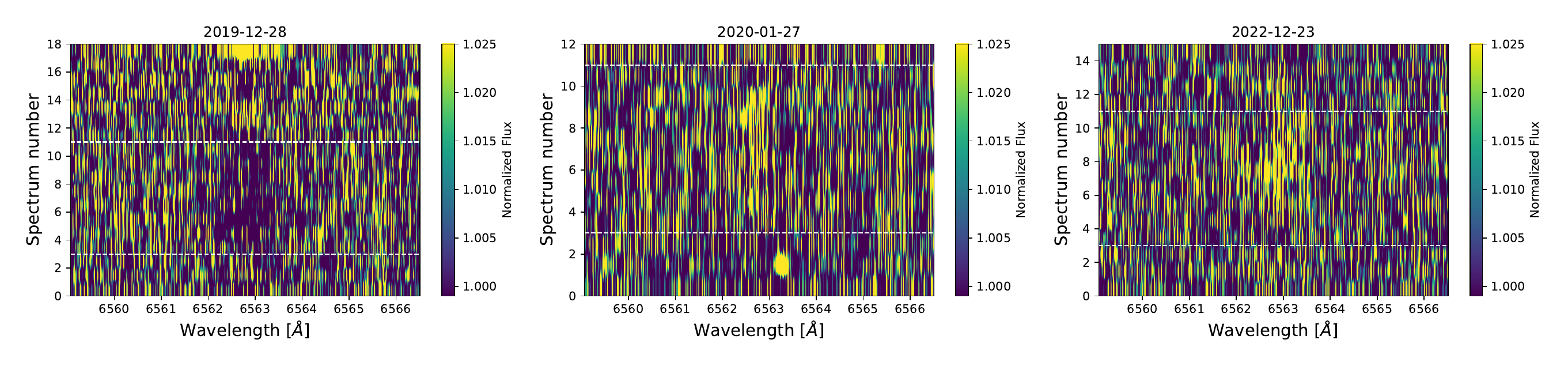}
		\includegraphics[scale=0.65]{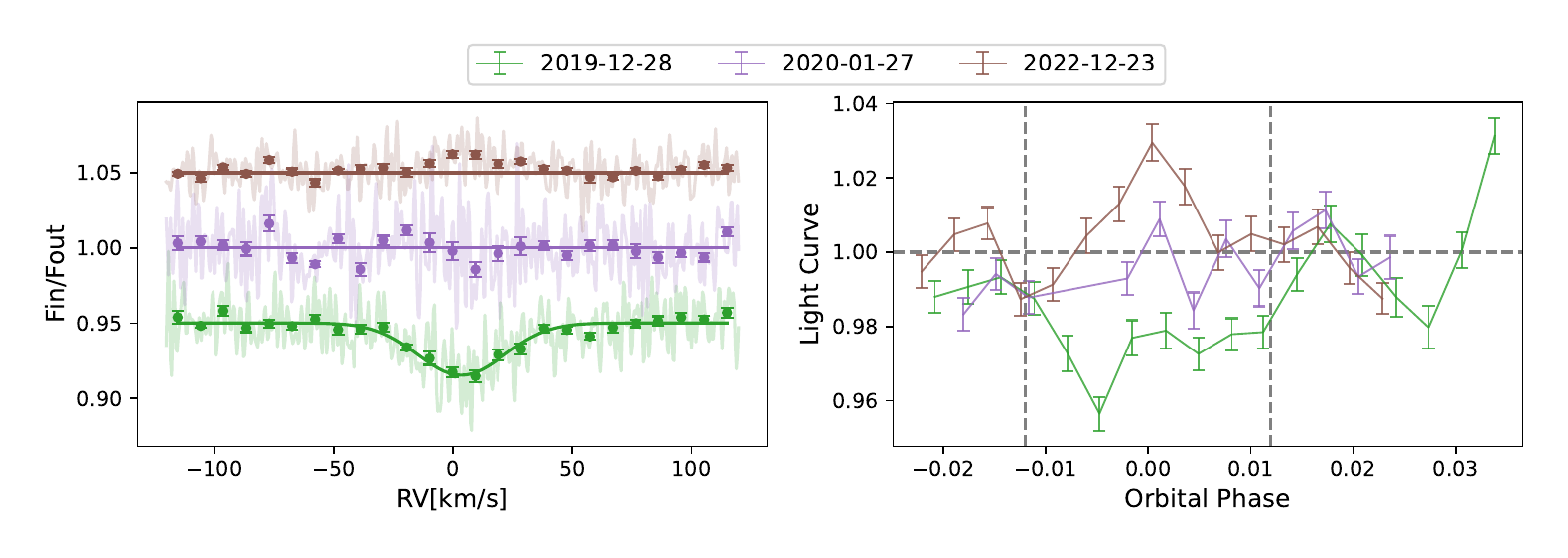}     
		\caption{H$\alpha$ transmission spectra and light curves for GJ-3470b. Top:  2D transmission spectroscopy maps in the stellar rest frame for transit 3, transit 4, and transit~5 in panels a, b, and c, respectively. Dotted red lines mark the position of the H$\alpha$ line in the planet rest frame. Bottom:  Spectroscopic light curves (d panel) and weighted average of the full in-transit spectra in the planet rest frame (e panel, the equivalent in the stellar rest frame is in the left panel of Fig.~\ref{tra_sp_halpha_W69star}).}
		\label{tra_sp_halpha_GJ}
	\end{figure*}

	Given the lack of homogeneity in the \Hei observation across the various investigated nights, we also analysed the H$\alpha$ line in a similar way to WASP-69\,b (Fig.\ref{tra_sp_halpha_GJ}). Out of the five nights under consideration, only three were usable in terms of S/N, namely transit~3, transit~4, and transit~5. We identified an absorption signal 
	(3.48$\pm$0.26\%, 13$\sigma$) at rest in the planet's reference frame during transit~3 (as depicted in panels a and e of Fig.\ref{tra_sp_halpha_GJ}). The BIC analysis confirms the absorption signal detected with a BIC value of 401.5 for the Gaussian and 462.5 for the linear fit. This particular night contributed to the majority of the \Hei signal.
	If this simultaneous absorption signal in both \Hei and H$\alpha$ was due to stellar activity, it could be explained by the planet's passage over a quiescent region of the stellar disc, during which time the star presents filaments. These regions being darker than the rest of the star in both helium and H$\alpha$ could have caused this pseudo-signal in absorption in both the stellar lines. On the other hand, it is also plausible that the H$\alpha$ absorption feature may be attributed to an extended planetary hydrogen atmosphere. 
	However, theoretical transmission profiles simulated using the ATES code \citep{Caldiroli2022} and the new add-on transmission probability module \citep{Biassoni2023} are not able to reproduce an absorption depth/profile consistent with our observations. For this simulation, we assumed a local thermodynamic equilibrium (LTE) profile and adopted the X-ray luminosity of log(L$_\mathrm{X}$)=27.58 measured by \citet{Spinelli2023} and two different X-ray-EUV relations, derived by \citet{Johnstone2021} and \citet{Sanz-Forcada2022}: log(L$_\mathrm{EUV}$)=27.98 and log(L$_\mathrm{EUV}$)=28.40, respectively. 
	Although the H$\alpha$ population process may not be in the LTE regime \citep[e.g.][]{Garcia2019,Huang2023}, our ATES analysis and the detection in only one of the three HARPS-N nights, make the hypothesis of a stellar origin for the feature detected in H$\alpha$ the most plausible. Further analyses that include more detailed radiative transfer models are needed to confirm this hypothesis.
	
	Concerning transit~4 (shown in violet in Fig.~\ref{tra_sp_halpha_GJ}), the one significantly impacted by the OH emission line in \ion{He}{I}, the HARPS-N data were affected by both noise and systematics (as evident in both the light curve and the 2D maps in panels b and d of Fig.\ref{tra_sp_halpha_GJ}) and we had to remove two spectra for which the telluric correction with {\fontfamily{pcr}\selectfont Molecfit} did not work resulting in an overcorrection.  In this case, the linear fit is favourite with respect to the Gaussian one (BIC 462.5 vs 447.4) consistently with the absence of features in the light curve. Hence, we do not detect any H$\alpha$ signal in this night.
	Finally, during transit~5 (depicted in brown in Fig.~\ref{tra_sp_halpha_GJ}), there appears to be an emission signal in H$\alpha$, as highlighted from both the 2D map (panel c) and the spectroscopy light curve (panel d). This signal seems more pronounced in the star's rest frame (right panel of Fig. \ref{tra_sp_halpha_W69star}) even if with a low significance of $\sim$4$\sigma$. Consequently, it may be attributed to stellar activity; for instance, the planet's transit over a dark region on the stellar disc, such as a filament, could potentially mimic a pseudo-signal in emission in both H$\alpha$ and \ion{He}{I}, thereby impacting our helium detection. 
	However, considering the low significance of this detection, the small number of spectra (i.e. 15), and the moderate S/N ($\sim$ 38), we cannot dismiss the possibility that this result may be solely attributed to statistical noise. This hypothesis is supported by the BIC test, which returns the same value for both the linear fit and the Gaussian one (BIC=398), suggesting that we should prefer the model with fewer parameters.     
	
	\subsection{Statistical analyses} 
	In this section, we looked for possible correlations between the \Hei absorption feature and the stellar and planetary parameters thought to be keys for the \Hei signal observability. To achieve this goal, in addition to the findings presented in this paper, we also considered previous \Hei studies performed with GIANO-B, namely HD189733b \citep{Guilluy2020}, WASP-80\,b \citep{Fossati2022}, and HAT-P-3b \citep{Guilluy2023}.
	
	For each planet in our sample, we derived the effective \Hei radius \citep[e.g.][]{Chen2018} that would produce the observed absorption contrast $c$ (see Table~\ref{tab_result_combined}). We then normalised it to the atmospheric $H_\mathrm{eq}$
	to compute the quantity $\delta_\mathrm{R_P}/H_\mathrm{eq}$ \citep{Nortmann2018}, which represents the number of scale heights probed by the atmosphere in the spectral range under consideration. Here, $H_\mathrm{eq}$ is defined as $\frac{k_\mathrm{B}T\mathrm{eq}}{\mu g_\mathrm{P}}$ (see Table~\ref{tab_result_combined}), where $k_\mathrm{B}$ is the Boltzmann constant, $T_\mathrm{eq}$ is the planetary equilibrium temperature (listed in Table~\ref{tab_par}), $g_\mathrm{P}$ is the planetary gravity calculated from the planetary mass and radius (see Table~\ref{tab_result_combined}), and $\mu$ is the mean molecular weight. According to \citet{Fossati2022}, we assumed a hydrogen-dominated atmosphere and hence $\mu$=1.3 times the mass of a hydrogen atom (we opted for a hydrogen-dominated atmosphere rather than a hydrogen-and-helium atmosphere to mitigate uncertainties arising from the unknown helium abundance\footnote{The value of $\mu$=1.3 is typically considered for hot Jupiters assuming solar abundances, but it also takes into account the presence of a mix of molecular hydrogen and electrons resulting from the ionisation of various elements, especially Na and K. Hydrogen and helium abundances are not the only factor to consider. We would like to emphasise that the adopted $\mu$ does not alter the final outcome, and does not influence the presence or absence of correlations.}).  The derived $\delta_\mathrm{R_P}/H_\mathrm{eq}$ values for each investigated planet are presented in Table~\ref{tab_result_combined}.
	
\begin{figure*}
		\centering
		\includegraphics[width=\linewidth]{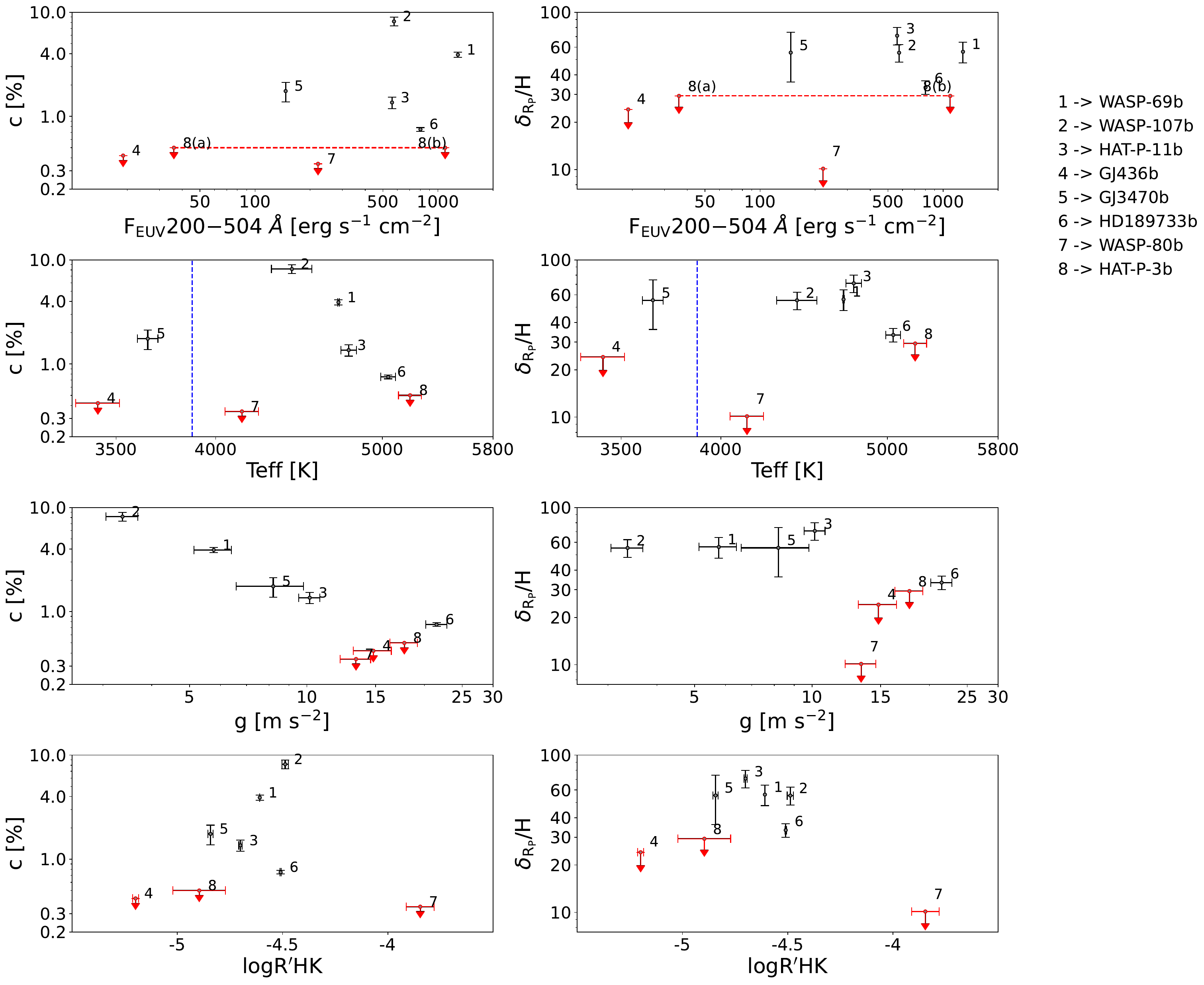}
		\caption{Correlation plots. Targets are from this work, \citet{Guilluy2020}, \citet{Fossati2022}, and \citet{Guilluy2023}. Contrast $c$ (left panels) and $\delta_\mathrm{R_P}/H_\mathrm{eq}$ (right panels) as a function of the EUV flux in the 200 to 504 \AA\ range, effective temperature and planetary gravity, and \logrhk. Red markers indicate targets with only an upper limit on the \Hei detection (reported at 1$\sigma$). The vertical dashed blue lines in the effective temperature diagrams indicate the transition from M-type to K-type stars. The dashed horizontal line connects the two possible locations for HAT-P-3b (for the two different coronal temperatures). }
		\label{final_results}
	\end{figure*}

	In Fig.~\ref{final_results}, we examined how these derived constraints vary with respect to the EUV flux in the 200 to 504 \AA~range (which controls the \Hei metastable production, and therefore absorption, in the planetary atmosphere, \citealt{Fossati2023}), effective temperature, planetary gravity, and \logrhk. We used the EUV flux derived in \citet{Fossati2023} from the scaling relations of \citet{Poppenhaeger2022}. We employed the same methodology to derive the EUV flux received by HAT-P-3b, the only one missing in the sample analysed by \citet{Fossati2023}. As no X-ray measurements were present in the literature, we estimated it starting from our derived \logrhk value (see Table~\ref{logrhk_table}) and the \citet{Mamajek&Hillenbrand2008} relation, obtaining a value of Lx=1.3$^{+2.2}_{-0.8}$~10$^{27}$~erg/s. We then employed the \citet{Poppenhaeger2022} relations to estimate the EUV flux. However, due to the large uncertainty on the derived X-ray luminosity, we were unable to constrain the coronal temperature (relation from \citealt{Johnstone2015}). As a result, considering a range of 1.25-2.1~MK for the coronal temperature, we obtained two different EUV flux values using the scaling relations of \citet{Poppenhaeger2022}: 36 erg s$^{-1}$ cm$^{-2}$ and 1095 erg s$^{-1}$ cm$^{-2}$.

	Trends related to the contrast $c$ and atmospheric extension $\delta_\mathrm{R_P}/H_\mathrm{eq}$ are shown in the left and right panels of Fig.~\ref{final_results}, respectively. 
	To evaluate the strength of the possible correlations between the investigated parameters, we ran an MCMC, varying the \Hei contrasts $c$ according to a normal distribution with a standard deviation equal to the error we found on $c$. In the case of non-detection, we used a half-Gaussian distribution centred at zero with a standard deviation equal to the 1$\sigma$ upper limit. We ran 10~000 iterations and at each MCMC step, we computed the Spearman rank-order correlation coefficient $\rho$ to measure the strength of our relationships. The final mean values, the corresponding errors, and the p-values are reported in Table~\ref{Spearman}. The p-values were computed using a t-test under the null hypothesis that the correlation coefficients are significantly different from 0.
	\begin{table}
		\small
		\centering 
		\caption{Correlations between the investigated planetary and stellar parameters expected to influence the \Hei observability.}.
		\resizebox{0.99\linewidth}{!}{
			\begin{tabular}{l |  c c  c }
				\hline  \hline 
				X, Y              &$\rho$ Spearman   & Correlation Strength  & p-value \\
				\hline 
				F$_\mathrm{EUV}$,$\delta_\mathrm{R_P}/H_\mathrm{eq}$ &   0.0116 $\pm$ 0.0029  & null  & 0.98 \\
				F$_\mathrm{EUV}$,$c$                                                            &   0.1663 $\pm$ 0.0029   & low   & 0.69 \\
				T$_\mathrm{eff}$,$\delta_\mathrm{R_P}/H_\mathrm{eq}$ &   -0.0302 $\pm$ 0.0028  & null  & 0.94 \\
				T$_\mathrm{eff}$,$c$                                                            & -0.1665 $\pm$ 0.0027 & low   & 0.70 \\
				g$_\mathrm{P}$,$\delta_\mathrm{R_P}/H_\mathrm{eq}$   &   0.0664 $\pm$ 0.0028 & null  & 0.88 \\
				g$_\mathrm{P}$,$c$                                                         & -0.1886 $\pm$ 0.0026 & low    & 0.65 \\
				\logrhk,$\delta_\mathrm{R_P}/H_\mathrm{eq}$          & -0.1303 $\pm$ 0.0028& low    & 0.76 \\
				\logrhk,$c$                                                                &  0.1461 $\pm$ 0.0030 & low    & 0.73 \\
				\hline                                  
			\end{tabular}
		}
		\tablefoot{From left to right: the parameters investigated for the correlation, the Spearman value, the strength of correlation from \citet{Kuckartz2013}, and the associated p-value.}
		\label{Spearman}
	\end{table}
	
	Our findings do not show significant correlations between the investigated parameters (p-values > 0.05). The null hypothesis that any two parameters X, Y in Table~\ref{Spearman} are uncorrelated is rejected with a confidence level not exceeding 90\%, corresponding to a $<2\sigma$ result.
	However, it is important to underline that we are still within the statistical limit of small numbers, so the results should be interpreted with caution.
	Furthermore, it is important to underscore that there might be variability among planets, attributable in part to possible differences in helium abundance in the atmosphere. This variability could introduce additional noise, potentially hindering our ability to establish statistical correlations. Finally, $H_\mathrm{eq}$ might not be the correct physical length scale to consider in such an analysis. For instance, \cite{Zhang2023} showed that the scale height does not correlate with the contrast measured for the deepest individual lines in a sample of exoplanets, attributing this effect to hydrodynamical effects arising in the upper part of the atmosphere. Even within the hydrostatic region of an atmosphere, in the presence of significant variations with altitude of temperature, gravity or even mean molecular weight, using a single scale height might be too simplistic.
	
	These potential trends do not highlight any explanations that could justify the detection and non-detection of metastable \Hei in the upper atmospheres of the investigated planets\footnote{The results remain consistent when we consider the contrast $c$ (and the corresponding $\delta_\mathrm{R_P}/H_\mathrm{eq}$) obtained by excluding the night of UT 24 July 2019 for WASP-69\,b, likely affected by stellar contamination.}. The three planets with only upper limits on the \Hei detection, namely GJ\,436\,b, WASP-80\,b, and HAT-P-3b do not seem to exhibit a clear correlation in the investigated parameter space. \citet{Fossati2023} emphasised that the low-EUV stellar flux, influenced by the low [Fe/O] coronal abundance \citep{Poppenhaeger2022}, is likely the primary factor behind the \Hei non-detection. \citet{Rumenskikh2022} attributed the weak \Hei signature reported by \citet[0.41\%, 2$\sigma$]{Nortmann2018} to the thinness of the region populated by the absorbing  \Hei (<3R$_\mathrm{P}$), the small R$_\mathrm{P}$/R$_\star$ ratio, and the radiation pressure force, which spreads \Hei atoms along the line of sight and around the planet. Yet, there is still no explanation for the non-detection in the atmosphere of HAT-P-3b.
	
	It is interesting to highlight that both HAT-P-3\,b and GJ\,436\,b are on a polar orbit with a 3D true obliquity ($\psi$) of 75.7$^{+8.5}_{-7.9}$ deg \citep{Bourrier2023} and 103.2$_{-11.5}^{+12.8}$ deg \citep{Bourrier2022}, respectively. On the other hand, WASP-80\,b is well known to be on a relatively aligned orbit  \citep{Triaud2015}. This may indicate that also the orbital obliquity does not influence the \Hei observability. 
	
	It is important to note that in our study, in \citet{Fossati2022}, and in \citet{Guilluy2020}, we analysed multiple transit events for each planet, investigating the repeatability of the \Hei signal. However, this is not the case for HAT-P-3b, as in \citet{Guilluy2023} only one night of observation was gathered and examined. Therefore, obtaining more data is essential to provide clarity on this non-detection.

	\section{Summary and conclusion}\label{conclusions}
	
	We employed the GIARPS mode of the TNG, focusing on GIANO-B observations to look for \Hei in the upper atmosphere of five planets hosted by the K and M dwarf stars of our sample, namely WASP-69\,b, WASP-107\,b, HAT-P-11\,b, GJ\,436\,b, and GJ\,3470\,b. 
	We measured a contrast, $c$, of the excess absorption of  3.91$\pm$0.22\% (17.6$\sigma$), 8.17$^{+ 0.80 }_{ -0.76 }$\% (10.5$\sigma$), and 1.36$\pm$0.17\% (8.0$\sigma$) for WASP-69\,b, WASP-107\,b, and HAT-P-11\,b, respectively, confirming the literature detections. Our analysis of WASP-69\,b showed a night-to-night variability in the helium absorption levels, with the first transit exhibiting a higher absorption value compared to the others. We thus inspected the H$\alpha$ line in HARPS-N spectra, finding an opposite behaviour in H$\alpha$ compared to that recorded in \Hei on UT 29 July 2019. We interpreted this as being due to the effect of stellar activity, and we speculated on the possible origins of this effect. 
	
	We report a detection of \Hei in the upper atmosphere of GJ\,3470\,b of 1.75$^{+ 0.39 }_{ -0.36 }$\% (4.7$\sigma$).
	Our result is in agreement with previous studies by \citet{Ninan2020} and \citet{Palle2020}. However, if we consider only the two nights not affected by the OH emission line, the signal seems to appear from only one transit, supporting the literature discrepancy and indicating the presence of variability, whereby the extended atmosphere is more evident on certain nights than on others. Additional observations are needed to unveil the origin of the \Hei signal. An inspection of the H$\alpha$ line reveals a hydrogen absorption signal during the same transit event. Our ATES simulations are not able to reproduce an absorption depth or profile consistent with our observations, thus making stellar activity the most plausible origin for this feature. Further analyses that include more detailed radiative transfer models are needed to confirm this hypothesis.
	
	In agreement with previous work \citep{Nortmann2018}, we did not detect a \Hei excess of absorption for GJ\,436\,b. Our finding is compatible with zero absorption.
	
	We finally placed our results in the context of other \Hei analyses of planets orbiting K- and MV-type stars obtained with GIANO-B, namely HD189733b \citep{Guilluy2020}, WASP-80\,b \citep{Fossati2022}, and HAT-P-3b \citep{Guilluy2023}. 
	We explored potential trends associated with stellar and planetary parameters believed to influence the \Hei detection, such as the EUV flux, the effective temperature, the planet's gravity, and the \logrhk. 
	Our analysis does not show any significant correlation, and our investigation did not reveal any relationship between GIANO-B detections and non-detections in the atmospheres of planets around M-K dwarf stars and the parameters we explored. We emphasise that this could also be a consequence of our small sample size. Moreover, both the stellar EUV flux and the helium abundance in the investigated atmospheres
	are highly uncertain, which may introduce noise when attempting to identify potential correlations. Additional observations are thus needed, together with the investigation of new parameters that could influence the \Hei observability.
	
	Our work underscores the importance of a \Hei survey homogeneous in both observation techniques and data analysis methods, both of which are essential to understanding the key parameters governing \Hei detectability. It also emphasises the necessity of simultaneous nIR and visible monitoring to investigate the potential presence of stellar activity pseudo-signals in \Hei measurements.
	
	
	\bibliographystyle{aa}
	\bibliography{ref_he}   

\newcommand{\noop}[1]{}
\begin{thebibliography}{112}
\expandafter\ifx\csname natexlab\endcsname\relax\def\natexlab#1{#1}\fi

\bibitem[{{Akeson} {et~al.}(2013){Akeson}, {Chen}, {Ciardi}, {Crane}, {Good},
  {Harbut}, {Jackson}, {Kane}, {Laity}, {Leifer}, {Lynn}, {McElroy}, {Papin},
  {Plavchan}, {Ram{\'\i}rez}, {Rey}, {von Braun}, {Wittman}, {Abajian}, {Ali},
  {Beichman}, {Beekley}, {Berriman}, {Berukoff}, {Bryden}, {Chan}, {Groom},
  {Lau}, {Payne}, {Regelson}, {Saucedo}, {Schmitz}, {Stauffer}, {Wyatt}, \&
  {Zhang}}]{Akeson2013}
{Akeson}, R.~L., {Chen}, X., {Ciardi}, D., {et~al.} 2013, \pasp, 125, 989

\bibitem[{{Allart} {et~al.}(2019){Allart}, {Bourrier}, {Lovis}, {Ehrenreich},
  {Aceituno}, {Guijarro}, {Pepe}, {Sing}, {Spake}, \&
  {Wyttenbach}}]{Allart2019}
{Allart}, R., {Bourrier}, V., {Lovis}, C., {et~al.} 2019, \aap, 623, A58

\bibitem[{{Allart} {et~al.}(2018){Allart}, {Bourrier}, {Lovis}, {Ehrenreich},
  {Spake}, {Wyttenbach}, {Pino}, {Pepe}, {Sing}, \& {Lecavelier des
  Etangs}}]{Allart2018}
{Allart}, R., {Bourrier}, V., {Lovis}, C., {et~al.} 2018, Science, 362, 1384

\bibitem[{{Allart} {et~al.}(2023){Allart}, {Lem{\'e}e-Joliecoeur}, {Jaziri},
  {Lafreni{\`e}re}, {Artigau}, {Cook}, {Darveau-Bernier}, {Dang}, {Cadieux},
  {Boucher}, {Bourrier}, {Deibert}, {Pelletier}, {Radica}, {Benneke},
  {Carmona}, {Cloutier}, {Cowan}, {Delfosse}, {Donati}, {Doyon}, {Figueira},
  {Forveille}, {Fouqu{\'e}}, {Gaidos}, {Gu}, {H{\'e}brard}, {Kiefer},
  {K{\'o}sp{\'a}l}, {Jayawardhana}, {Martioli}, {Dos Santos}, {Shang},
  {Turner}, \& {Vidotto}}]{Allart2023}
{Allart}, R., {Lem{\'e}e-Joliecoeur}, P.~B., {Jaziri}, A.~Y., {et~al.} 2023,
  \aap, 677, A164

\bibitem[{{Anderson} {et~al.}(2014){Anderson}, {Collier Cameron}, {Delrez},
  {Doyle}, {Faedi}, {Fumel}, {Gillon}, {G{\'o}mez Maqueo Chew}, {Hellier},
  {Jehin}, {Lendl}, {Maxted}, {Pepe}, {Pollacco}, {Queloz}, {S{\'e}gransan},
  {Skillen}, {Smalley}, {Smith}, {Southworth}, {Triaud}, {Turner}, {Udry}, \&
  {West}}]{Anderson2014}
{Anderson}, D.~R., {Collier Cameron}, A., {Delrez}, L., {et~al.} 2014, \mnras,
  445, 1114

\bibitem[{{Anderson} {et~al.}(2017){Anderson}, {Collier Cameron}, {Delrez},
  {Doyle}, {Gillon}, {Hellier}, {Jehin}, {Lendl}, {Maxted}, {Madhusudhan},
  {Pepe}, {Pollacco}, {Queloz}, {S{\'e}gransan}, {Smalley}, {Smith}, {Triaud},
  {Turner}, {Udry}, \& {West}}]{Anderson2017}
{Anderson}, D.~R., {Collier Cameron}, A., {Delrez}, L., {et~al.} 2017, \aap,
  604, A110

\bibitem[{{Bakos} {et~al.}(2010){Bakos}, {Torres}, {P{\'a}l}, {Hartman},
  {Kov{\'a}cs}, {Noyes}, {Latham}, {Sasselov}, {Sip{\H{o}}cz}, {Esquerdo},
  {Fischer}, {Johnson}, {Marcy}, {Butler}, {Isaacson}, {Howard}, {Vogt},
  {Kov{\'a}cs}, {Fernandez}, {Mo{\'o}r}, {Stefanik}, {L{\'a}z{\'a}r}, {Papp},
  \& {S{\'a}ri}}]{bakos2010}
{Bakos}, G.~{\'A}., {Torres}, G., {P{\'a}l}, A., {et~al.} 2010, \apj, 710, 1724

\bibitem[{Basilicata {et~al.}(2024)Basilicata, Giacobbe, Bonomo, Scandariato,
  Brogi, Singh, Paola, Mancini, Sozzetti, Lanza, Cubillos, Damasso, Desidera,
  Biazzo, Bignamini, Borsa, Cabona, Carleo, Ghedina, Guilluy, Maggio, Mainella,
  Micela, Molinari, Molinaro, Nardiello, Pedani, Pino, Poretti, Southworth,
  Stangret, \& Turrini}]{Basilicata2023}
Basilicata, M., Giacobbe, P., Bonomo, A.~S., {et~al.} 2024
  [\eprint[arXiv]{2403.01527}]

\bibitem[{{Beaug{\'e}} \& {Nesvorn{\'y}}(2013)}]{Beauge2013}
{Beaug{\'e}}, C. \& {Nesvorn{\'y}}, D. 2013, \apj, 763, 12

\bibitem[{{Ben-Jaffel} {et~al.}(2022){Ben-Jaffel}, {Ballester}, {Mu{\~n}oz},
  {Lavvas}, {Sing}, {Sanz-Forcada}, {Cohen}, {Kataria}, {Henry}, {Buchhave},
  {Mikal-Evans}, {Wakeford}, \& {L{\'o}pez-Morales}}]{Jaffel2022}
{Ben-Jaffel}, L., {Ballester}, G.~E., {Mu{\~n}oz}, A.~G., {et~al.} 2022, Nature
  Astronomy, 6, 141

\bibitem[{{Benneke} {et~al.}(2019){Benneke}, {Knutson}, {Lothringer},
  {Crossfield}, {Moses}, {Morley}, {Kreidberg}, {Fulton}, {Dragomir}, {Howard},
  {Wong}, {D{\'e}sert}, {McCullough}, {Kempton}, {Fortney}, {Gilliland},
  {Deming}, \& {Kammer}}]{Benneke2019}
{Benneke}, B., {Knutson}, H.~A., {Lothringer}, J., {et~al.} 2019, Nature
  Astronomy, 3, 813

\bibitem[{{Biassoni} {et~al.}(2024){Biassoni}, {Caldiroli}, {Gallo}, {Haardt},
  {Spinelli}, \& {Borsa}}]{Biassoni2023}
{Biassoni}, F., {Caldiroli}, A., {Gallo}, E., {et~al.} 2024, \aap, 682, A115

\bibitem[{{Biddle} {et~al.}(2014){Biddle}, {Pearson}, {Crossfield}, {Fulton},
  {Ciceri}, {Eastman}, {Barman}, {Mann}, {Henry}, {Howard}, {Williamson},
  {Sinukoff}, {Dragomir}, {Vican}, {Mancini}, {Southworth}, {Greenberg},
  {Turner}, {Thompson}, {Taylor}, {Levine}, \& {Webber}}]{Biddle2014}
{Biddle}, L.~I., {Pearson}, K.~A., {Crossfield}, I. J.~M., {et~al.} 2014,
  \mnras, 443, 1810

\bibitem[{{Bonfils} {et~al.}(2012){Bonfils}, {Gillon}, {Udry}, {Armstrong},
  {Bouchy}, {Delfosse}, {Forveille}, {Fumel}, {Jehin}, {Lendl}, {Lovis},
  {Mayor}, {McCormac}, {Neves}, {Pepe}, {Perrier}, {Pollaco}, {Queloz}, \&
  {Santos}}]{Bonfils2012}
{Bonfils}, X., {Gillon}, M., {Udry}, S., {et~al.} 2012, \aap, 546, A27

\bibitem[{{Bonomo} {et~al.}(2023){Bonomo}, {Dumusque}, {Massa}, {Mortier},
  {Bongiolatti}, {Malavolta}, {Sozzetti}, {Buchhave}, {Damasso}, {Haywood},
  {Morbidelli}, {Latham}, {Molinari}, {Pepe}, {Poretti}, {Udry}, {Affer},
  {Boschin}, {Charbonneau}, {Cosentino}, {Cretignier}, {Ghedina}, {Lega},
  {L{\'o}pez-Morales}, {Margini}, {Mart{\'\i}nez Fiorenzano}, {Mayor},
  {Micela}, {Pedani}, {Pinamonti}, {Rice}, {Sasselov}, {Tronsgaard}, \&
  {Vanderburg}}]{Bonomo2023}
{Bonomo}, A.~S., {Dumusque}, X., {Massa}, A., {et~al.} 2023, \aap, 677, A33

\bibitem[{{Bourrier} {et~al.}(2023){Bourrier}, {Attia}, {Mallonn}, {Marret},
  {Lendl}, {Konig}, {Krenn}, {Cretignier}, {Allart}, {Henry}, {Bryant},
  {Leleu}, {Nielsen}, {Hebrard}, {Hara}, {Ehrenreich}, {Seidel}, {dos Santos},
  {Lovis}, {Bayliss}, {Cegla}, {Dumusque}, {Boisse}, {Boucher}, {Bouchy},
  {Pepe}, {Lavie}, {Rey Cerda}, {S{\'e}gransan}, {Udry}, \&
  {Vrignaud}}]{Bourrier2023}
{Bourrier}, V., {Attia}, O., {Mallonn}, M., {et~al.} 2023, \aap, 669, A63

\bibitem[{{Bourrier} \& {des Etangs}(2018)}]{Bourrier2018_lib}
{Bourrier}, V. \& {des Etangs}, A.~L. 2018, in Handbook of Exoplanets, ed.
  H.~J. {Deeg} \& J.~A. {Belmonte}, 148

\bibitem[{{Bourrier} {et~al.}(2018){Bourrier}, {Ehrenreich}, {Allan}, {Sing},
  {Wakeford}, {Griffith}, {D{\'e}sert}, {Dinelli}, {L{\'o}pez-Morales},
  {Garc{\'\i}a Mu{\~n}oz}, {Louden}, {Lavvas}, {Nikolov}, {Yurchenko},
  {Wheatley}, {Goyal}, {Vidal-Madjar}, {Zarka}, {Ben-Jaffel}, {Howarth},
  {Kirk}, {Knutson}, {Meadows}, {Murokh}, {Rimmer}, {Tsiaras}, {Bezard}, \&
  {Mayne}}]{Bourrier2018}
{Bourrier}, V., {Ehrenreich}, D., {Allan}, A., {et~al.} 2018, \aap, 619, A1

\bibitem[{{Bourrier} {et~al.}(2022){Bourrier}, {Zapatero Osorio}, {Allart},
  {Attia}, {Cretignier}, {Dumusque}, {Lovis}, {Adibekyan}, {Borsa}, {Figueira},
  {Gonz{\'a}lez Hern{\'a}ndez}, {Mehner}, {Santos}, {Schmidt}, {Seidel},
  {Sozzetti}, {Alibert}, {Casasayas-Barris}, {Ehrenreich}, {Lo Curto},
  {Martins}, {Di Marcantonio}, {M{\'e}gevand}, {Nunes}, {Palle}, {Poretti}, \&
  {Sousa}}]{Bourrier2022}
{Bourrier}, V., {Zapatero Osorio}, M.~R., {Allart}, R., {et~al.} 2022, \aap,
  663, A160

\bibitem[{{Brown} {et~al.}(2001){Brown}, {Charbonneau}, {Gilliland}, {Noyes},
  \& {Burrows}}]{Brown2001}
{Brown}, T.~M., {Charbonneau}, D., {Gilliland}, R.~L., {Noyes}, R.~W., \&
  {Burrows}, A. 2001, \apj, 552, 699

\bibitem[{{Butler} {et~al.}(2004){Butler}, {Vogt}, {Marcy}, {Fischer},
  {Wright}, {Henry}, {Laughlin}, \& {Lissauer}}]{Butler2004}
{Butler}, R.~P., {Vogt}, S.~S., {Marcy}, G.~W., {et~al.} 2004, \apj, 617, 580

\bibitem[{{Caldiroli} {et~al.}(2022){Caldiroli}, {Haardt}, {Gallo}, {Spinelli},
  {Malsky}, \& {Rauscher}}]{Caldiroli2022}
{Caldiroli}, A., {Haardt}, F., {Gallo}, E., {et~al.} 2022, \aap, 663, A122

\bibitem[{{Casasayas-Barris} {et~al.}(2017){Casasayas-Barris}, {Palle},
  {Nowak}, {Yan}, {Nortmann}, \& {Murgas}}]{Casasayas-Barris2017}
{Casasayas-Barris}, N., {Palle}, E., {Nowak}, G., {et~al.} 2017, \aap, 608,
  A135

\bibitem[{Chachan {et~al.}(2019)Chachan, Knutson, Gao, \& et~al.}]{Chachan2019}
Chachan, Y., Knutson, H.~A., Gao, P., \& et~al. 2019, The Astronomical Journal,
  158, 244

\bibitem[{{Chen} {et~al.}(2017){Chen}, {Guenther}, {Pall{\'e}}, {Nortmann},
  {Nowak}, {Kunz}, {Parviainen}, \& {Murgas}}]{Chen2017}
{Chen}, G., {Guenther}, E.~W., {Pall{\'e}}, E., {et~al.} 2017, \aap, 600, A138

\bibitem[{{Chen} {et~al.}(2018){Chen}, {Pall{\'e}}, {Welbanks},
  {Prieto-Arranz}, {Madhusudhan}, {Gandhi}, {Casasayas-Barris}, {Murgas},
  {Nortmann}, {Crouzet}, {Parviainen}, \& {Gandolfi}}]{Chen2018}
{Chen}, G., {Pall{\'e}}, E., {Welbanks}, L., {et~al.} 2018, \aap, 616, A145

\bibitem[{{Claudi} {et~al.}(2017){Claudi}, {Benatti}, {Carleo}, {Ghedina},
  {Guerra}, {Micela}, {Molinari}, {Oliva}, {Rainer}, {Tozzi}, {Baffa},
  {Baruffolo}, {Buchschacher}, {Cecconi}, {Cosentino}, {Fantinel}, {Fini},
  {Ghinassi}, {Giani}, {Gonzalez}, {Gonzalez}, {Gratton}, {Harutyunyan},
  {Hernandez}, {Lodi}, {Malavolta}, {Maldonado}, {Origlia}, {Sanna}, {Sanjuan},
  {Scuderi}, {Seemann}, {Sozzetti}, {Perez Ventura}, {Hernandez Diaz}, {Galli},
  {Gonzalez}, {Riverol}, \& {Riverol}}]{GIARPS_claudi}
{Claudi}, R., {Benatti}, S., {Carleo}, I., {et~al.} 2017, European Physical
  Journal Plus, 132, 364

\bibitem[{{Cosentino} {et~al.}(2012){Cosentino}, {Lovis}, {Pepe}, {Collier
  Cameron}, {Latham}, {Molinari}, {Udry}, {Bezawada}, {Black}, {Born},
  {Buchschacher}, {Charbonneau}, {Figueira}, {Fleury}, {Galli}, {Gallie},
  {Gao}, {Ghedina}, {Gonzalez}, {Gonzalez}, {Guerra}, {Henry}, {Horne},
  {Hughes}, {Kelly}, {Lodi}, {Lunney}, {Maire}, {Mayor}, {Micela}, {Ordway},
  {Peacock}, {Phillips}, {Piotto}, {Pollacco}, {Queloz}, {Rice}, {Riverol},
  {Riverol}, {San Juan}, {Sasselov}, {Segransan}, {Sozzetti}, {Sosnowska},
  {Stobie}, {Szentgyorgyi}, {Vick}, \& {Weber}}]{Cosentino2012}
{Cosentino}, R., {Lovis}, C., {Pepe}, F., {et~al.} 2012, Society of
  Photo-Optical Instrumentation Engineers (SPIE) Conference Series, Vol. 8446,
  {Harps-N: the new planet hunter at TNG}, 84461V

\bibitem[{{Cutri} {et~al.}(2003){Cutri}, {Skrutskie}, {van Dyk}, {Beichman},
  {Carpenter}, {Chester}, {Cambresy}, {Evans}, {Fowler}, {Gizis}, {Howard},
  {Huchra}, {Jarrett}, {Kopan}, {Kirkpatrick}, {Light}, {Marsh}, {McCallon},
  {Schneider}, {Stiening}, {Sykes}, {Weinberg}, {Wheaton}, {Wheelock}, \&
  {Zacarias}}]{Cutri2003}
{Cutri}, R.~M., {Skrutskie}, M.~F., {van Dyk}, S., {et~al.} 2003, VizieR Online
  Data Catalog, II/246

\bibitem[{{Davis} \& {Wheatley}(2009)}]{Davis2009}
{Davis}, T.~A. \& {Wheatley}, P.~J. 2009, \mnras, 396, 1012

\bibitem[{{Dos Santos}(2023)}]{DosSantos2023}
{Dos Santos}, L.~A. 2023, in Winds of Stars and Exoplanets, ed. A.~A.
  {Vidotto}, L.~{Fossati}, \& J.~S. {Vink}, Vol. 370, 56--71

\bibitem[{{Dyrek} {et~al.}(2024){Dyrek}, {Min}, {Decin}, {Bouwman}, {Crouzet},
  {Molli{\`e}re}, {Lagage}, {Konings}, {Tremblin}, {G{\"u}del}, {Pye},
  {Waters}, {Henning}, {Vandenbussche}, {Ardevol Martinez}, {Argyriou},
  {Ducrot}, {Heinke}, {van Looveren}, {Absil}, {Barrado}, {Baudoz},
  {Boccaletti}, {Cossou}, {Coulais}, {Edwards}, {Gastaud}, {Glasse}, {Glauser},
  {Greene}, {Kendrew}, {Krause}, {Lahuis}, {Mueller}, {Olofsson}, {Patapis},
  {Rouan}, {Royer}, {Scheithauer}, {Waldmann}, {Whiteford}, {Colina}, {van
  Dishoeck}, {{\"O}stlin}, {Ray}, \& {Wright}}]{Dyrek2023}
{Dyrek}, A., {Min}, M., {Decin}, L., {et~al.} 2024, \nat, 625, 51

\bibitem[{{D’Arpa} {et~al.}(2023){D’Arpa}, {Saba}, {Borsa}, \&
  et~al}]{Darpa}
{D’Arpa}, M.~C., {Saba}, A., {Borsa}, F., \& et~al. 2023, \aap, submitted

\bibitem[{{Ehrenreich} {et~al.}(2014){Ehrenreich}, {Bonfils}, {Lovis},
  {Delfosse}, {Forveille}, {Mayor}, {Neves}, {Santos}, {Udry}, \&
  {S{\'e}gransan}}]{Ehrenreich2014}
{Ehrenreich}, D., {Bonfils}, X., {Lovis}, C., {et~al.} 2014, \aap, 570, A89

\bibitem[{Ehrenreich {et~al.}(2015)Ehrenreich, Bourrier, Wheatley, \&
  et~al.}]{Ehrenreich2015}
Ehrenreich, D., Bourrier, V., Wheatley, P.~J., \& et~al. 2015, Nature, 522, 459

\bibitem[{{Estrela} {et~al.}(2021){Estrela}, {Swain}, {Roudier}, {West},
  {Sedaghati}, \& {Valio}}]{Estrela2021}
{Estrela}, R., {Swain}, M.~R., {Roudier}, G.~M., {et~al.} 2021, \aj, 162, 91

\bibitem[{{Fossati} {et~al.}(2022){Fossati}, {Guilluy}, {Shaikhislamov},
  {Carleo}, {Borsa}, {Bonomo}, {Giacobbe}, {Rainer}, {Cecchi-Pestellini},
  {Khodachenko}, {Efimov}, {Rumenskikh}, {Miroshnichenko}, {Berezutsky},
  {Nascimbeni}, {Brogi}, {Lanza}, {Mancini}, {Affer}, {Benatti}, {Biazzo},
  {Bignamini}, {Carosati}, {Claudi}, {Cosentino}, {Covino}, {Desidera},
  {Fiorenzano}, {Harutyunyan}, {Maggio}, {Malavolta}, {Maldonado}, {Micela},
  {Molinari}, {Pagano}, {Pedani}, {Piotto}, {Poretti}, {Scandariato},
  {Sozzetti}, \& {Stoev}}]{Fossati2022}
{Fossati}, L., {Guilluy}, G., {Shaikhislamov}, I.~F., {et~al.} 2022, \aap, 658,
  A136

\bibitem[{{Fossati} {et~al.}(2023){Fossati}, {Pillitteri}, {Shaikhislamov},
  {Bonfanti}, {Borsa}, {Carleo}, {Guilluy}, \& {Rumenskikh}}]{Fossati2023}
{Fossati}, L., {Pillitteri}, I., {Shaikhislamov}, I.~F., {et~al.} 2023, \aap,
  673, A37

\bibitem[{{Fouqu{\'e}} {et~al.}(2018){Fouqu{\'e}}, {Moutou}, {Malo},
  {Martioli}, {Lim}, {Rajpurohit}, {Artigau}, {Delfosse}, {Donati},
  {Forveille}, {Morin}, {Allard}, {Delage}, {Doyon}, {H{\'e}brard}, \&
  {Neves}}]{Fouqu2018}
{Fouqu{\'e}}, P., {Moutou}, C., {Malo}, L., {et~al.} 2018, \mnras, 475, 1960

\bibitem[{Fraine {et~al.}(2014)Fraine, Deming, Benneke, \& et~al.}]{Fraine2014}
Fraine, J.~D., Deming, D., Benneke, B., \& et~al. 2014, Nature, 513, 526

\bibitem[{{Fulton} \& {Petigura}(2018)}]{Fulton2018}
{Fulton}, B.~J. \& {Petigura}, E.~A. 2018, \aj, 156, 264

\bibitem[{{Fulton} {et~al.}(2017){Fulton}, {Petigura}, {Howard}, {Isaacson},
  {Marcy}, {Cargile}, {Hebb}, {Weiss}, {Johnson}, {Morton}, {Sinukoff},
  {Crossfield}, \& {Hirsch}}]{Fulton2017}
{Fulton}, B.~J., {Petigura}, E.~A., {Howard}, A.~W., {et~al.} 2017, \aj, 154,
  109

\bibitem[{{Gaia Collaboration} {et~al.}(2018){Gaia Collaboration}, {Brown},
  {Vallenari}, {Prusti}, {de Bruijne}, {Babusiaux}, {Bailer-Jones}, {Biermann},
  {Evans}, {Eyer}, {Jansen}, {Jordi}, {Klioner}, {Lammers}, {Lindegren},
  {Luri}, {Mignard}, {Panem}, {Pourbaix}, {Randich}, {Sartoretti}, {Siddiqui},
  {Soubiran}, {van Leeuwen}, {Walton}, {Arenou}, {Bastian}, {Cropper},
  {Drimmel}, {Katz}, {Lattanzi}, {Bakker}, {Cacciari}, {Casta{\~n}eda},
  {Chaoul}, {Cheek}, {De Angeli}, {Fabricius}, {Guerra}, {Holl}, {Masana},
  {Messineo}, {Mowlavi}, {Nienartowicz}, {Panuzzo}, {Portell}, {Riello},
  {Seabroke}, {Tanga}, {Th{\'e}venin}, {Gracia-Abril}, {Comoretto},
  {Garcia-Reinaldos}, {Teyssier}, {Altmann}, {Andrae}, {Audard},
  {Bellas-Velidis}, {Benson}, {Berthier}, {Blomme}, {Burgess}, {Busso},
  {Carry}, {Cellino}, {Clementini}, {Clotet}, {Creevey}, {Davidson}, {De
  Ridder}, {Delchambre}, {Dell'Oro}, {Ducourant},
  {Fern{\'a}ndez-Hern{\'a}ndez}, {Fouesneau}, {Fr{\'e}mat}, {Galluccio},
  {Garc{\'\i}a-Torres}, {Gonz{\'a}lez-N{\'u}{\~n}ez}, {Gonz{\'a}lez-Vidal},
  {Gosset}, {Guy}, {Halbwachs}, {Hambly}, {Harrison}, {Hern{\'a}ndez},
  {Hestroffer}, {Hodgkin}, {Hutton}, {Jasniewicz}, {Jean-Antoine-Piccolo},
  {Jordan}, {Korn}, {Krone-Martins}, {Lanzafame}, {Lebzelter}, {L{\"o}ffler},
  {Manteiga}, {Marrese}, {Mart{\'\i}n-Fleitas}, {Moitinho}, {Mora}, {Muinonen},
  {Osinde}, {Pancino}, {Pauwels}, {Petit}, {Recio-Blanco}, {Richards},
  {Rimoldini}, {Robin}, {Sarro}, {Siopis}, {Smith}, {Sozzetti}, {S{\"u}veges},
  {Torra}, {van Reeven}, {Abbas}, {Abreu Aramburu}, {Accart}, {Aerts},
  {Altavilla}, {{\'A}lvarez}, {Alvarez}, {Alves}, {Anderson}, {Andrei},
  {Anglada Varela}, {Antiche}, {Antoja}, {Arcay}, {Astraatmadja}, {Bach},
  {Baker}, {Balaguer-N{\'u}{\~n}ez}, {Balm}, {Barache}, {Barata}, {Barbato},
  {Barblan}, {Barklem}, {Barrado}, {Barros}, {Barstow}, {Bartholom{\'e}
  Mu{\~n}oz}, {Bassilana}, {Becciani}, {Bellazzini}, {Berihuete}, {Bertone},
  {Bianchi}, {Bienaym{\'e}}, {Blanco-Cuaresma}, {Boch}, {Boeche}, {Bombrun},
  {Borrachero}, {Bossini}, {Bouquillon}, {Bourda}, {Bragaglia}, {Bramante},
  {Breddels}, {Bressan}, {Brouillet}, {Br{\"u}semeister}, {Brugaletta},
  {Bucciarelli}, {Burlacu}, {Busonero}, {Butkevich}, {Buzzi}, {Caffau},
  {Cancelliere}, {Cannizzaro}, {Cantat-Gaudin}, {Carballo}, {Carlucci},
  {Carrasco}, {Casamiquela}, {Castellani}, {Castro-Ginard}, {Charlot},
  {Chemin}, {Chiavassa}, {Cocozza}, {Costigan}, {Cowell}, {Crifo}, {Crosta},
  {Crowley}, {Cuypers}, {Dafonte}, {Damerdji}, {Dapergolas}, {David}, {David},
  {de Laverny}, {De Luise}, {De March}, {de Martino}, {de Souza}, {de Torres},
  {Debosscher}, {del Pozo}, {Delbo}, {Delgado}, {Delgado}, {Di Matteo},
  {Diakite}, {Diener}, {Distefano}, {Dolding}, {Drazinos}, {Dur{\'a}n},
  {Edvardsson}, {Enke}, {Eriksson}, {Esquej}, {Eynard Bontemps}, {Fabre},
  {Fabrizio}, {Faigler}, {Falc{\~a}o}, {Farr{\`a}s Casas}, {Federici},
  {Fedorets}, {Fernique}, {Figueras}, {Filippi}, {Findeisen}, {Fonti},
  {Fraile}, {Fraser}, {Fr{\'e}zouls}, {Gai}, {Galleti}, {Garabato},
  {Garc{\'\i}a-Sedano}, {Garofalo}, {Garralda}, {Gavel}, {Gavras}, {Gerssen},
  {Geyer}, {Giacobbe}, {Gilmore}, {Girona}, {Giuffrida}, {Glass}, {Gomes},
  {Granvik}, {Gueguen}, {Guerrier}, {Guiraud}, {Guti{\'e}rrez-S{\'a}nchez},
  {Haigron}, {Hatzidimitriou}, {Hauser}, {Haywood}, {Heiter}, {Helmi}, {Heu},
  {Hilger}, {Hobbs}, {Hofmann}, {Holland}, {Huckle}, {Hypki}, {Icardi},
  {Jan{\ss}en}, {Jevardat de Fombelle}, {Jonker}, {Juh{\'a}sz}, {Julbe},
  {Karampelas}, {Kewley}, {Klar}, {Kochoska}, {Kohley}, {Kolenberg},
  {Kontizas}, {Kontizas}, {Koposov}, {Kordopatis}, {Kostrzewa-Rutkowska},
  {Koubsky}, {Lambert}, {Lanza}, {Lasne}, {Lavigne}, {Le Fustec}, {Le
  Poncin-Lafitte}, {Lebreton}, {Leccia}, {Leclerc}, {Lecoeur-Taibi},
  {Lenhardt}, {Leroux}, {Liao}, {Licata}, {Lindstr{\o}m}, {Lister}, {Livanou},
  {Lobel}, {L{\'o}pez}, {Managau}, {Mann}, {Mantelet}, {Marchal}, {Marchant},
  {Marconi}, {Marinoni}, {Marschalk{\'o}}, {Marshall}, {Martino}, {Marton},
  {Mary}, {Massari}, {Matijevi{\v{c}}}, {Mazeh}, {McMillan}, {Messina},
  {Michalik}, {Millar}, {Molina}, {Molinaro}, {Moln{\'a}r}, {Montegriffo},
  {Mor}, {Morbidelli}, {Morel}, {Morris}, {Mulone}, {Muraveva}, {Musella},
  {Nelemans}, {Nicastro}, {Noval}, {O'Mullane}, {Ord{\'e}novic},
  {Ord{\'o}{\~n}ez-Blanco}, {Osborne}, {Pagani}, {Pagano}, {Pailler},
  {Palacin}, {Palaversa}, {Panahi}, {Pawlak}, {Piersimoni}, {Pineau}, {Plachy},
  {Plum}, {Poggio}, {Poujoulet}, {Pr{\v{s}}a}, {Pulone}, {Racero}, {Ragaini},
  {Rambaux}, {Ramos-Lerate}, {Regibo}, {Reyl{\'e}}, {Riclet}, {Ripepi}, {Riva},
  {Rivard}, {Rixon}, {Roegiers}, {Roelens}, {Romero-G{\'o}mez}, {Rowell},
  {Royer}, {Ruiz-Dern}, {Sadowski}, {Sagrist{\`a} Sell{\'e}s}, {Sahlmann},
  {Salgado}, {Salguero}, {Sanna}, {Santana-Ros}, {Sarasso}, {Savietto},
  {Schultheis}, {Sciacca}, {Segol}, {Segovia}, {S{\'e}gransan}, {Shih},
  {Siltala}, {Silva}, {Smart}, {Smith}, {Solano}, {Solitro}, {Sordo}, {Soria
  Nieto}, {Souchay}, {Spagna}, {Spoto}, {Stampa}, {Steele},
  {Steidelm{\"u}ller}, {Stephenson}, {Stoev}, {Suess}, {Surdej}, {Szabados},
  {Szegedi-Elek}, {Tapiador}, {Taris}, {Tauran}, {Taylor}, {Teixeira},
  {Terrett}, {Teyssandier}, {Thuillot}, {Titarenko}, {Torra Clotet}, {Turon},
  {Ulla}, {Utrilla}, {Uzzi}, {Vaillant}, {Valentini}, {Valette}, {van Elteren},
  {Van Hemelryck}, {van Leeuwen}, {Vaschetto}, {Vecchiato}, {Veljanoski},
  {Viala}, {Vicente}, {Vogt}, {von Essen}, {Voss}, {Votruba}, {Voutsinas},
  {Walmsley}, {Weiler}, {Wertz}, {Wevers}, {Wyrzykowski}, {Yoldas},
  {{\v{Z}}erjal}, {Ziaeepour}, {Zorec}, {Zschocke}, {Zucker}, {Zurbach}, \&
  {Zwitter}}]{Gaia2018}
{Gaia Collaboration}, {Brown}, A.~G.~A., {Vallenari}, A., {et~al.} 2018, \aap,
  616, A1

\bibitem[{{Garc{\'\i}a Mu{\~n}oz} \& {Schneider}(2019)}]{Garcia2019}
{Garc{\'\i}a Mu{\~n}oz}, A. \& {Schneider}, P.~C. 2019, \apjl, 884, L43

\bibitem[{{Giacobbe} {et~al.}(2021){Giacobbe}, {Brogi}, {Gandhi}, {Cubillos},
  {Bonomo}, {Sozzetti}, {Fossati}, {Guilluy}, {Carleo}, {Rainer},
  {Harutyunyan}, {Borsa}, {Pino}, {Nascimbeni}, {Benatti}, {Biazzo},
  {Bignamini}, {Chubb}, {Claudi}, {Cosentino}, {Covino}, {Damasso}, {Desidera},
  {Fiorenzano}, {Ghedina}, {Lanza}, {Leto}, {Maggio}, {Malavolta}, {Maldonado},
  {Micela}, {Molinari}, {Pagano}, {Pedani}, {Piotto}, {Poretti}, {Scandariato},
  {Yurchenko}, {Fantinel}, {Galli}, {Lodi}, {Sanna}, \& {Tozzi}}]{Giacobbe2021}
{Giacobbe}, P., {Brogi}, M., {Gandhi}, S., {et~al.} 2021, \nat, 592, 205

\bibitem[{{Gillon} {et~al.}(2007){Gillon}, {Pont}, {Demory}, {Mallmann},
  {Mayor}, {Mazeh}, {Queloz}, {Shporer}, {Udry}, \& {Vuissoz}}]{Gillon2007}
{Gillon}, M., {Pont}, F., {Demory}, B.~O., {et~al.} 2007, \aap, 472, L13

\bibitem[{{Guilluy} {et~al.}(2020){Guilluy}, {Andretta}, {Borsa}, {Giacobbe},
  {Sozzetti}, {Covino}, {Bourrier}, {Fossati}, {Bonomo}, {Esposito},
  {Giampapa}, {Harutyunyan}, {Rainer}, {Brogi}, {Bruno}, {Claudi}, {Frustagli},
  {Lanza}, {Mancini}, {Pino}, {Poretti}, {Scandariato}, {Affer}, {Baffa},
  {Baruffolo}, {Benatti}, {Biazzo}, {Bignamini}, {Boschin}, {Carleo},
  {Cecconi}, {Cosentino}, {Damasso}, {Desidera}, {Falcini}, {Martinez
  Fiorenzano}, {Ghedina}, {Gonz{\'a}lez-{\'A}lvarez}, {Guerra}, {Hernandez},
  {Leto}, {Maggio}, {Malavolta}, {Maldonado}, {Micela}, {Molinari},
  {Nascimbeni}, {Pagano}, {Pedani}, {Piotto}, \& {Reiners}}]{Guilluy2020}
{Guilluy}, G., {Andretta}, V., {Borsa}, F., {et~al.} 2020, \aap, 639, A49

\bibitem[{{Guilluy} {et~al.}(2023){Guilluy}, {Bourrier}, {Jaziri}, {Dethier},
  {Mounzer}, {Giacobbe}, {Attia}, {Allart}, {Bonomo}, {Dos Santos}, {Rainer},
  \& {Sozzetti}}]{Guilluy2023}
{Guilluy}, G., {Bourrier}, V., {Jaziri}, Y., {et~al.} 2023, \aap, 676, A130

\bibitem[{{Guilluy} {et~al.}(2022){Guilluy}, {Giacobbe}, {Carleo}, {Cubillos},
  {Sozzetti}, {Bonomo}, {Brogi}, {Gandhi}, {Fossati}, {Nascimbeni}, {Turrini},
  {Schisano}, {Borsa}, {Lanza}, {Mancini}, {Maggio}, {Malavolta}, {Micela},
  {Pino}, {Rainer}, {Bignamini}, {Claudi}, {Cosentino}, {Covino}, {Desidera},
  {Fiorenzano}, {Harutyunyan}, {Lorenzi}, {Knapic}, {Molinari}, {Pacetti},
  {Pagano}, {Pedani}, {Piotto}, \& {Poretti}}]{Guilluy2022}
{Guilluy}, G., {Giacobbe}, P., {Carleo}, I., {et~al.} 2022, \aap, 665, A104

\bibitem[{{Gully-Santiago} {et~al.}(2023){Gully-Santiago}, {Morley}, {Luna},
  {MacLeod}, {Oklop{\v{c}}i{\'c}}, {Ganesh}, {Tran}, {Zhang}, {Bowler},
  {Cochran}, {Krolikowski}, {Mahadevan}, {Ninan}, {Stef{\'a}nsson},
  {Vanderburg}, {Zalesky}, \& {Zeimann}}]{Gully2023}
{Gully-Santiago}, M., {Morley}, C.~V., {Luna}, J., {et~al.} 2023, arXiv
  e-prints, arXiv:2307.08959

\bibitem[{{Huang} {et~al.}(2023){Huang}, {Koskinen}, {Lavvas}, \&
  {Fossati}}]{Huang2023}
{Huang}, C., {Koskinen}, T., {Lavvas}, P., \& {Fossati}, L. 2023, \apj, 951,
  123

\bibitem[{{Hunter} {et~al.}(2012){Hunter}, {Macgregor}, {Szabo}, {Wellington},
  \& {Bellgard}}]{yabi}
{Hunter}, A., {Macgregor}, A.~B., {Szabo}, T.~O., {Wellington}, C.~A., \&
  {Bellgard}, M.~I. 2012, Source Code for Biology and Medicine

\bibitem[{{Johnstone} {et~al.}(2021){Johnstone}, {Bartel}, \&
  {G{\"u}del}}]{Johnstone2021}
{Johnstone}, C.~P., {Bartel}, M., \& {G{\"u}del}, M. 2021, \aap, 649, A96

\bibitem[{{Johnstone} \& {G{\"u}del}(2015)}]{Johnstone2015}
{Johnstone}, C.~P. \& {G{\"u}del}, M. 2015, \aap, 578, A129

\bibitem[{Kass \& Raftery(1995)}]{kass_bic}
Kass, R.~E. \& Raftery, A.~E. 1995, Journal of the American Statistical
  Association, 90, 773

\bibitem[{{Kausch} {et~al.}(2015){Kausch}, {Noll}, {Smette}, {Kimeswenger},
  {Barden}, {Szyszka}, {Jones}, {Sana}, {Horst}, \& {Kerber}}]{Kausch2015}
{Kausch}, W., {Noll}, S., {Smette}, A., {et~al.} 2015, \aap, 576, A78

\bibitem[{{Khalafinejad} {et~al.}(2021){Khalafinejad}, {Molaverdikhani},
  {Blecic}, {Mallonn}, {Nortmann}, {Caballero}, {Rahmati}, {Kaminski},
  {Sadegi}, {Nagel}, {Carone}, {Amado}, {Azzaro}, {Bauer}, {Casasayas-Barris},
  {Czesla}, {von Essen}, {Fossati}, {G{\"u}del}, {Henning},
  {L{\'o}pez-Puertas}, {Lendl}, {L{\"u}ftinger}, {Montes}, {Oshagh},
  {Pall{\'e}}, {Quirrenbach}, {Reffert}, {Reiners}, {Ribas}, {Stock}, {Yan},
  {Zapatero Osorio}, \& {Zechmeister}}]{Khalafinejad2021}
{Khalafinejad}, S., {Molaverdikhani}, K., {Blecic}, J., {et~al.} 2021, \aap,
  656, A142

\bibitem[{{Kirk} {et~al.}(2020){Kirk}, {Alam}, {L{\'o}pez-Morales}, \&
  {Zeng}}]{Kirk2020}
{Kirk}, J., {Alam}, M.~K., {L{\'o}pez-Morales}, M., \& {Zeng}, L. 2020, \aj,
  159, 115

\bibitem[{{Knutson} {et~al.}(2014){Knutson}, {Fulton}, {Montet}, {Kao}, {Ngo},
  {Howard}, {Crepp}, {Hinkley}, {Bakos}, {Batygin}, {Johnson}, {Morton}, \&
  {Muirhead}}]{Knutson2014}
{Knutson}, H.~A., {Fulton}, B.~J., {Montet}, B.~T., {et~al.} 2014, \apj, 785,
  126

\bibitem[{{Kokori} {et~al.}(2023){Kokori}, {Tsiaras}, {Edwards}, {Jones},
  {Pantelidou}, {Tinetti}, {Bewersdorff}, {Iliadou}, {Jongen}, {Lekkas},
  {Nastasi}, {Poultourtzidis}, {Sidiropoulos}, {Walter}, {W{\"u}nsche},
  {Abraham}, {Agnihotri}, {Albanesi}, {Arce-Mansego}, {Arnot}, {Audejean},
  {Aumasson}, {Bachschmidt}, {Baj}, {Barroy}, {Belinski}, {Bennett}, {Benni},
  {Bernacki}, {Betti}, {Biagini}, {Bosch}, {Brandebourg}, {Br{\'a}t},
  {Bretton}, {Brincat}, {Brouillard}, {Bruzas}, {Bruzzone}, {Buckland},
  {Cal{\'o}}, {Campos}, {Carre{\~n}o}, {Carrion Rodrigo}, {Casali},
  {Casalnuovo}, {Cataneo}, {Chang}, {Changeat}, {Chowdhury}, {Ciantini},
  {Cilluffo}, {Coliac}, {Conzo}, {Correa}, {Coulon}, {Crouzet}, {Crow},
  {Curtis}, {Daniel}, {Dauchet}, {Dawes}, {Deldem}, {Deligeorgopoulos},
  {Dransfield}, {Dymock}, {Eenm{\"a}e}, {Esseiva}, {Evans}, {Falco},
  {Farf{\'a}n}, {Fern{\'a}ndez-Laj{\'u}s}, {Ferratfiat}, {Ferreira},
  {Ferretti}, {Fio{\l}ka}, {Fowler}, {Futcher}, {Gabellini}, {Gainey},
  {Gaitan}, {Gajdo{\v{s}}}, {Garc{\'\i}a-S{\'a}nchez}, {Garlitz}, {Gillier},
  {Gison}, {Gonzales}, {Gorshanov}, {Grau Horta}, {Grivas}, {Guerra},
  {Guillot}, {Haswell}, {Haymes}, {Hentunen}, {Hills}, {Hose}, {Humbert},
  {Hurter}, {Hynek}, {Irzyk}, {Jacobsen}, {Jannetta}, {Johnson},
  {J{\'o}{\'z}wik-Wabik}, {Kaeouach}, {Kang}, {Kiiskinen}, {Kim}, {Kivila},
  {Koch}, {Kolb}, {Ku{\v{c}}{\'a}kov{\'a}}, {Lai}, {Laloum}, {Lasota}, {Lewis},
  {Liakos}, {Libotte}, {Lomoz}, {Lopresti}, {Majewski}, {Malcher}, {Mallonn},
  {Mannucci}, {Marchini}, {Mari}, {Marino}, {Marino}, {Mario}, {Marquette},
  {Mart{\'\i}nez-Bravo}, {Ma{\v{s}}ek}, {Matassa}, {Michel}, {Michelet},
  {Miller}, {Miny}, {Molina}, {Mollier}, {Monteleone}, {Montigiani},
  {Morales-Aimar}, {Mortari}, {Morvan}, {Mugnai}, {Murawski}, {Naponiello},
  {Naudin}, {Naves}, {N{\'e}el}, {Neito}, {Neveu}, {Noschese},
  {{\"O}{\u{g}}men}, {Ohshima}, {Orbanic}, {Pace}, {Pantacchini}, {Paschalis},
  {Pereira}, {Peretto}, {Perroud}, {Phillips}, {Pintr}, {Pioppa}, {Plazas},
  {Poelarends}, {Popowicz}, {Purcell}, {Quinn}, {Raetz}, {Rees}, {Regembal},
  {Rocchetto}, {Rocci}, {Rockenbauer}, {Roth}, {Rousselot}, {Rubia}, {Ruocco},
  {Russo}, {Salisbury}, {Salvaggio}, {Santos}, {Savage}, {Scaggiante},
  {Sedita}, {Shadick}, {Silva}, {Sioulas}, {{\v{S}}koln{\'\i}k}, {Smith},
  {Smolka}, {Solmaz}, {Stanbury}, {Stouraitis}, {Tan}, {Theusner}, {Thurston},
  {Tifner}, {Tomacelli}, {Tomatis}, {Trnka}, {Tyl{\v{s}}ar}, {Valeau},
  {Vignes}, {Villa}, {Vives Sureda}, {Vora}, {Vra{\v{s}}t'{\'a}k}, {Walliang},
  {Wenzel}, {Wright}, {Zambelli}, {Zhang}, \& {Z{\'\i}bar}}]{Kokori2022}
{Kokori}, A., {Tsiaras}, A., {Edwards}, B., {et~al.} 2023, \apjs, 265, 4

\bibitem[{{Kosiarek} {et~al.}(2019){Kosiarek}, {Crossfield},
  {Hardegree-Ullman}, {Livingston}, {Benneke}, {Henry}, {Howard}, {Berardo},
  {Blunt}, {Fulton}, {Hirsch}, {Howard}, {Isaacson}, {Petigura}, {Sinukoff},
  {Weiss}, {Bonfils}, {Dressing}, {Knutson}, {Schlieder}, {Werner}, {Gorjian},
  {Krick}, {Morales}, {Astudillo-Defru}, {Almenara}, {Delfosse}, {Forveille},
  {Lovis}, {Mayor}, {Murgas}, {Pepe}, {Santos}, {Udry}, {Corbett}, {Fors},
  {Law}, {Ratzloff}, \& {del Ser}}]{Kosiarek2019}
{Kosiarek}, M.~R., {Crossfield}, I. J.~M., {Hardegree-Ullman}, K.~K., {et~al.}
  2019, \aj, 157, 97

\bibitem[{{Kreidberg} {et~al.}(2018){Kreidberg}, {Line}, {Thorngren}, {Morley},
  \& {Stevenson}}]{Kreidberg2018}
{Kreidberg}, L., {Line}, M.~R., {Thorngren}, D., {Morley}, C.~V., \&
  {Stevenson}, K.~B. 2018, \apjl, 858, L6

\bibitem[{{Krishnamurthy} {et~al.}(2023){Krishnamurthy}, {Hirano}, {Gaidos},
  {Sato}, {Kopparapu}, {Barclay}, {Garcia-Sage}, {Harakawa}, {Hodapp},
  {Jacobson}, {Konishi}, {Kotani}, {Kudo}, {Kurokawa}, {Kuzuhara}, {Lopez},
  {Nishikawa}, {Omiya}, {Schlieder}, {Serizawa}, {Tamura}, {Ueda}, \&
  {Vievard}}]{Krishnamurthy2023}
{Krishnamurthy}, V., {Hirano}, T., {Gaidos}, E., {et~al.} 2023, \mnras, 521,
  1210

\bibitem[{{Kuckartz} {et~al.}(2013){Kuckartz}, {R\..{a}diker}, {Ebert}, \&
  Julia}]{Kuckartz2013}
{Kuckartz}, U., {R\..{a}diker}, S., {Ebert}, T., \& Julia, S. 2013, 2nd ed.
  Wiesbaden: Springer Fachmedien

\bibitem[{{Kulow} {et~al.}(2014){Kulow}, {France}, {Linsky}, \&
  {Loyd}}]{Kulow2014}
{Kulow}, J.~R., {France}, K., {Linsky}, J., \& {Loyd}, R.~O.~P. 2014, \apj,
  786, 132

\bibitem[{{Langeveld} {et~al.}(2022){Langeveld}, {Madhusudhan}, \&
  {Cabot}}]{Langeveld2022}
{Langeveld}, A.~B., {Madhusudhan}, N., \& {Cabot}, S. H.~C. 2022, \mnras, 514,
  5192

\bibitem[{{Lanotte} {et~al.}(2014){Lanotte}, {Gillon}, {Demory}, {Fortney},
  {Astudillo}, {Bonfils}, {Magain}, {Delfosse}, {Forveille}, {Lovis}, {Mayor},
  {Neves}, {Pepe}, {Queloz}, {Santos}, \& {Udry}}]{Lanotte2014}
{Lanotte}, A.~A., {Gillon}, M., {Demory}, B.~O., {et~al.} 2014, \aap, 572, A73

\bibitem[{{Lavie} {et~al.}(2017){Lavie}, {Ehrenreich}, {Bourrier}, {Lecavelier
  des Etangs}, {Vidal-Madjar}, {Delfosse}, {Gracia Berna}, {Heng}, {Thomas},
  {Udry}, \& {Wheatley}}]{Lavie2017}
{Lavie}, B., {Ehrenreich}, D., {Bourrier}, V., {et~al.} 2017, \aap, 605, L7

\bibitem[{{Lecavelier Des Etangs}(2007)}]{Lecavelier2007}
{Lecavelier Des Etangs}, A. 2007, \aap, 461, 1185

\bibitem[{{Lovis} {et~al.}(2011){Lovis}, {Dumusque}, {Santos}, {Bouchy},
  {Mayor}, {Pepe}, {Queloz}, {S{\'e}gransan}, \& {Udry}}]{Lovis2011}
{Lovis}, C., {Dumusque}, X., {Santos}, N.~C., {et~al.} 2011, arXiv e-prints,
  arXiv:1107.5325

\bibitem[{{Lundkvist} {et~al.}(2016){Lundkvist}, {Kjeldsen}, {Albrecht},
  {Davies}, {Basu}, {Huber}, {Justesen}, {Karoff}, {Silva Aguirre}, {van
  Eylen}, {Vang}, {Arentoft}, {Barclay}, {Bedding}, {Campante}, {Chaplin},
  {Christensen-Dalsgaard}, {Elsworth}, {Gilliland}, {Handberg}, {Hekker},
  {Kawaler}, {Lund}, {Metcalfe}, {Miglio}, {Rowe}, {Stello}, {Tingley}, \&
  {White}}]{Lundkvist2016}
{Lundkvist}, M.~S., {Kjeldsen}, H., {Albrecht}, S., {et~al.} 2016, Nature
  Communications, 7, 11201

\bibitem[{{Mamajek} \& {Hillenbrand}(2008)}]{Mamajek&Hillenbrand2008}
{Mamajek}, E.~E. \& {Hillenbrand}, L.~A. 2008, \apj, 687, 1264

\bibitem[{{Mansfield} {et~al.}(2018){Mansfield}, {Bean}, {Oklop{\v{c}}i{\'c}},
  {Kreidberg}, {D{\'e}sert}, {Kempton}, {Line}, {Fortney}, {Henry}, {Mallonn},
  {Stevenson}, {Dragomir}, {Allart}, \& {Bourrier}}]{Mansfield2018}
{Mansfield}, M., {Bean}, J.~L., {Oklop{\v{c}}i{\'c}}, A., {et~al.} 2018, \apjl,
  868, L34

\bibitem[{Morello {et~al.}(2020)Morello, Claret, Martin-Lagarde, Cossou,
  Tsiaras, \& Lagage}]{morello2020exotethys}
Morello, G., Claret, A., Martin-Lagarde, M., {et~al.} 2020, The Astronomical
  Journal, 159, 75

\bibitem[{{Nail} {et~al.}(2023){Nail}, {Oklop{\v{c}}i{\'c}}, \&
  {MacLeod}}]{Nail2023}
{Nail}, F., {Oklop{\v{c}}i{\'c}}, A., \& {MacLeod}, M. 2023, arXiv e-prints,
  arXiv:2312.04682

\bibitem[{{Nascimbeni} {et~al.}(2013){Nascimbeni}, {Piotto}, {Pagano},
  {Scandariato}, {Sani}, \& {Fumana}}]{Nascimbeni2013}
{Nascimbeni}, V., {Piotto}, G., {Pagano}, I., {et~al.} 2013, \aap, 559, A32

\bibitem[{{Ninan} {et~al.}(2020){Ninan}, {Stefansson}, {Mahadevan}, {Bender},
  {Robertson}, {Ramsey}, {Terrien}, {Wright}, {Diddams}, {Kanodia}, {Cochran},
  {Endl}, {Ford}, {Fredrick}, {Halverson}, {Hearty}, {Jennings}, {Kaplan},
  {Lubar}, {Metcalf}, {Monson}, {Nitroy}, {Roy}, \& {Schwab}}]{Ninan2020}
{Ninan}, J.~P., {Stefansson}, G., {Mahadevan}, S., {et~al.} 2020, \apj, 894, 97

\bibitem[{{Nortmann} {et~al.}(2018){Nortmann}, {Pall{\'e}}, {Salz},
  {Sanz-Forcada}, {Nagel}, {Alonso-Floriano}, {Czesla}, {Yan}, {Chen},
  {Snellen}, {Zechmeister}, {Schmitt}, {L{\'o}pez-Puertas}, {Casasayas-Barris},
  {Bauer}, {Amado}, {Caballero}, {Dreizler}, {Henning}, {Lamp{\'o}n}, {Montes},
  {Molaverdikhani}, {Quirrenbach}, {Reiners}, {Ribas}, {S{\'a}nchez-L{\'o}pez},
  {Schneider}, \& {Zapatero Osorio}}]{Nortmann2018}
{Nortmann}, L., {Pall{\'e}}, E., {Salz}, M., {et~al.} 2018, Science, 362, 1388

\bibitem[{{Noyes} {et~al.}(1984){Noyes}, {Weiss}, \& {Vaughan}}]{Noyes1984}
{Noyes}, R.~W., {Weiss}, N.~O., \& {Vaughan}, A.~H. 1984, \apj, 287, 769

\bibitem[{{Oklop{\v c}i{\'c}} \& {Hirata}(2018)}]{Oklop2018}
{Oklop{\v c}i{\'c}}, A. \& {Hirata}, C.~M. 2018, \apjl, 855, L11

\bibitem[{{Oklop{\v{c}}i{\'c}}(2019)}]{Oklop2019}
{Oklop{\v{c}}i{\'c}}, A. 2019, \apj, 881, 133

\bibitem[{{Oliva} {et~al.}(2006){Oliva}, {Origlia}, {Baffa}, {Biliotti},
  {Bruno}, {D'Amato}, {Del Vecchio}, {Falcini}, {Gennari}, {Ghinassi}, {Giani},
  {Gonzalez}, {Leone}, {Lolli}, {Lodi}, {Maiolino}, {Mannucci}, {Marcucci},
  {Mochi}, {Montegriffo}, {Rossetti}, {Scuderi}, \& {Sozzi}}]{Oliva2006}
{Oliva}, E., {Origlia}, L., {Baffa}, C., {et~al.} 2006, in Society of
  Photo-Optical Instrumentation Engineers (SPIE) Conference Series, Vol. 6269,
  Society of Photo-Optical Instrumentation Engineers (SPIE) Conference Series,
  ed. I.~S. {McLean} \& M.~{Iye}, 626919

\bibitem[{{Oliva} {et~al.}(2013){Oliva}, {Origlia}, {Maiolino}, {Baffa},
  {Biliotti}, {Bruno}, {Falcini}, {Gavriousev}, {Ghinassi}, {Giani},
  {Gonzalez}, {Leone}, {Lodi}, {Massi}, {Montegriffo}, {Mochi}, {Pedani},
  {Rossetti}, {Scuderi}, {Sozzi}, {Tozzi}, \& {Valenti}}]{Oliva2013}
{Oliva}, E., {Origlia}, L., {Maiolino}, R., {et~al.} 2013, \aap, 555, A78

\bibitem[{{Palle} {et~al.}(2020){Palle}, {Nortmann}, {Casasayas-Barris},
  {Lamp{\'o}n}, {L{\'o}pez-Puertas}, {Caballero}, {Sanz-Forcada}, {Lara},
  {Nagel}, {Yan}, {Alonso-Floriano}, {Amado}, {Chen}, {Cifuentes},
  {Cort{\'e}s-Contreras}, {Czesla}, {Molaverdikhani}, {Montes}, {Passegger},
  {Quirrenbach}, {Reiners}, {Ribas}, {S{\'a}nchez-L{\'o}pez}, {Schweitzer},
  {Stangret}, {Zapatero Osorio}, \& {Zechmeister}}]{Palle2020}
{Palle}, E., {Nortmann}, L., {Casasayas-Barris}, N., {et~al.} 2020, \aap, 638,
  A61

\bibitem[{{Poppenhaeger}(2022)}]{Poppenhaeger2022}
{Poppenhaeger}, K. 2022, \mnras, 512, 1751

\bibitem[{{Rainer} {et~al.}(2018){Rainer}, {Harutyunyan}, {Carleo}, {Oliva},
  {Benatti}, {Bignamini}, {Claudi}, {Gonzalez-Alvarez}, {Sanna}, \&
  {Ghedina}}]{Rainer2018}
{Rainer}, M., {Harutyunyan}, A., {Carleo}, I., {et~al.} 2018, in Society of
  Photo-Optical Instrumentation Engineers (SPIE) Conference Series, Vol. 10702,
  Ground-based and Airborne Instrumentation for Astronomy VII, 1070266

\bibitem[{{Rumenskikh} {et~al.}(2023){Rumenskikh}, {Khodachenko},
  {Shaikhislamov}, {Miroshnichenko}, {Berezutsky}, {Shepelin}, \&
  {Dwivedi}}]{Rumenskikh2022}
{Rumenskikh}, M.~S., {Khodachenko}, M.~L., {Shaikhislamov}, I.~F., {et~al.}
  2023, \mnras, 526, 4120

\bibitem[{{Salz} {et~al.}(2018){Salz}, {Czesla}, {Schneider}, {Nagel},
  {Schmitt}, {Nortmann}, {Alonso-Floriano}, {L{\'o}pez-Puertas}, {Lamp{\'o}n},
  {Bauer}, {Snellen}, {Pall{\'e}}, {Caballero}, {Yan}, {Chen}, {Sanz-Forcada},
  {Amado}, {Quirrenbach}, {Ribas}, {Reiners}, {B{\'e}jar}, {Casasayas-Barris},
  {Cort{\'e}s-Contreras}, {Dreizler}, {Guenther}, {Henning}, {Jeffers},
  {Kaminski}, {K{\"u}rster}, {Lafarga}, {Lara}, {Molaverdikhani}, {Montes},
  {Morales}, {S{\'a}nchez-L{\'o}pez}, {Seifert}, {Zapatero Osorio}, \&
  {Zechmeister}}]{Salz2018}
{Salz}, M., {Czesla}, S., {Schneider}, P.~C., {et~al.} 2018, \aap, 620, A97

\bibitem[{{Sanz-Forcada}(2022)}]{Sanz-Forcada2022}
{Sanz-Forcada}, J. 2022, Astronomische Nachrichten, 343, e20008

\bibitem[{{Seager} \& {Sasselov}(2000)}]{Seager2000}
{Seager}, S. \& {Sasselov}, D.~D. 2000, \apj, 537, 916

\bibitem[{{Sicilia} {et~al.}(2022){Sicilia}, {Malavolta}, {Pino},
  {Scandariato}, {Nascimbeni}, {Piotto}, \& {Pagano}}]{Sicilia2022}
{Sicilia}, D., {Malavolta}, L., {Pino}, L., {et~al.} 2022, \aap, 667, A19

\bibitem[{{Sicilia} {et~al.}(\noop{3001}in prep.){Sicilia}, {Malavolta},
  {Scandariato}, \& et~al}]{Siciliainprep}
{Sicilia}, D., {Malavolta}, L., {Scandariato}, G., \& et~al. \noop{3001}in
  prep.

\bibitem[{{Smette} {et~al.}(2015){Smette}, {Sana}, {Noll}, {Horst}, {Kausch},
  {Kimeswenger}, {Barden}, {Szyszka}, {Jones}, {Gallenne}, {Vinther},
  {Ballester}, \& {Taylor}}]{Smette2015}
{Smette}, A., {Sana}, H., {Noll}, S., {et~al.} 2015, \aap, 576, A77

\bibitem[{{Spake} {et~al.}(2021){Spake}, {Oklop{\v{c}}i{\'c}}, \&
  {Hillenbrand}}]{Spake2021}
{Spake}, J.~J., {Oklop{\v{c}}i{\'c}}, A., \& {Hillenbrand}, L.~A. 2021, \aj,
  162, 284

\bibitem[{{Spake} {et~al.}(2018){Spake}, {Sing}, {Evans}, {Oklop{\v{c}}i{\'c}},
  {}, {Bourrier}, {Kreidberg}, {Rackham}, {Irwin}, {Ehrenreich}, {Wyttenbach},
  {Wakeford}, {Zhou}, {Chubb}, {Nikolov}, {Goyal}, {Henry}, {Williamson},
  {Blumenthal}, {Anderson}, {Hellier}, {Charbonneau}, {Udry}, \&
  {Madhusudhan}}]{Spake2018}
{Spake}, J.~J., {Sing}, D.~K., {Evans}, T.~M., {et~al.} 2018, \nat, 557, 68

\bibitem[{{Spinelli} {et~al.}(2023){Spinelli}, {Gallo}, {Haardt}, {Caldiroli},
  {Biassoni}, {Borsa}, \& {Rauscher}}]{Spinelli2023}
{Spinelli}, R., {Gallo}, E., {Haardt}, F., {et~al.} 2023, \aj, 165, 200

\bibitem[{{Stassun} {et~al.}(2017){Stassun}, {Collins}, \&
  {Gaudi}}]{Stassun2017}
{Stassun}, K.~G., {Collins}, K.~A., \& {Gaudi}, B.~S. 2017, \aj, 153, 136

\bibitem[{{Stef{\`a}nsson} {et~al.}(2022){Stef{\`a}nsson}, {Mahadevan},
  {Petrovich}, {Winn}, {Kanodia}, {Millholland}, {Maney}, {Ca{\~n}as},
  {Wisniewski}, {Robertson}, {Ninan}, {Ford}, {Bender}, {Blake}, {Cegla},
  {Cochran}, {Diddams}, {Dong}, {Endl}, {Fredrick}, {Halverson}, {Hearty},
  {Hebb}, {Hirano}, {Lin}, {Logsdon}, {Lubar}, {McElwain}, {Metcalf}, {Monson},
  {Rajagopal}, {Ramsey}, {Roy}, {Schwab}, {Schweiker}, {Terrien}, \&
  {Wright}}]{2022}
{Stef{\`a}nsson}, G., {Mahadevan}, S., {Petrovich}, C., {et~al.} 2022, \apjl,
  931, L15

\bibitem[{Stevenson {et~al.}(2010)Stevenson, Harrington, Nymeyer, \&
  et~al.}]{Stevenson2010}
Stevenson, K.~B., Harrington, J., Nymeyer, S., \& et~al. 2010, Nature, 464,
  1161

\bibitem[{{Su{\'a}rez Mascare{\~n}o} {et~al.}(2015){Su{\'a}rez Mascare{\~n}o},
  {Rebolo}, {Gonz{\'a}lez Hern{\'a}ndez}, \& {Esposito}}]{suarez2015}
{Su{\'a}rez Mascare{\~n}o}, A., {Rebolo}, R., {Gonz{\'a}lez Hern{\'a}ndez},
  J.~I., \& {Esposito}, M. 2015, \mnras, 452, 2745

\bibitem[{{Szab{\'o}} \& {Kiss}(2011)}]{Szabo2011}
{Szab{\'o}}, G.~M. \& {Kiss}, L.~L. 2011, \apjl, 727, L44

\bibitem[{{Ter Braak}(2006)}]{TerBraak2006}
{Ter Braak}, C. J.~F. 2006, Statistics and Computing, 16, 239

\bibitem[{{Triaud} {et~al.}(2015){Triaud}, {Gillon}, {Ehrenreich}, {Herrero},
  {Lendl}, {Anderson}, {Collier Cameron}, {Delrez}, {Demory}, {Hellier},
  {Heng}, {Jehin}, {Maxted}, {Pollacco}, {Queloz}, {Ribas}, {Smalley}, {Smith},
  \& {Udry}}]{Triaud2015}
{Triaud}, A. H.~M.~J., {Gillon}, M., {Ehrenreich}, D., {et~al.} 2015, \mnras,
  450, 2279

\bibitem[{{Tsiaras} {et~al.}(2018){Tsiaras}, {Waldmann}, {Zingales},
  {Rocchetto}, {Morello}, {Damiano}, {Karpouzas}, {Tinetti}, {McKemmish},
  {Tennyson}, \& {Yurchenko}}]{Tsiaras2018}
{Tsiaras}, A., {Waldmann}, I.~P., {Zingales}, T., {et~al.} 2018, \aj, 155, 156

\bibitem[{{Turner} {et~al.}(2016){Turner}, {Pearson}, {Biddle}, {Smart},
  {Zellem}, {Teske}, {Hardegree-Ullman}, {Griffith}, {Leiter}, {Cates},
  {Nieberding}, {Smith}, {Thompson}, {Hofmann}, {Berube}, {Nguyen}, {Small},
  {Guvenen}, {Richardson}, {McGraw}, {Raphael}, {Crawford}, {Robertson},
  {Tombleson}, {Carleton}, {Towner}, {Walker-LaFollette}, {Hume}, {Watson},
  {Jones}, {Lichtenberger}, {Hoglund}, {Cook}, {Crossen}, {Jorgensen},
  {Romine}, {Thompson}, {Villegas}, {Wilson}, {Sanford}, {Taylor}, \&
  {Henz}}]{Turner2016}
{Turner}, J.~D., {Pearson}, K.~A., {Biddle}, L.~I., {et~al.} 2016, \mnras, 459,
  789

\bibitem[{{Tyler} {et~al.}(2024){Tyler}, {Petigura}, {Oklop{\v{c}}i{\'c}}, \&
  {David}}]{Tyler2023}
{Tyler}, D., {Petigura}, E.~A., {Oklop{\v{c}}i{\'c}}, A., \& {David}, T.~J.
  2024, \apj, 960, 123

\bibitem[{{Vidal-Madjar} {et~al.}(2004){Vidal-Madjar}, {D{\'e}sert},
  {Lecavelier des Etangs}, {H{\'e}brard}, {Ballester}, {Ehrenreich}, {Ferlet},
  {McConnell}, {Mayor}, \& {Parkinson}}]{Vidal-Madjar2004}
{Vidal-Madjar}, A., {D{\'e}sert}, J.~M., {Lecavelier des Etangs}, A., {et~al.}
  2004, \apjl, 604, L69

\bibitem[{{Vidal-Madjar} {et~al.}(2003){Vidal-Madjar}, {Lecavelier des Etangs},
  {D{\'e}sert}, {Ballester}, {Ferlet}, {H{\'e}brard}, \&
  {Mayor}}]{Vidal-Madjar2003}
{Vidal-Madjar}, A., {Lecavelier des Etangs}, A., {D{\'e}sert}, J.~M., {et~al.}
  2003, \nat, 422, 143

\bibitem[{{Vissapragada} {et~al.}(2022){Vissapragada}, {Knutson},
  {Greklek-McKeon}, {Oklop{\v{c}}i{\'c}}, {Dai}, {dos Santos}, {Jovanovic},
  {Mawet}, {Millar-Blanchaer}, {Paragas}, {Spake}, {Tinyanont}, \&
  {Vasisht}}]{Vissapragada2022}
{Vissapragada}, S., {Knutson}, H.~A., {Greklek-McKeon}, M., {et~al.} 2022, \aj,
  164, 234

\bibitem[{{Vissapragada} {et~al.}(2020){Vissapragada}, {Knutson}, {Jovanovic},
  {Harada}, {Oklop{\v{c}}i{\'c}}, {Eriksen}, {Mawet}, {Millar-Blanchaer},
  {Tinyanont}, \& {Vasisht}}]{Vissapragada2020}
{Vissapragada}, S., {Knutson}, H.~A., {Jovanovic}, N., {et~al.} 2020, \aj, 159,
  278

\bibitem[{Yan {et~al.}(2017)Yan, Pall{\'e}, Fosbury, Petr-Gotzens, \&
  Henning}]{yan2017effect}
Yan, F., Pall{\'e}, E., Fosbury, R.~A., Petr-Gotzens, M.~G., \& Henning, T.
  2017, Astronomy \& Astrophysics, 603, A73

\bibitem[{{Zhang} {et~al.}(2023){Zhang}, {Dai}, {Bean}, {Knutson}, \&
  {Rescigno}}]{Zhang2023}
{Zhang}, M., {Dai}, F., {Bean}, J.~L., {Knutson}, H.~A., \& {Rescigno}, F.
  2023, \apjl, 953, L25

\end{thebibliography}
	\begin{acknowledgements}
		We want to thank the anonymous referee for the constructive comments which helped to improve the quality of the manuscript. The authors acknowledge financial contribution from the European Union - Next Generation EU RRF M4C2 1.1 PRIN MUR 2022 project 2022CERJ49 (ESPLORA), from PRIN INAF 2019, and from the ASI-INAF agreement n.2018-16-HH.0 (THE StellaR PAth project).        
	\end{acknowledgements}
	
	\begin{appendix}
		\onecolumn
		\section{Additional figures and tables} \label{add}

		\begin{longtable}{lcc}
			\caption{Adopted parameters.}    \\
			\hline\hline
			\small
			Parameter &Value & Reference \\ 
			\hline
			\endfirsthead
			\caption{Continued.}\\
			\hline
			Parameter &Value & Reference \\ 
			\hline
			\endhead
			\hline
			\endfoot
			\hline
			\endlastfoot
			\textbf{WASP-69}  & &  \\ [2pt] %
			\underline{Stellar Parameters} &  & \\
			Spectral type \dotfill& K5 &  \citet{Anderson2014}\\
			Stellar mass, M$_\star$ [M$_\sun$]\dotfill &  0.826$\pm$0.029 & \citet{Anderson2014}\\
			Stellar radius, R$_\star$ [R$_\sun$]\dotfill & 0.813$\pm$0.028 & \citet{Anderson2014}\\
			Effective temperature, T$_\mathrm{eff}$ [K]\dotfill & 4715$\pm$50 & \citet{Anderson2014}\\
			Metallicity, [Fe/H] [dex] \dotfill & 0.144$\pm$0.077 & \citet{Anderson2014}\\
			log~g$_\star$ (log$_{10}$[\cms])\dotfill & 4.535$\pm$0.023 & \citet{Anderson2014}\\
			Systemic velocity, v$_\mathrm{sys}$ (\kms )\dotfill & -9.37$\pm$0.20 & Gaia DR2 \citep{Gaia2018}\\
			Magnitude (J-band) \dotfill&  8.032$\pm$0.023 & \citet{Cutri2003} \\
			\\
			\underline{WASP-69b Parameters} & & \\
			Orbital period, P [days]\dotfill & 3.86813888$\pm$9.1e-07 & \citet{Kokori2022}\\
			Transit epoch, T$_0$ [BJD$_\mathrm{TDB}$] \dotfill& 2457269.01322$\pm$0.00027 & \citet{Kokori2022}\\
			Eccentricity, e \dotfill& 0 (fixed ) &  \citet{Anderson2014} \\
			Argument of periastron, $\omega$ [deg] \dotfill& 90(fixed ) &\\
			Stellar reflex velocity, K$_\star$ [\ms] \dotfill& 38.1$\pm$2.4 & \citet{Anderson2014}\\
			Orbital major semi-axis, a [au]\dotfill & 0.04525$\pm$0.00053 & \citet{Anderson2014}\\
			Orbital inclination, $i$ [deg] \dotfill& 86.71$\pm$ 0.20 & \citet{Anderson2014}\\
			Planetary mass, M$_\mathrm{pl}$  \dotfill& 0.260$\pm$0.017 & \citet{Anderson2014}\\
			Planetary radius, R$_\mathrm{pl}$ [\rjup] \dotfill& 1.057$\pm$0.047 & \citet{Anderson2014}\\
			Impact parameter, b \dotfill& 0.686$\pm$0.023 & \citet{Anderson2014}\\
			Equilibrium temperature, T$_\mathrm{eq}$ [K]\dotfill& 963$\pm$18 & \citet{Anderson2014}\\ 
			Planet radial-velocity semi-amplitude, K$_\mathrm{p}$ [\ms] \dotfill& 127.1$\pm$1.5 & This paper \footnotemark[2] \\
			\hline         \\
			
			\textbf{ WASP-107}  &  & \\
			\underline{Stellar Parameters} &  & \\
			Spectral type \dotfill& K6 & \citet{Anderson2017} \\
			Stellar mass, M$_\star$ [M$_\sun$] \dotfill& 0.69$\pm$0.05& \citet{Anderson2017}\\
			Stellar radius, R$_\star$ [R$_\sun$] \dotfill& 0.66$\pm$0.02 & \citet{Anderson2017} \\
			Effective temperature, T$_\mathrm{eff}$ [K] \dotfill& 4430$\pm$120 & \citet{Anderson2014} \\
			Metallicity, [Fe/H] [dex]  \dotfill & 0.020$\pm$0.100  & \citet{Anderson2014}\\
			log~g$_\star$ (log$_{10}$[\cms]) \dotfill& 4.5$\pm$0.1 & \citet{Anderson2014}\\
			Systemic velocity, v$_\mathrm{sys}$ [\kms] \dotfill& 13.74$\pm$0.31 & Gaia DR2 \citep{Gaia2018}\\
			Magnitude (J-band) \dotfill& 9.378 $\pm$0.021 & \citet{Cutri2003} \\
			\\
			\underline{WASP-107b Parameters} & & \\
			Orbital period, P [days] \dotfill& 5.72148926 $\pm$ 8.5e-07 & \citet{Kokori2022}\\
			Transit epoch, T$_0$ [BJD$_\mathrm{TDB}$] \dotfill& 2457515.672118$\pm$7.5e-05 &  \citet{Kokori2022}\\
			Eccentricity, e \dotfill& 0 (fixed) & \citet{Anderson2017}\\ 
			Argument of periastron, $\omega$ [deg]\dotfill&  90 (adopted) &\\
			Stellar reflex velocity, K$_\star$ (\ms) \dotfill& 17$\pm$2 & \citet{Anderson2017}\\
			Orbital major semi-axis, a [au] \dotfill& 0.055 $\pm$0.001 &\citet{Anderson2017}\\
			Orbital inclination, $i$ [deg] \dotfill& 89.7$\pm$0.2 &\citet{Anderson2017}\\
			Planetary mass, M$_\mathrm{pl}$ [\mjup] \dotfill& 0.12$\pm$0.01 &\citet{Anderson2017}\\
			Planetary radius, R$_\mathrm{pl}$ [\rjup] \dotfill& 0.94$\pm$0.02 &\citet{Anderson2017}\\
			Impact parameter, b \dotfill& 0.09$\pm$0.07 &\citet{Anderson2017}\\
			Equilibrium temperature, T$_\mathrm{eq}$ [K]\dotfill& 770$\pm$60 & \citet{Anderson2014}\\ 
			Planet radial-velocity semi-amplitude, K$_\mathrm{p}$ [\ms] \dotfill& 105.2$\pm$2.5 & This paper \footnotemark[2] \\
			\hline         \\
			
			\multicolumn{1}{l}{\textbf{HAT-P-11}} \\ [2pt] %
			\underline{Stellar Parameters} & &\\  
			Spectral\ type\dotfill & K4\,V & \citet{bakos2010}\\
			Stellar mass, M$_\star$ [M$_\sun$]\dotfill& $0.86\pm0.06$ & \citet{Lundkvist2016} \\
			Stellar radius, R$_\star$ [R$_\sun$] \dotfill& $0.76\pm0.01$ & \citet{Lundkvist2016}\\
			Effective temperature, T$_\mathrm{eff}$ [K]\dotfill& $4780\pm50$ & \citet{bakos2010}\\
			Metallicity, [Fe/H] [dex]\dotfill& $0.31\pm0.05$ & \citet{bakos2010}\\
			log~g$_\star$ (log$_{10}$[\cms]) & 4.37$\pm$0.22 & \citet{Stassun2017}\\
			Systemic velocity, v$_\mathrm{sys}$ [\kms] \dotfill& $-63.24\pm0.26$ & Gaia DR2 \citep{Gaia2018}\\
			Magnitude (J-band) \dotfill& 7.608$\pm$0.029 & \citet{Cutri2003} \\
			\\
			\underline{HAT-P-11b Parameters} & &\\  %
			Orbital period, P [days]\dotfill& 4.88780201$\pm$1.7e-07 & \citet{Kokori2022}\\
			Transit epoch, T$_0$ [BJD$_\mathrm{TDB}$] \dotfill & 2455798.515261$\pm$2.3e-05 & \citet{Kokori2022}\\
			Eccentricity, e \dotfill&$0.2577^{+0.0033}_{-0.0025}$ & \citet{Basilicata2023} \\
			Argument of periastron, $\omega$ [deg] \dotfill& $19.0^{+2.9}_{-3.0}$ & \citet{Basilicata2023} \\
			Stellar reflex velocity, K$_\star$ [\ms] \dotfill& 11.21$\pm$0.36  & \citet{Basilicata2023}\\
			Orbital major semi-axis, a [au] \dotfill& $0.0532\pm0.0010$ & \citet{Basilicata2023}\\
			Orbital inclination, $i$ [deg] \dotfill& $89.027\pm0.068$ & \citet{Basilicata2023} \\ [2pt]
			Planetary mass, M$_\mathrm{pl}$ [\mjup]\dotfill& $0.0818\pm0.0046$ & \citet{Basilicata2023}\\
			Planetary radius, R$_\mathrm{pl}$ [\rjup]\dotfill& $0.4466\pm0.0059$ & \citet{Basilicata2023} \\
			Impact parameter, b \dotfill& $0.227^{+0.013}_{-0.015}$& \citet{Basilicata2023} \\
			Equilibrium temperature T$_\mathrm{eq}$ [K] \dotfill& 847$^{+46}_{-54}$& \citet{Basilicata2023}\\
			Planet radial-velocity semi-amplitude, K$_\mathrm{p}$ [\ms] \dotfill& 123.5$\pm$2.9 & This paper \footnotemark[2] \\
			\hline         \\
			
			\multicolumn{1}{l}{\textbf{GJ436}} \\ [2pt] %
			\underline{Stellar Parameters} & &\\  
			Spectral\ type\dotfill & M2.5V & \citet{Butler2004}\\
			Stellar mass, M$_\star$ [M$_\sun$]\dotfill& 0.556$^{+0.071}_{-0.065}$ & \citet{Lanotte2014} \\
			Stellar radius, R$_\star$ [R$_\sun$] \dotfill& 0.455 $\pm$0.018 & \citet{Lanotte2014}\\
			Effective temperature, T$_\mathrm{eff}$ [K]\dotfill& 3416 $\pm$100 & \citet{Lanotte2014}\\
			Metallicity, [Fe/H] [dex]\dotfill& 0.02$\pm$0.20 & \citet{Lanotte2014}\\
			log~g$_\star$ (log$_{10}$[\cms]) & 4.843$\pm$0.018 & \citet{Lanotte2014}\\
			Systemic velocity, v$_\mathrm{sys}$ [\kms]\dotfill& 9.59 $\pm$0.0008& \citet{Fouqu2018}\\
			Magnitude (J-band) \dotfill&  6.900$\pm$0.024 & \citet{Cutri2003} \\
			\\
			\underline{GJ436b Parameters} & &\\  
			Orbital period, P [days]\dotfill& 2.643897621 $\pm$ 9.6e-08 & \citet{Kokori2022}\\
			Transit epoch, T$_0$ (BJD$_\mathrm{TT}$) \dotfill & 2455290.751684 $\pm$ 5.2e-05 & \citet{Kokori2022}\\
			Eccentricity, e \dotfill& 0.1616$^{0.0041}_{-0.0032}$& \citet{Lanotte2014} \\
			Argument of periastron, $\omega$ [deg] \dotfill& 327.2$^{+1.8}_{-2.2}$ & \citet{Lanotte2014} \\
			Stellar reflex velocity, K$_\star$ [\ms] \dotfill& 17.59 $\pm$0.25  & \citet{Lanotte2014}\\
			Orbital major semi-axis, a [au] \dotfill& 0.0308$^{+0.0013}_{-0.0012}$ & \citet{Lanotte2014}\\
			Orbital inclination, $i$ [deg] \dotfill& 86.858$^{+0.049}_{-0.052}$ & \citet{Lanotte2014} \\ [2pt]
			Planetary mass, M$_\mathrm{pl}$ [\mjup]\dotfill& 0.080$^{+0.007}_{-0.006}$ & \citet{Lanotte2014}\\
			Planetary radius, R$_\mathrm{pl}$ [\rjup]\dotfill& 0.366$\pm$0.014 & \citet{Lanotte2014} \\
			Impact parameter, b \dotfill& 0.7972$^{+0.0053}_{-0.0055}$ & \citet{Lanotte2014} \\
			Equilibrium temperature T$_\mathrm{eq}$ [K] \dotfill& 686$\pm$10 & \citet{Turner2016} \\
			Planet radial-velocity semi-amplitude, K$_\mathrm{p}$ [\ms] \dotfill& 128.1$^{+5.5}_{-5.0}$ & This paper \footnotemark[2] \\
			
			\hline         \\
			\multicolumn{1}{l}{\textbf{GJ3470}} \\ [2pt] %
			\underline{Stellar Parameters} & &\\  
			Spectral\ type\dotfill &        M1.5 & \citet{Kosiarek2019}\\
			Stellar mass, M$_\star$ [M$_\sun$]\dotfill& 0.51$\pm$0.06 & \citet{Kosiarek2019}\\
			Stellar radius, R$_\star$ [R$_\sun$] \dotfill& 0.48$\pm$0.04 & \citet{Kosiarek2019}\\
			Effective temperature, T$_\mathrm{eff}$ [K]\dotfill& 3652$\pm$50 & \citet{Kosiarek2019}\\
			Metallicity, [Fe/H] [dex]\dotfill& 0.20$\pm$0.10 & \citet{Kosiarek2019}\\
			log~g$_\star$ (log$_{10}$[\cms]) & 4.658$\pm$0.035 & \citet{Kosiarek2019}\\
			Systemic velocity, v$_\mathrm{sys}$ [\kms] \dotfill& 26.09$\pm$0.25& Gaia DR2 \citep{Gaia2018}\\
			Magnitude (J-band) \dotfill& 8.794$\pm$0.026 & \citet{Cutri2003} \\
			\\
			\underline{GJ3470b Parameters} & &\\  
			Orbital period, P [days]\dotfill& 3.33665266$^{+0.0000003}_{-0.0000003}$&\citet{2022}\\ 
			Transit epoch, T$_0$ [BJD$_\mathrm{TDB}$] \dotfill & 2456340.72559$^{+0.00011}_{-0.0001}$  & \citet{2022}\\ 
			Eccentricity, e \dotfill& 0.125$^{+0.043}_{-0.042}$ & \citet{2022} \\
			Argument of periastron, $\omega$ [deg] \dotfill&  -83.4$^{+3.4}_{-1.7}$ &  \citet{2022}\\
			Stellar reflex velocity, K$_\star$ [\ms] \dotfill&  8.03$^{+0.38}_{-0.37}$ & \citet{2022}\\
			Orbital major semi-axis, a [au] \dotfill& 0.0288$^{+0.0029}_{-0.0028}$ & derived from \citet{Kosiarek2019}\\
			Orbital inclination, $i$ [deg] \dotfill& 88.88$^{+0.62}_{-0.45}$  &  \citet{Biddle2014}\\ [2pt]
			Planetary mass, M$_\mathrm{pl}$ [\mjup]\dotfill& 0.03958$^{+0.00412}_{-0.00403}$ & \citet{Kosiarek2019}\\
			Planetary radius, R$_\mathrm{pl}$ [\rjup]\dotfill& 0.346$\pm$0.029  &  \citet{Kosiarek2019}\\
			Impact parameter, b \dotfill& 0.29$\pm$0.14 &  \citet{Biddle2014}\\
			Equilibrium temperature T$_\mathrm{eq}$ [K] \dotfill& 604$\pm$98 & \citet{Biddle2014}\\
			Planet radial-velocity semi-amplitude, K$_\mathrm{p}$ [\ms] \dotfill& 114.7$\pm$0.5 & This paper \footnotemark[2] \\
			\label{tab_par}
		\end{longtable}
		\tablefoot{     
			K$_\mathrm{pl}=\frac{2\pi a}{P}\frac{\sin{i}}{\sqrt{1-e^2}}=(\frac{2\pi G}{P})^{\frac{1}{3}}\frac{(M_\star+M_\mathrm{pl})^{\frac{1}{3}}\sin{i}}{\sqrt{1-e^2}}$.}

		\vspace{3cm}

		\begin{figure}[h]
			\includegraphics[width=\textwidth,height=11cm]{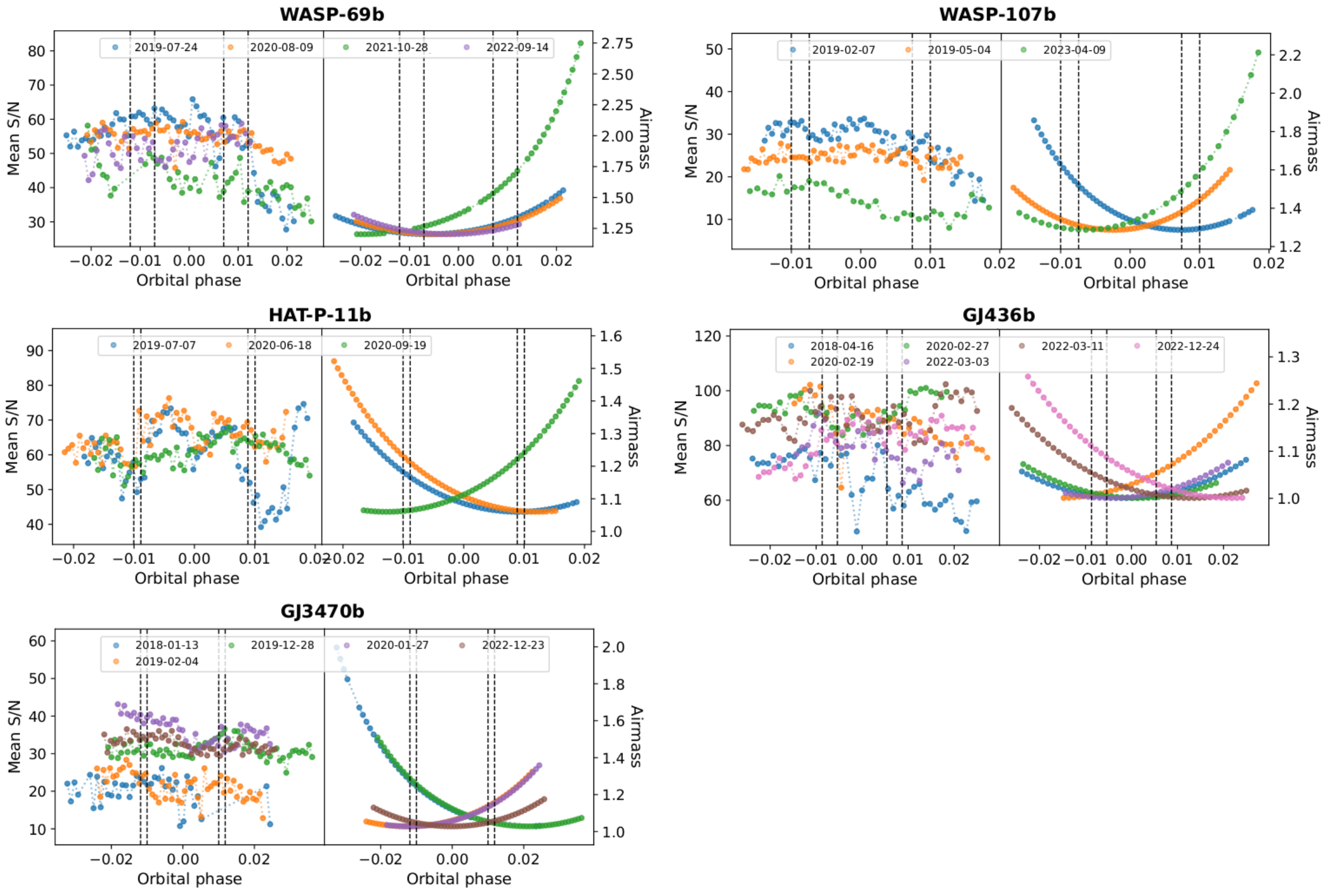}
			\caption{S/N in the region of interest (1082.49-1085.5\,nm; left
				panel) and airmass (right panel) measured during the GIANO-B
				observations for each investigated target. The vertical dashed lines mark the t$_1$, t$_2$, t$_3$, and
				t$_4$ contact points (from left to right). The dashed green line for WASP-107b indicates the transit we had to discard for adverse seeing conditions.}
			\label{snr_ph_am}
		\end{figure}

		\begin{figure*}
			\includegraphics[width=\linewidth]{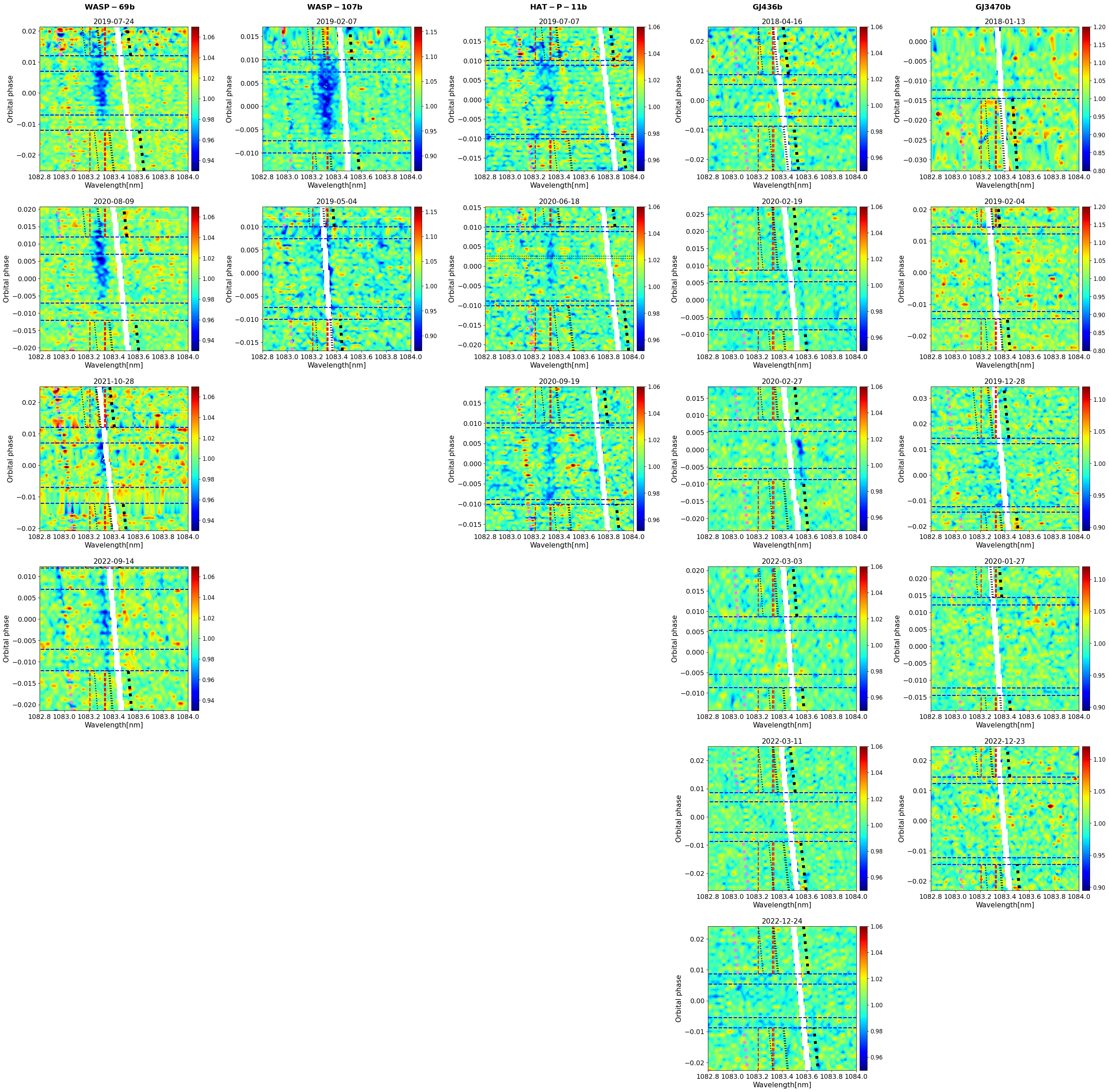}
			\caption{Same as Fig.~\ref{MAPS} but for all the investigated nights.}
			\label{MAPS_app}
		\end{figure*}

		\begin{figure*} [ht!]
			\centering
			\includegraphics[width=\linewidth]{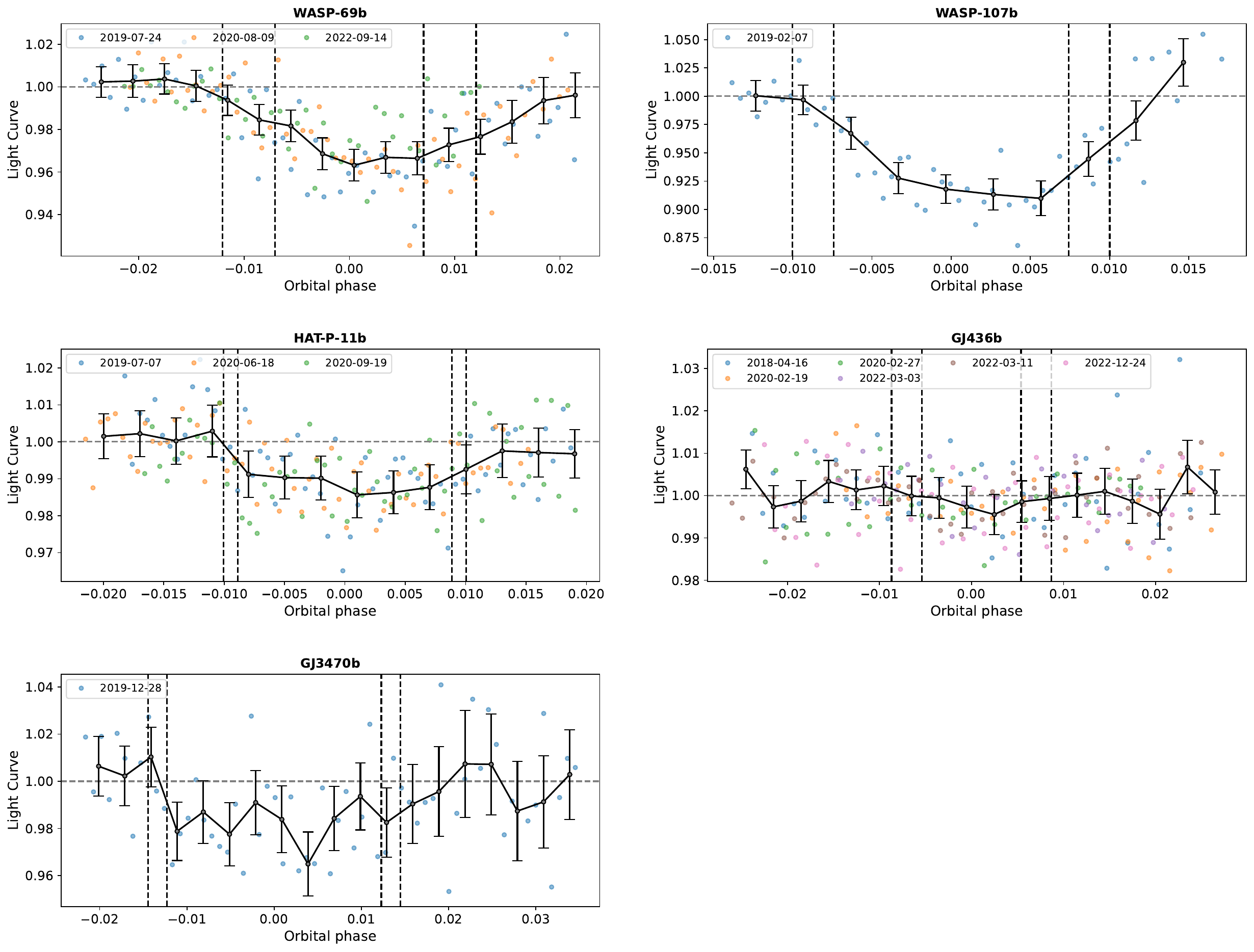}      
			\caption{Transmission light curve of \ion{He}{I} in the planetary rest frame for the investigated targets.  The vertical dashed lines mark the t$_1$, t$_2$, t$_3$, and t$_4$ contact points (from left to right). The grey horizontal dashed line is the continuum level.}
			\label{LC}
			\vspace{2cm}
			\captionof{table}{Priors used in the DE-MCMC analysis.}
			\begin{tabular}{l c }
				\hline \hline
				Value  &  Priors  \\    
				\hline
				Peak Pos$_\mathrm{gauss}$ [nm] & U[1083.29,1083.36]  \\
				$\sigma_\mathrm{jit}$ & U[0,+$\infty$] \\
				Ampl [\%] & U[0,3] \\
				$\lambda$ [nm] & U[0.01,0.09] \\
				\hline
			\end{tabular}
			\label{priors_tab}
		\end{figure*}
		\FloatBarrier   
		\twocolumn
		\begin{landscape}
			\begin{figure}
				\includegraphics[width=0.42\textwidth]{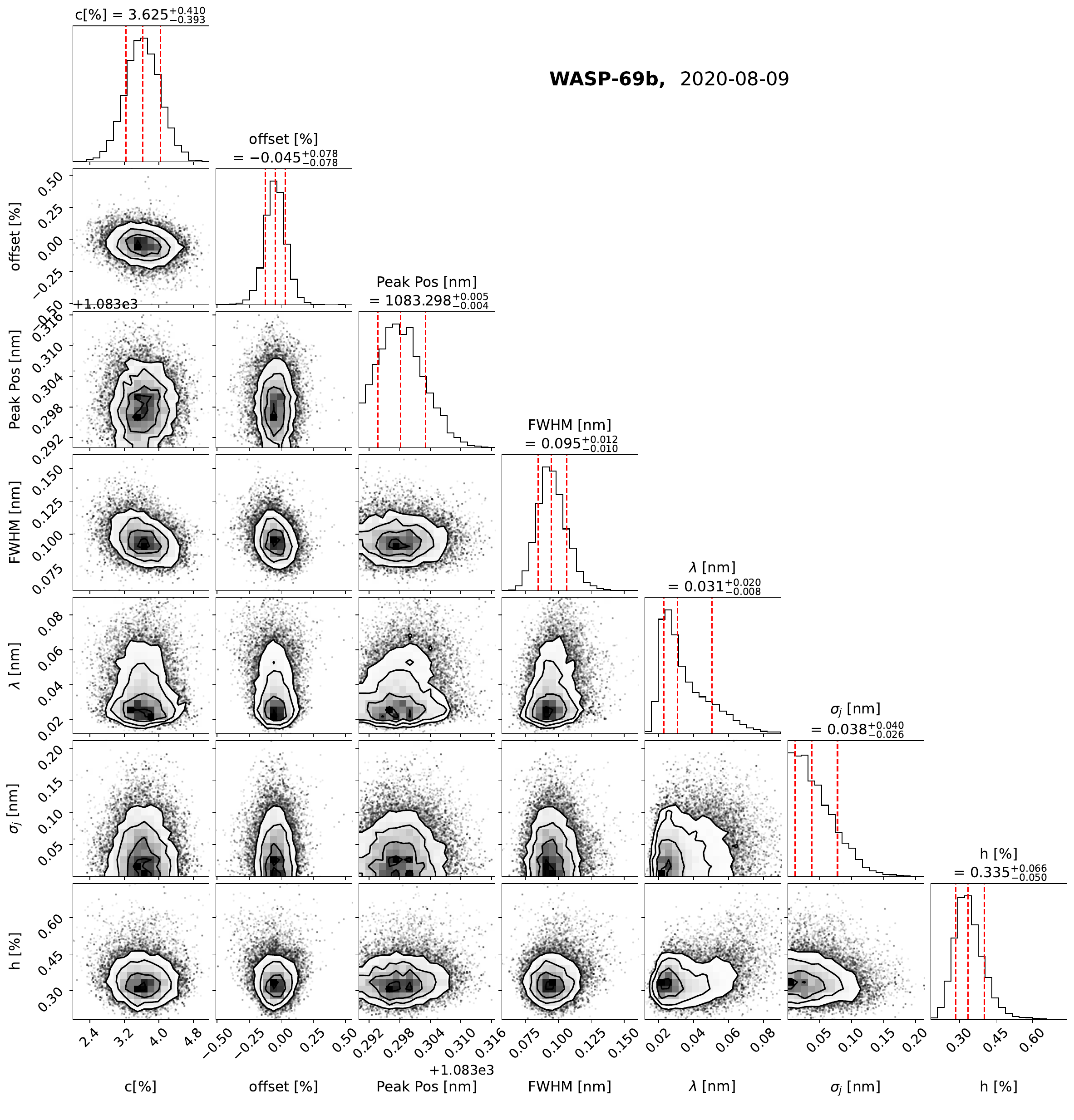}
				\includegraphics[width=0.42\textwidth]{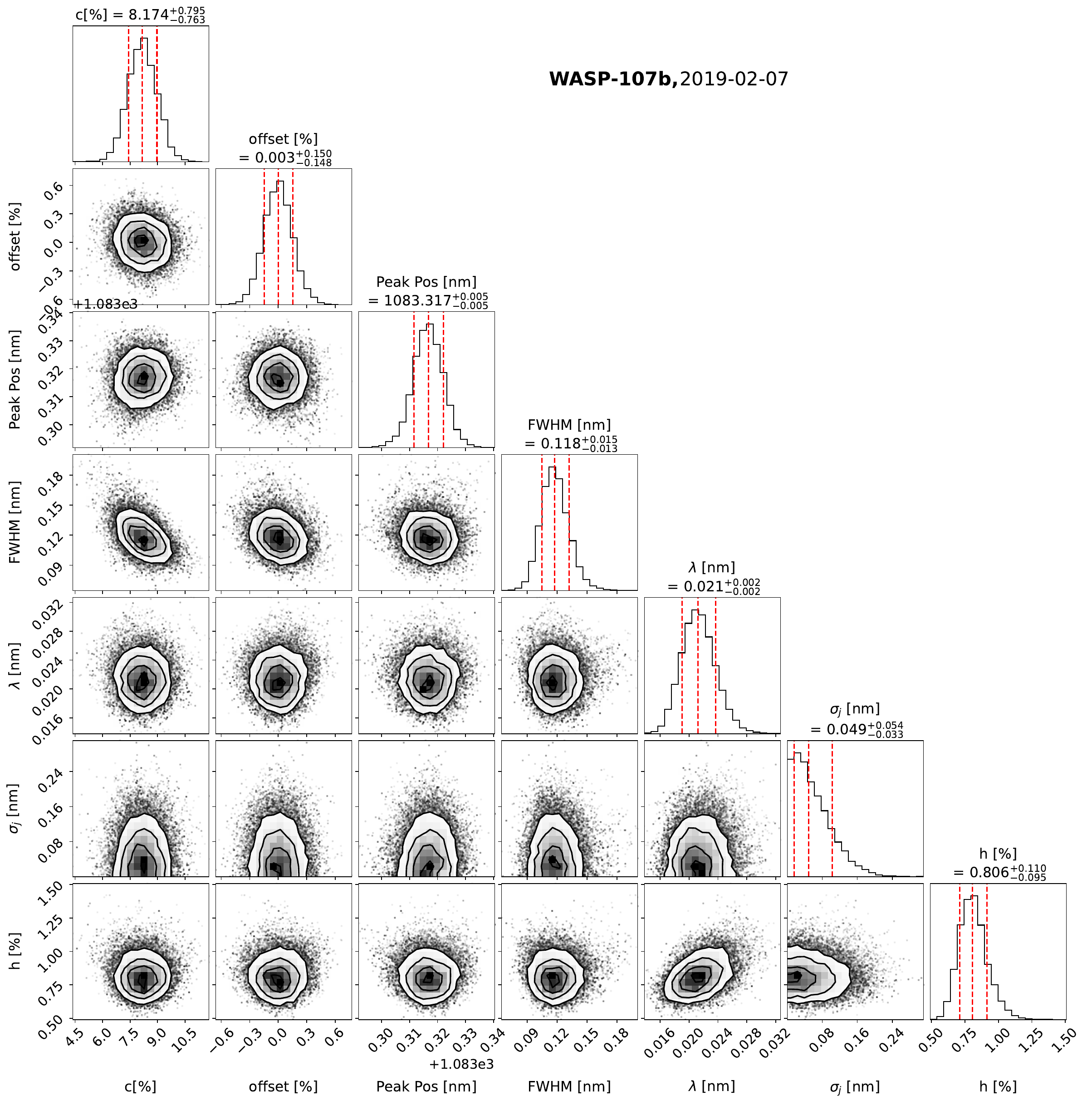}
				\includegraphics[width=0.42\textwidth]{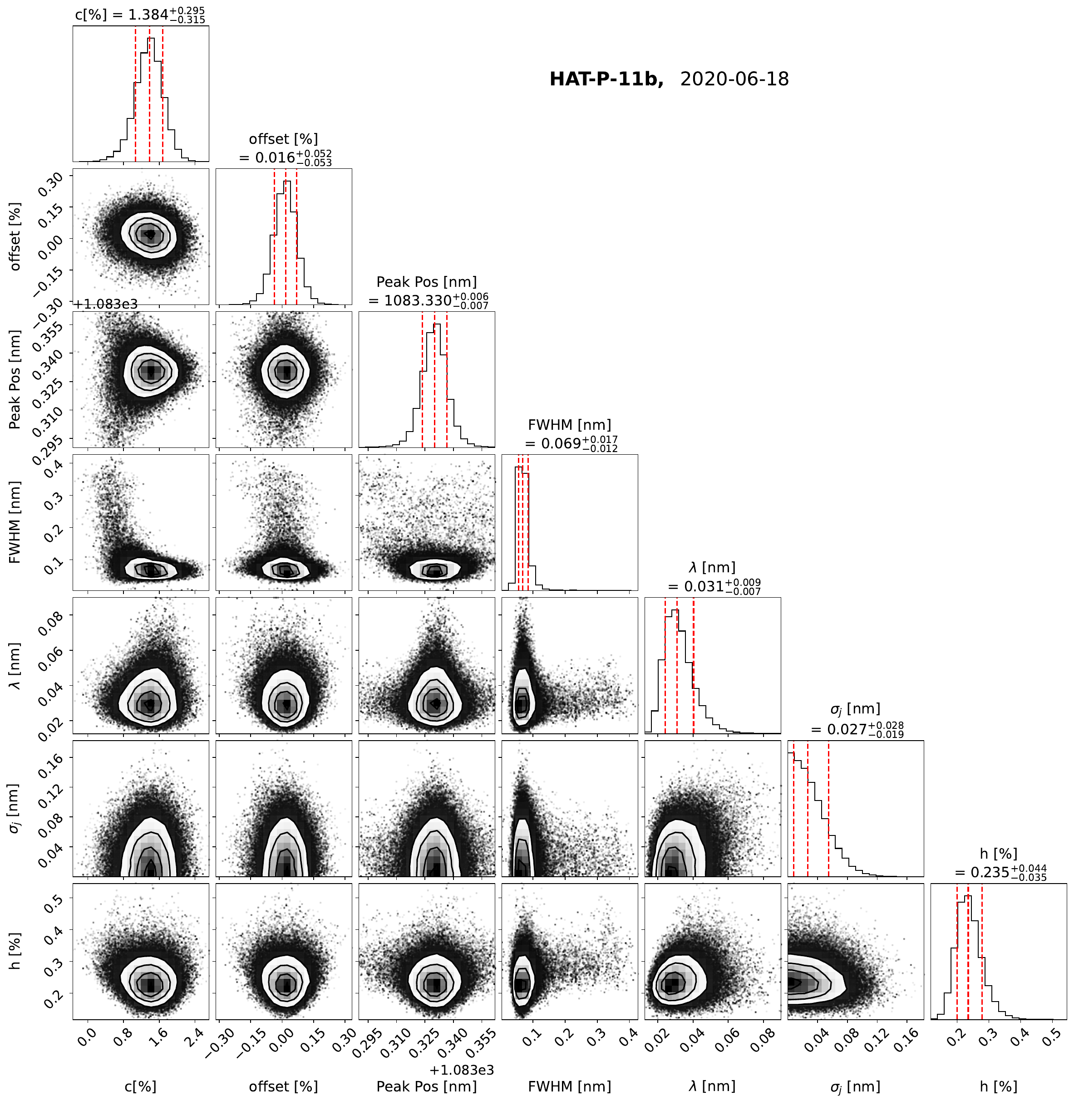}\\
				\includegraphics[width=0.42\textwidth]{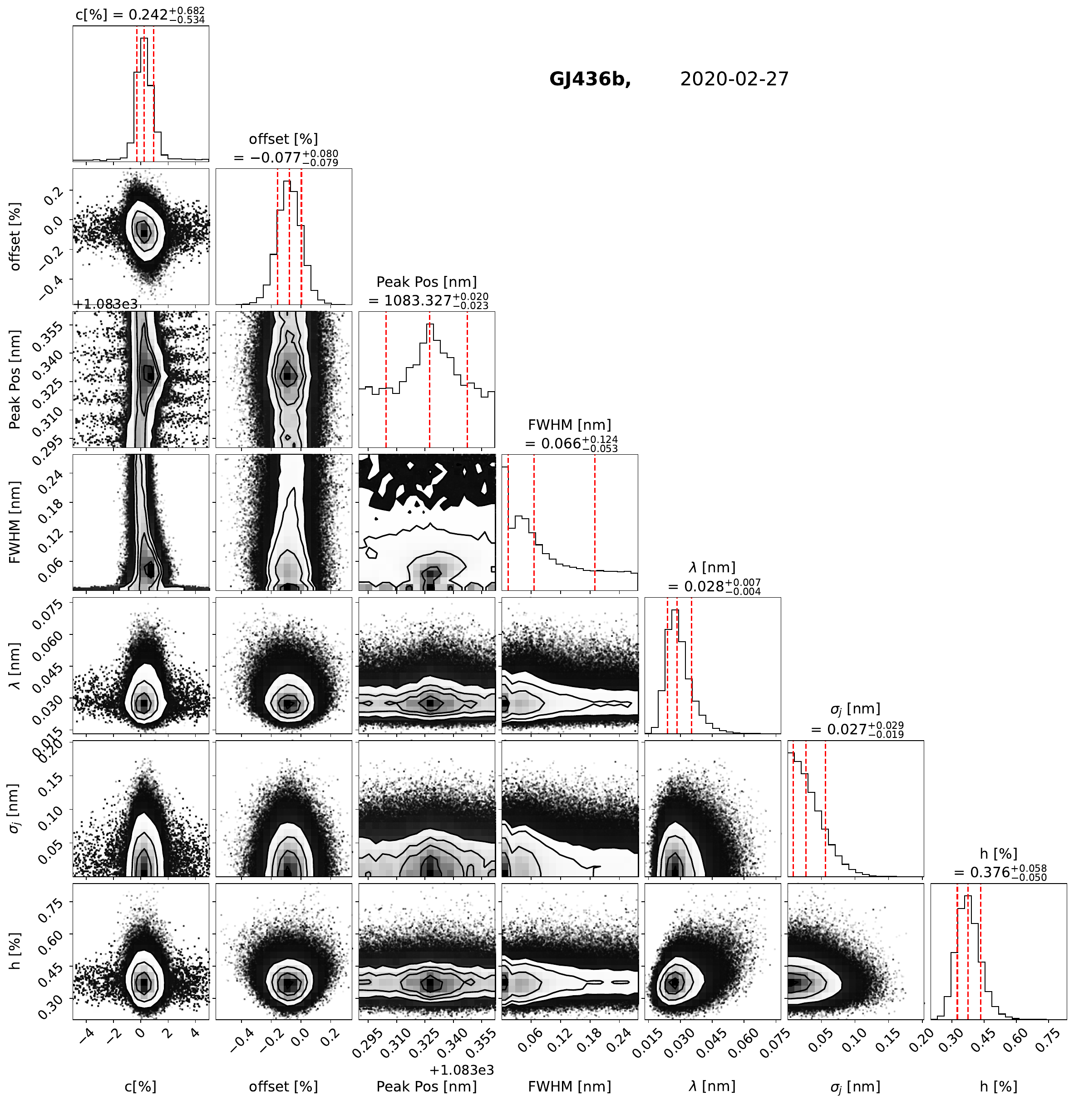}
				\includegraphics[width=0.42\textwidth]{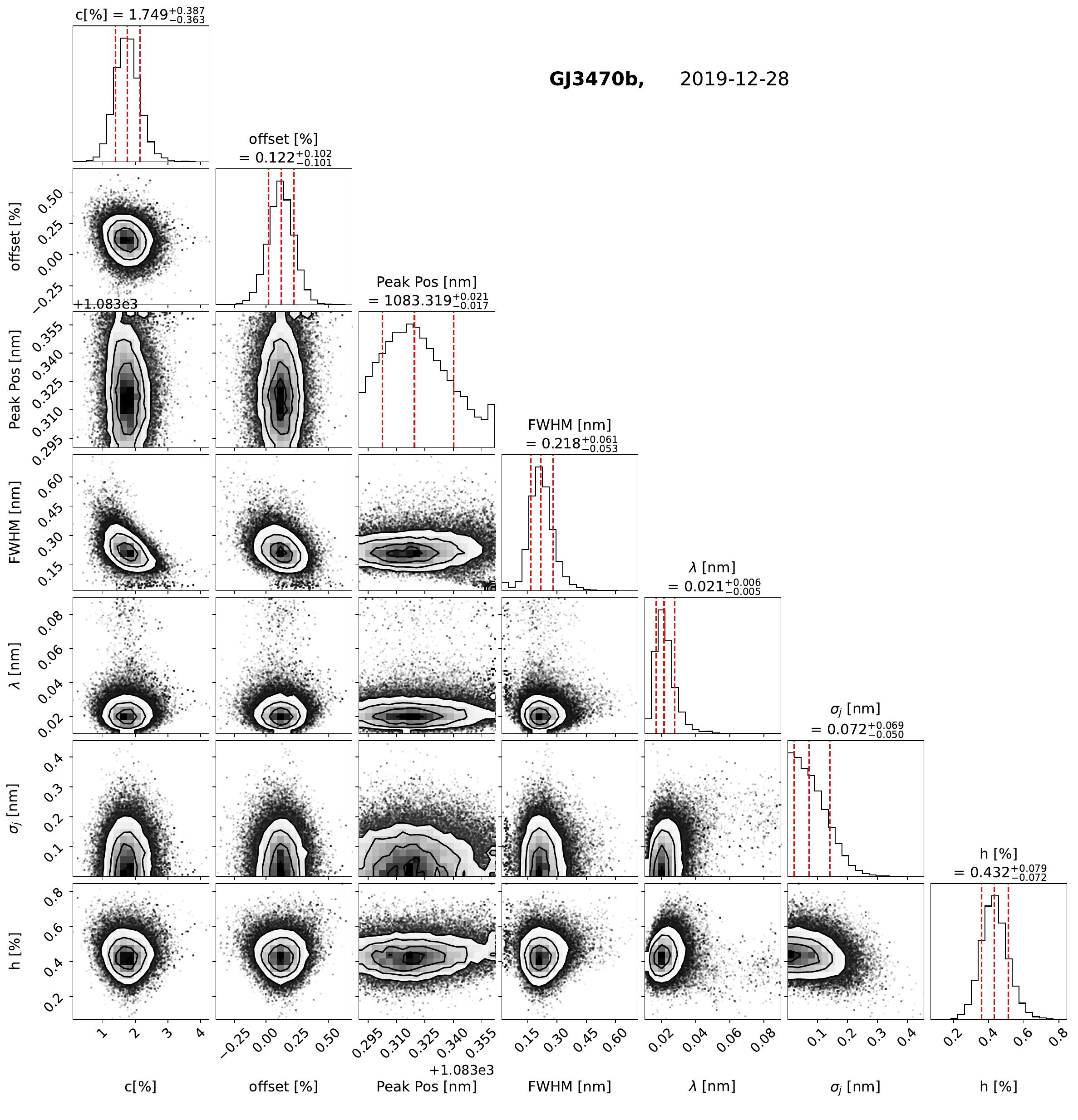}
				\caption{Example of the posterior distribution of the investigated parameters in the DE-MCMC analysis. The excess of absorption $c$ [\%], offset [\%], peak position, and FWHM correspond to the parameters we used in the Gaussian fit, while the jitter term $\sigma_\mathrm{j}$, the semi-amplitude of the correlated noise $h$, the correlation length $\lambda$ were used to parametrise the SE kernel within the GP. The larger FWHM for GJ\,3470\,b is believed to be attributed to residuals from OH line contamination.}
				\label{Cornerplots}
			\end{figure}
		\end{landscape}
		
		\begin{figure*}[h]
			\centering
			\captionof{table}{Result night  by night.}
			\begin{tabular}{c | c | c | c| c| c }
				\hline \hline
				&       Night             & Peak position & Excess of absorption $c$ & FWHM   & Significance \\
				&                          & [nm]          & [\%]                     & [nm]   & [$\sigma$]  \\
				\hline
				\multirow{3}{*}{\textbf{WASP-69b}}  & 24 July 2019                    & 1083.2974 $^{+ 0.0032 }_{ -0.0031 }$                  & 4.31 $^{+ 0.30 }_{ -0.31 }$                 & 0.0960$^{+ 0.008 }_{ -0.006 }$            & 14.2    \\
				& 09 August 2020                    & 1083.2982 $^{+ 0.0049 }_{ -0.0044 }$& 3.63 $^{+ 0.41 }_{ -0.39 }$                & 0.0948$^{+ 0.012 }_{ -0.010 }$            & 9.0     \\
				& 14 September 2022                    & 1083.3187 $^{+ 0.0069 }_{ -0.0073 }$& 3.14 $^{+ 0.54 }_{ -0.55 }$ & 0.0764 $^{+ 0.015 }_{ -0.010 }$           & 5.8     \\
				\hline
				\multirow{1}{*}{\textbf{WASP-107b}} & 07 February 2019                    & 1083.3167 $^{+ 0.0053 }_{ -0.0052 }$                 & 8.17$^{+ 0.80 }_{ -0.76 }$& 0.1176$^{+ 0.015 }_{ -0.013 }$              & 10.5    \\
				\hline
				\multirow{3}{*}{\textbf{HAT-P-11b}}& 07 July 2019                    & 1083.3189 $^{+ 0.0092 }_{ -0.0098 }$& 1.35 $^{+ 0.27 }_{ -0.27 }$             & 0.1059$^{+ 0.023 }_{ -0.018 }$             & 4.9     \\
				& 18 June 2020                                                                                    & 1083.3302 $^{+ 0.0064 }_{ -0.0066 }$& 1.38 $^{+ 0.29 }_{ -0.31 }$       & 0.0686$^{+ 0.017 }_{ -0.012 }$             & 4.5      \\
				& 19 September 2020                                                                                   & 1083.3271 $^{+ 0.0085 }_{ -0.0088 }$                  & 1.35 $^{+ 0.31 }_{ -0.32 }$           & 0.0829$^{+ 0.018 }_{ -0.015 }$             & 4.3      \\
				\hline
				\multirow{6}{*}{\textbf{GJ436b}}  & 16 April 2018                     &                                     & <0.24(0.34)                   &                                            &             \\
				& 19 February 2020                   &                                     &<0.34(0.44)                 &                                            &             \\
				& 27 February 2020                   &                                     &<0.52(0.66)               &                                            &             \\
				& 03 March2023                   &                                     & <0.38(0.52)                  &                                            &  \\
				& 11 March2023                   &                                     & <0.63(0.68)                 &                                              &          \\
				&24 December 2022                  &                                     &<0.39(0.45)                 &                                            &              \\  
				\hline
				\multirow{2}{*}{\textbf{GJ3470b}}   &28 December 2019                    & 1083.3194 $^{+ 0.0207 }_{ -0.0170 }$     & 1.75 $^{+ 0.39 }_{ -0.36 }$ & 0.2180$^{+ 0.061 }_{ -0.053 }$ & 4.7 \\
				& 23 December 2022                    &                                     & <0.10(0.22)              &                                            &           \\
				\hline
			\end{tabular}
			\tablefoot{From left to right: the investigated night, the peak position, excess of absorption, and FWHM obtained from the DE-MCMC analysis, and the significance of the detection. The upper limits are reported at 1$\sigma$ and at 2$\sigma$ (in brackets).}
			\label{tab_result_single}
			
			\vspace{3cm}
			
			\includegraphics[width=0.49\textwidth,height=6cm]{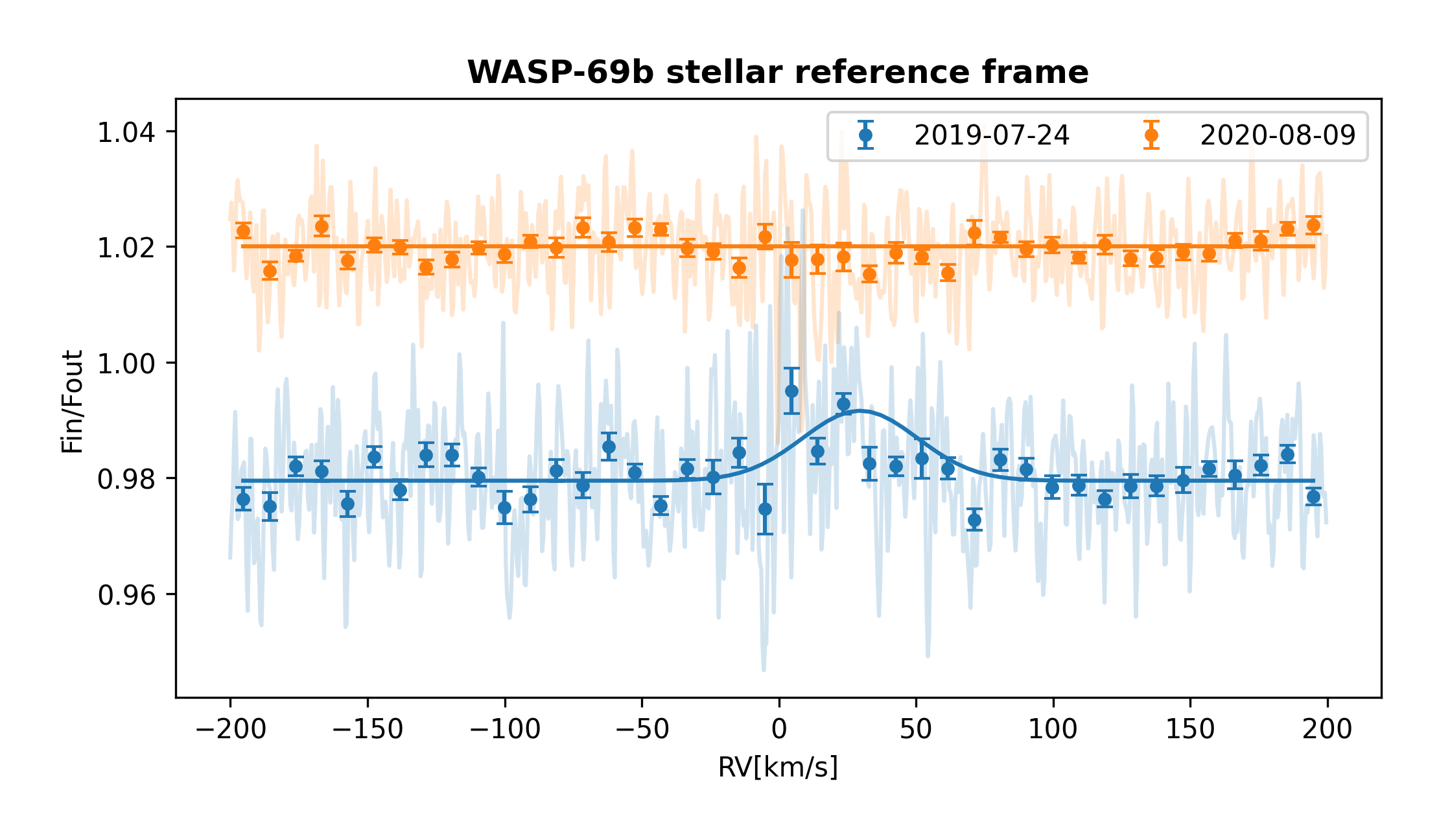}
			\includegraphics[width=0.49\textwidth,height=6cm]{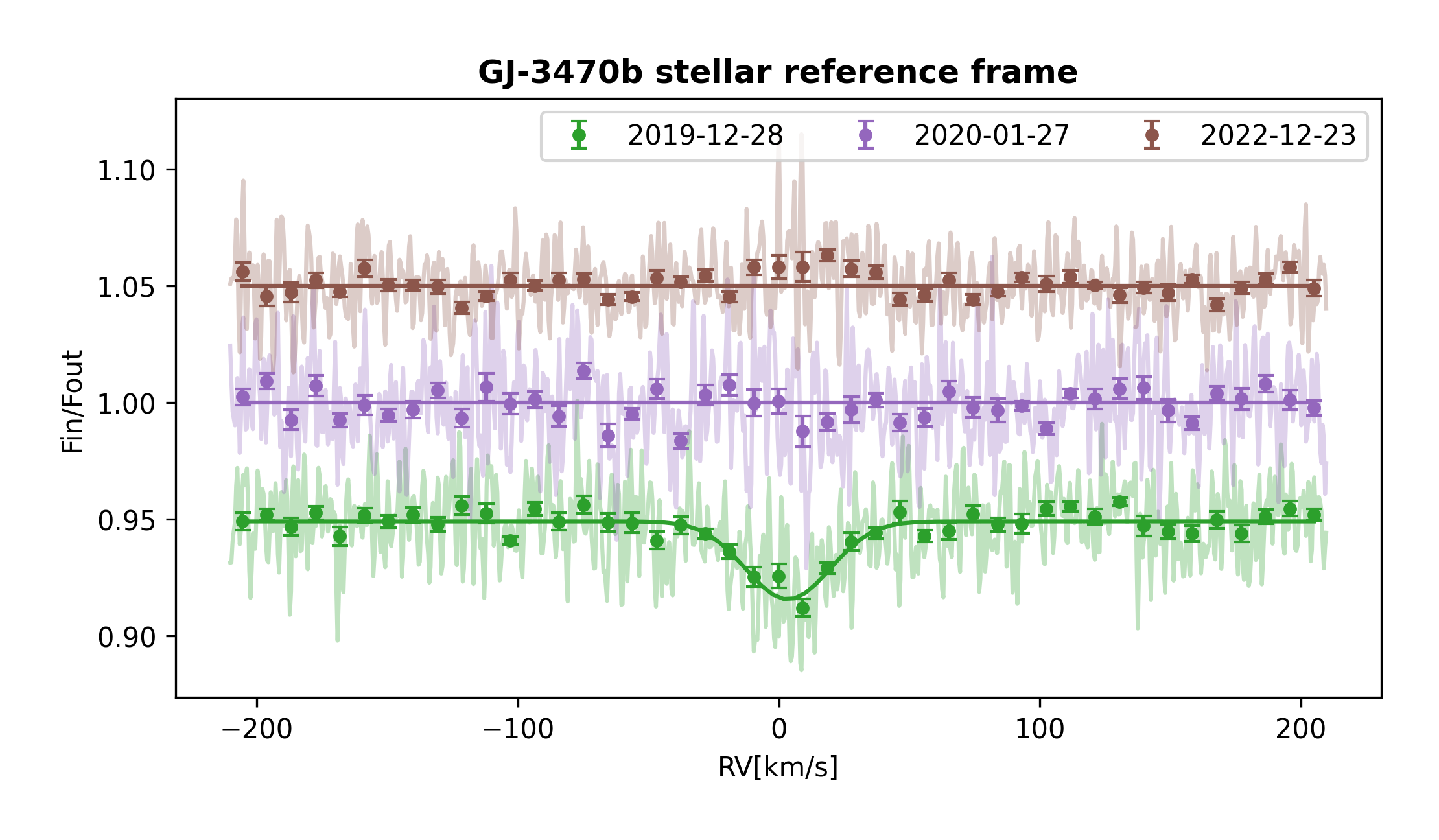}
			\captionof{figure}{Weighted average of the full in-transit spectra in the star rest frame for WASP-69b (left panel) and GJ3470b (right panel) centred on the H$\alpha$ line. Light colours indicate the not binned transmission spectra, while dots show the transmission spectra binned 20$\times$(in RV). The overplotted fit is performed on the not-binned spectra and represents the model favourite by the BIC test.}
			\label{tra_sp_halpha_W69star}
		\end{figure*}
		\FloatBarrier
		\begin{table}[]
			\caption{\logrhk values.}
			\label{logrhk_table}
			\centering
			\begin{tabular}{c| c | c   }
				\hline \hline
				& Night & \logrhk \\    
				\hline
				
				\multirow{6}{*}{\textbf{HD189733b}} & 30 May 2017 & -4.461$\pm$0.006\\
				&19 June 2017 & -4.546$\pm$0.005\\
				&20 July 2017 & -4.499$\pm$0.006\\
				&29 July 2017 & -4.500$\pm$0.005\\
				&18 October 2018 & -4.509$\pm$0.003\\
				&all        & -4.502$\pm$0.029\\
				\hline
				\multirow{6}{*}{\textbf{WASP-80b}} & 9 August 2019 & -3.836$\pm$0.015 \\
				&21 September 2019 & -3.870$\pm$0.010 \\
				& 26 June 2020 & -3.859$\pm$0.016 \\
				&17 September 2020 & -3.846$\pm$0.068 \\
				&all        & -3.853$\pm$0.037 \\
				\hline
				\multirow{4}{*}{\textbf{WASP-69b}} & 24 July 2019 & -4.561$\pm$0.008 \\
				&09 August 2020 & -4.571$\pm$0.007  \\
				&28 October 2021 & -4.608$\pm$0.006 \\
				&all        & -4.567$\pm$0.011 \\
				\hline
				\multirow{3}{*}{\textbf{WASP-107b}} & 07 February 2019 & -4.431$\pm$0.017\\
				&04 May 2019& -4.488$\pm$0.015\\
				&all        & -4.461$\pm$0.032\\
				\hline
				\multirow{4}{*}{\textbf{HAT-P-11b}} & 07 July 2019 & -4.712$\pm$0.008\\
				& 18 June 2020 & -4.675$\pm$0.005\\
				& 19 September 2020 & -4.700$\pm$0.008\\
				&all        & -4.696$\pm$0.017\\
				\hline
				\multirow{6}{*}{\textbf{GJ3470b}} & 13 January 2018 & -4.814$\pm$0.092\\
				&04 February 2019 & -4.787$\pm$0.078\\
				&28 December 2019  & -4.795$\pm$0.019\\
				&27 January 2020 & -4.815$\pm$0.019\\
				&23 December 2022 & -4.842$\pm$0.013\\
				&all        & -4.809$\pm$0.054\\
				\hline
				\multirow{6}{*}{\textbf{GJ436b}}      & 16 April 2018 & -5.223$\pm$0.028\\
				& 19 February 2020 & -5.183$\pm$0.010\\
				& 27 February 2020 & -5.173$\pm$0.018\\
				& 11 March2023 & -5.182$\pm$0.015\\
				& 24 December 2022 & -5.199$\pm$0.015\\
				& all        & -5.191$\pm$0.024\\
				\hline
				\multirow{1}{*}{\textbf{HAT-P-3b}}    &29 January 2020  & -4.896$\pm$0.128\\
				\hline
			\end{tabular}
			\tablefoot{ The \logrhk measurements have been obtained from the CaII H$\&$K lines through the offline version of HARPS-N DRS available through the Yabi web application \citep{yabi}.[$^{\mathrm{o}}$] For GJ436 and GJ3470, which have an index colour B-V$>$1.1, we applied the \citet{suarez2015} formalism.  }
		\end{table}
	\end{appendix}

\end{document}